\documentclass{IEEEoj}
\usepackage{cite}
\usepackage{amsmath,amssymb,amsfonts}
\usepackage{algorithmic}
\usepackage{graphicx,color}
\usepackage{textcomp}
\usepackage{multirow}
\usepackage{adforn}
\usepackage{fontawesome}
\usepackage{tikz}
\usetikzlibrary{arrows.meta,shapes,positioning,shadows.blur,trees}

\usepackage{hyperref}
\hypersetup{%
  colorlinks=true,
  linkcolor = {blue!70!black},
  urlcolor  = blue,
  citecolor = {blue!70!black},
  anchorcolor = blue,
  linkbordercolor={blue!30!black},
  pdfborderstyle={/S/U/W 1}
}

\definecolor{maroon}{RGB}{186,0,0}
\definecolor{pembe}{RGB}{96,26,149}
\definecolor{mavi}{RGB}{46,76,255}
\definecolor{haki}{RGB}{38,99,33}

\def\BibTeX{{\rm B\kern-.05em{\sc i\kern-.025em b}\kern-.08em
    T\kern-.1667em\lower.7ex\hbox{E}\kern-.125emX}}
\AtBeginDocument{\definecolor{ojcolor}{cmyk}{0.93,0.59,0.15,0.02}}
\def\OJlogo{\vspace{-10pt}\includegraphics[height=20pt]{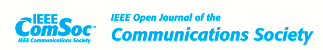}}

\begin{document}
\receiveddate{15 December, 2023}
\reviseddate{13 January, 2024}
\accepteddate{01 February, 2024}
\doiinfo{10.1109/OJCOMS.2024.3362271}

\title{At the Dawn of Generative AI Era: \\ An Exploratory Survey of New Frontiers in 6G Wireless Intelligence}

\title{At the Dawn of Generative AI Era: \\ A Tutorial-cum-Survey on New Frontiers in 6G Wireless Intelligence}

\author{ABDULKADIR CELIK, SENIOR MEMBER, IEEE, \\ AND AHMED M. ELTAWIL,
SENIOR MEMBER, IEEE}
\affil{Computer, Electrical and Mathematical Sciences and Engineering (CEMSE) Division,\\ King Abdullah University of Science and Technology (KAUST),  Thuwal, KSA 23955-6900.}
\corresp{CORRESPONDING AUTHOR: Abdulkadir Celik (e-mail: abdulkadir.celik@kaust.edu.sa).}

\authornote{This work was supported by the Office of Sponsored Research (OSR) at King Abdullah University of Science and Technology (KAUST).}

\markboth{At the Dawn of Generative AI Era: A Tutorial-cum-Survey on New Frontiers in 6G Wireless Intelligence}{Celik \textit{et al.}}

\begin{abstract}
As we transition from the 5G epoch, a new horizon beckons with the advent of 6G, seeking a profound fusion with novel communication paradigms and emerging technological trends, bringing once-futuristic visions to life along with added technical intricacies. Although analytical models lay the foundations and offer systematic insights, we have recently witnessed a noticeable surge in research suggesting machine learning (ML) and artificial intelligence (AI) can efficiently deal with complex problems by complementing or replacing model-based approaches. The majority of data-driven wireless research leans heavily on discriminative AI (DAI) that requires vast real-world datasets. Unlike the DAI, Generative AI (GenAI) pertains to generative models (GMs) capable of discerning the underlying data distribution, patterns, and features of the input data. This makes GenAI a crucial asset in wireless domain wherein real-world data is often scarce, incomplete, costly to acquire, and hard to model or comprehend. With these appealing attributes, GenAI can replace or supplement DAI methods in various capacities. \\

Accordingly, this combined tutorial-survey paper commences with preliminaries of 6G and wireless intelligence by outlining candidate 6G applications and services, presenting a taxonomy of state-of-the-art DAI models, exemplifying prominent DAI use cases, and elucidating the multifaceted ways through which GenAI enhances DAI. Subsequently, we present a tutorial on GMs by spotlighting seminal examples such as generative adversarial networks, variational autoencoders, flow-based GMs, diffusion-based GMs, generative transformers, large language models, to name a few. Contrary to the prevailing belief that GenAI is a nascent trend, our exhaustive review of approximately 120 technical papers demonstrates the scope of research across core wireless research areas, including physical layer design; network optimization, organization, and management; network traffic analytics; cross-layer network security; and localization \& positioning. Furthermore, we outline the central role of GMs in pioneering areas of 6G network research, including semantic communications, integrated sensing and communications, THz communications, extremely large antenna arrays, near-field communications, digital twins, AI-generated content services, mobile edge computing and edge AI, adversarial ML, and trustworthy AI. Lastly, we shed light on the multifarious challenges ahead, suggesting potential strategies and promising remedies. Given its depth and breadth, we are confident that this tutorial-cum-survey will serve as a pivotal reference for researchers and professionals delving into this dynamic and promising domain.
\end{abstract}



\maketitle

\begin{table*}
\centering
\caption{List of Abbreviations}
\label{table:abbreviations}
\resizebox{1\textwidth}{!}{%
\begin{tabular}{llll}
\hline
\textbf{Abbreviation} & \textbf{Full Form} & \textbf{Abbreviation} & \textbf{Full Form} \\
\hline
AI & Artificial Intelligence & LIS & Large Intelligent Surfaces \\
AIGC & AI-Generated Content & LLMs & Large Language Models \\
AML & Adversarial ML & Li-Fi & Light-Fidelity \\
AAEs & Adversarial Autoencoders & LoS & Line-Of-Sight \\
AE & Autoencoders & LSTM & Long Short-Term Memory \\
APs & Access Points & LTE & Long-Term Evolution \\
AR & Augmented Reality & MEC & Mobile Edge Computing \\
CIR & Channel Impulse Response & ML & Machine Learning \\
CNNs & Convolutional Neural Networks & MLPs & Multilayer Perceptrons \\
CR & Cognitive Radio & mMIMO & Massive Multi-Input Multi-Output \\
CSI & Channel State Information & mMTC & Massive Machine Type Of Communications \\
DAI & Discriminative AI & mmWave & Millimeter-Wave \\
DBNs & Deep Belief Networks & NFC & Near-Field Communication \\
DDPG & Deep Deterministic Policy Gradient & NIDS & Network-Based Intrusion Detection Systems \\
DGMs & Diffusion-Based Generative Models & NLoS & Non-Line-Of-Sight \\
DL & Deep Learning & NTNs & Non-Terrestrial Networks \\
DNN & Deep Neural Networks & OCC & Optical Camera Communications \\
DQN & Deep Q-Learning & OWC & Optical Wireless Communications \\
DRL & Deep Reinforcement Learning & PDF & Probability Density Function \\
EAI & Edge AI & QoE & Quality-Of-User-Experience \\
E2E & End-To-End & QoS & Quality Of Service \\
ELAA & Extremely Large Antenna Arrays & RBMs & Restricted Boltzmann Machines \\
ELBO & Evidence Lower Bound & RF & Radio Frequency \\
eMBB & Enhanced Mobile Broadband & RIS & Reconfigurable Intelligent Surfaces \\
FNNs & Feedforward Neural Networks & RL & Reinforcement Learning \\
FSO & Free Space Optics & RSSI & Received Signal Strength Indicator \\
GenAI & Generative AI & RNNs & Recurrent Neural Networks \\
GAN & Generative Adversarial Network & SL & Supervised Learning \\
GAMs & Generative Autoregressive Models & SON & Self-Organizing Network \\
GPT & Generative Pretrained Transformer & SVM & Support Vector Machines \\
GM & Generative Model & TAI & Trustworthy AI \\
GNSS & Global Navigation Satellite System & THz & Terahertz \\
GRUs & Gated Recurrent Units & UAVs & Unmanned Aerial Vehicles \\
GTMs & Generative Transformer Models & UEs & User Equipments \\
HAPs & High-Altitude Platforms & URLLC & Ultra-Reliable Low-Latency Communications \\
I/Q & In-Phase And Quadrature & USL & Unsupervised Learning \\
IoT & Internet Of Things & VAEs & Variational Autoencoders \\
ISAC & Integrated Sensing And Communications & VLC & Visible Light Communication \\
ISI & Inter-Symbol Interference & VR & Virtual Reality \\
JSCC & Joint Source-Channel Coding & WLAN & Wireless Local Area Network \\
K-NN & K-Nearest Neighbors & 5G & Fifth Generation \\
KL-D & Kullback-Leibler Divergence & 6G & Sixth Generation \\
KPIs & Key Performance Indicators &  &  \\
\hline
\end{tabular}
}
\end{table*}

\section{INTRODUCTION} 
\label{sec:introduction}
\IEEEPARstart{T}{he} trajectory of wireless networks has been marked by leaps of innovation, with each generation heralding transformative waves that recalibrate our engagement with digital realms. As we transition from the 5G epoch, characterized by impressive data rates and robust connectivity \cite{akyildiz20195g}, a new horizon beckons with the advent of 6G.  While 5G was celebrated for its massive machine type of communications (mMTC), ultra-reliable low-latency communications (URLLC), and enhanced mobile broadband (eMBB), 6G is not just an enhanced iteration of its forerunner; it represents a paradigm shift, geared to reframe the foundations of wireless connectivity \cite{dang2020what}. That is, the envisaged 6G doesn't just aim to elevate key performance indicators (KPIs) but seeks a profound fusion with novel communication paradigms and emerging technological trajectories, bringing once-futuristic visions to life \cite{zhang20206G}. Therefore, 6G is not expected to be just another step in the evolutionary ladder of wireless technology; but an ambitious leap towards an interconnected, intelligent, and immersive future. Below, we spotlight groundbreaking technologies that set the stage for an era where communication is not just about connectivity, but about fostering deeper, more meaningful interactions in an increasingly digital world.

Delving into 6G's prospective technological marvels, semantic communications emerge as a cornerstone by aiming at transcending the conventional data transfer paradigm to enable networks that comprehend and process content semantics, ensuring communication that's not just ultra-speed but also contextually intelligent \cite{luo2022semantic}. In parallel with aerospace companies' (e.g., SpaceX, OneWeb, etc.) large scale deployment of satellite internet constellations, seamless integration of non-terrestrial networks (NTNs) with terrestrial infrastructure becomes paramount, promising unmatched global coverage and even stretching connectivity's frontiers beyond earth \cite{giordani2020toward}. At this very point, as fiber links submerged in oceans  establish transcontinental connectivity, free space optical (FSO) links assume critical importance \cite{khalighi2014survey}, endowing NTN platforms with high-capacity feeder services and cross-links, thus bridging the digital divide to uplift under-served communities in under-developed regions.

Concurrently, integrated sensing and communications (ISAC) chart a transformative path for 6G networks, where multi-modal sensory data from various sources (e.g., GPS, radar, lidar, and cameras) converge to enhance communication quality and birth diverse applications \cite{demirhan2023integrated}. This sensor fusion can enrich networks with a holistic environmental perception, catalyzing smarter, adaptive responses. At the heart of 6G evolution lies the synergy between high-frequency spectrum (e.g., mmWave \cite{AbdallahCME22} and terahertz \cite{sarieddeen2020next}), reconfigurable intelligent surfaces (RIS) \cite{Abdallah2023RIS}, and  extremely large antenna arrays (ELAA) \cite{bjornson2021rethinking} to intensify connection density and efficacy. As we gear up in the spectrum chart and condense more antennas per unit area, we venture into the near-field communication (NFC) domain \cite{cui2022near}, gaining granular control over beam direction, shape, and structure. Holographic communication leverages this control, especially in the near-field, to create detailed ``hologram'' of electromagnetic fields \cite{zhang2020holographic}. It employs the massive array to generate beams with unprecedented spatial resolution, reshaping them in complex ways, similar to how optical holography manipulates light. This nexus of ELAA, NFC dynamics, and holographic beamforming fortifies advanced wireless communication's vanguard, yet exacerbating the intricacies of channel estimation and beamforming. The performance boost offered by these cutting-edge technologies will also enable disruptive 6G services and applications such as large scale IoT and digital twins, edge AI and mobile computing, AI-generated contents, metaverse, intelligent services, etc.  

{In addition to hosting groundbreaking advancements, 6G entails multi-faceted environments, heterogeneous network architectures, as well as dynamic and diverse user behaviors. These complexities challenge the capabilities of conventional analytical models, which offer a systematic approaches grounded in well-defined parameters and equations that may not wholly capture the intricacies of real-world scenarios \cite{zappone2019wireless}.  There's a risk these models might render an oversimplified picture or lag behind the swiftly evolving network conditions, compromising their reliability in real-world performance prediction. This gap underscores the growing importance of machine learning (ML) and artificial intelligence (AI)\footnote{AI is an overarching concept that encompasses various machine learning methods, including (un)supervised learning, reinforcement learning, deep learning, generative models, etc. Therefore, throughout the paper, we will prefer `AI' over `ML' when discussing the broader scope of intelligent systems and algorithms.}. The recent decade has witnessed a surge in publications unveiling the merits of data-driven techniques, suggesting their potential to either augment or even replace model-based approaches \cite{buzzi2021_survey}. While analytical models lay the foundations and offer systematic insights, AI models can leverage real-life datasets for evolving to be adaptive and precise in capturing and responding to real-world complexities.}

{The majority of contemporary data-driven wireless research leans heavily on discriminative AI (DAI) models, which prioritize discerning differences between various data classes or categories. DAI models predominantly harness three foundational learning paradigms: 1) \textit{supervised learning} techniques are trained using labeled data to understand and predict the predefined categories, 2) \textit{unsupervised learning} seeks patterns in unlabeled data to  identify clusters or groups and subsequently applying discriminative techniques to classify future data points, and 3) \textit{reinforcement learning} centers on agents that seek optimal decisions by distinguishing between a plethora of actions depending on states and interaction with the environment. Despite the versatility of these learning paradigms and their variations, securing vast real-world training datasets is often a costly endeavor in terms of both time and computational resources. Beyond depending on dataset to perform classical ML tasks (e.g., classification, regression, clustering, pattern search, dimensionality reduction, etc.), DAI models are not fully capable of interpreting the data and can sometimes miss nuanced patterns/states, posing challenges in complex, real-world scenarios. Contrary to the DAI, Generative AI (GenAI) pertains to generative models (GMs) capable of discerning the underlying data distribution, patterns, and features of the input data \cite{goodfellow2014generative}. Subsequently, GMs can model the underlying data distribution to generate novel data instances--that were not present in the original dataset--reminiscent of the training data and examples. This becomes a pivotal asset in wireless domain wherein real-world data is often scarce, incomplete, or costly to gather, making them indispensable for  data augmentation, data imputation,  disentanglement, anomaly detection, and so on. 
}

Contrary to the prevailing belief that GenAI is a nascent trend, our meticulous review of around 120 technical works underscores the depth and breadth of research across primary wireless research trajectories. Notwithstanding, it's undeniable that the full spotlight on GenAI was only realized post the release of large language model-based chatbots by tech giants such as OpenAI's ChatGPT in Nov. 2022, Google's BARD in Mar. 2023, and Microsoft's Bing Chat in Feb. 2023. These breakthroughs significantly piqued interest and aroused intense curiosity in both the industrial and academic spheres \footnote{An inspiring  vision blog is shared here: https://petarpopovski-51271.medium.com/communication-engineering-in-the-era-of-generative-ai-703f44211933}. Given this renewed interest, a flourishing era in GenAI-driven wireless communication and networking research is palpably on the horizon. Owing to its comprehensive nature and rich content, we are confident that this survey will serve as a pivotal reference for researchers and professionals delving into this dynamic and promising domain.

\begin{figure*}[t!]
\frame{\includegraphics[width=2\columnwidth]{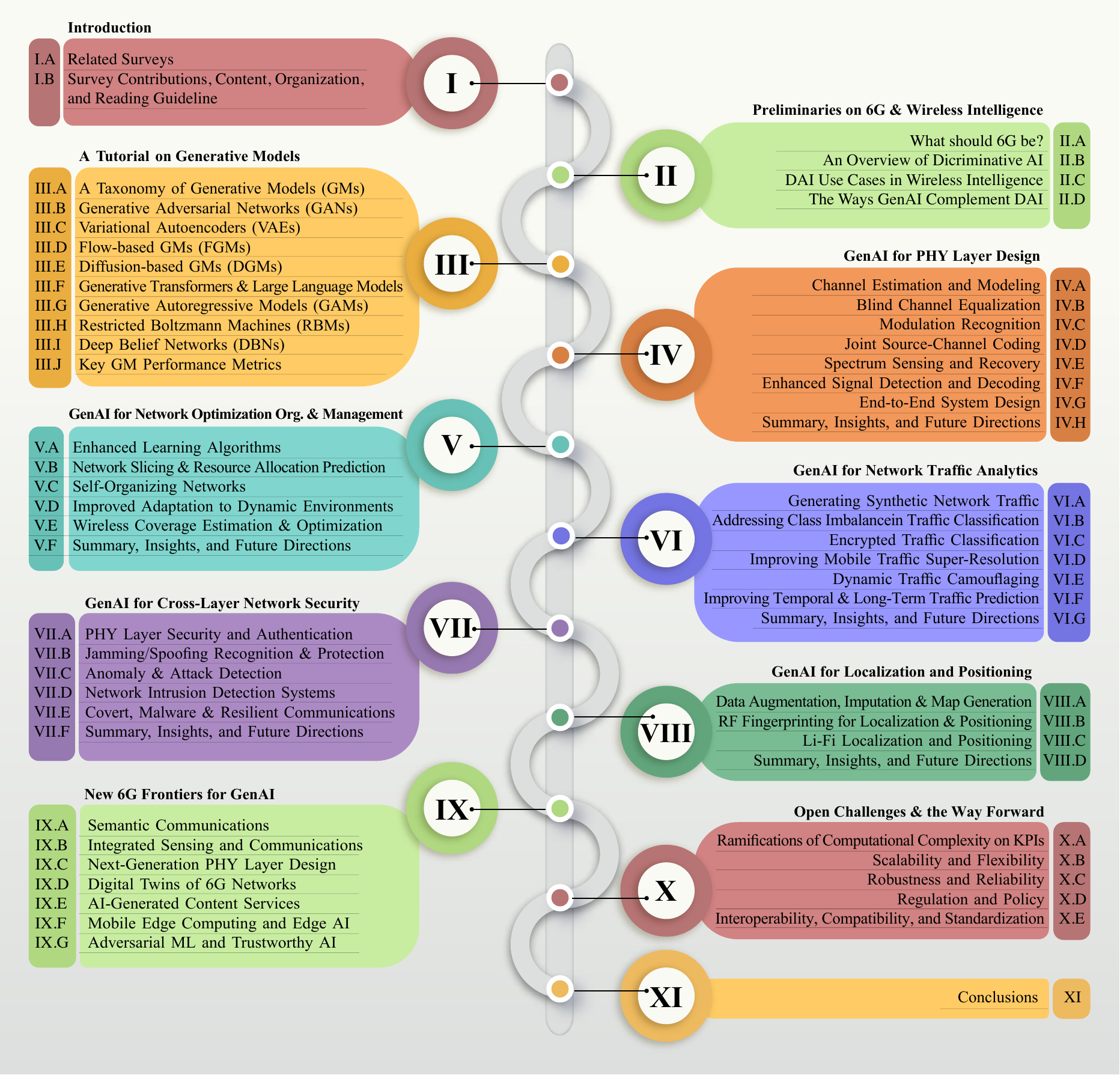}}
    \caption{Survey Content and Organization.}
    \label{fig:org}
\end{figure*}

\subsection{RELATED SURVEYS}
Generative models (GMs) have recently gained significant attention in the realm of wireless communications and networks. Over a very short course of time, a plethora of studies have been conducted to unveil the potential of these models across various domains, which has recently been partially reviewed by following works: 

Yang et al. provides a comprehensive perspective on the opportunities and challenges inherent in leveraging GMs for wireless channel modeling \cite{yang2019generative}. Traditional channel modeling approaches necessitate intricate domain-specific expertise and tend to be technically challenging. To address this issue, they introduced a Generative Adversarial Network (GAN)-based approach that endeavors to model wireless channels autonomously by training on raw data. 

Liu et. al. offers a comprehensive overview of harnessing GenAI for wireless networks, with a special focus on three representative GMs; GANs, variational autoencoders (VAEs), and diffusion-based generative models (DGMs) \cite{liu2023deep}. Moreover, a new GenAI-based framework for wireless network management is introduced, critiquing traditional methods and advocating for GM solutions. A case study harnesses the DGMs for optimizing contracts in mobile AI-Generated Content (AIGC) services. Another review on AIGC is presented in \cite{xu2023unleashing}, which underscores the emergence of mobile AIGC networks, focusing on real-time, personalized AIGC services that prioritize user privacy. A detailed exploration of the infrastructure, technologies, and challenges pertinent to this domain is presented.

Moreover, Navidan et. al. reviews how GMs can benefit multiple aspects of computer and communication networks, including mobile networks, network analysis, the Internet of Things (IoT), physical (PHY) layer, and cybersecurity \cite{NAVIDAN2021108149}. Given the challenges faced by DAI models in the industrial IoT paradigm, \cite{Suparna2022survey} suggests the adoption of GMs and reviews the state-of-the-art GMs, categorizing them based on their relevance to IIoT: anomaly detection, trust-boundary protection, network traffic prediction, and platform monitoring. Karapantelakis \cite{Karapantelakis2023survey} offers a detailed review of  GenAI's application in mobile telecommunications networks by categorizing literature based on GM types, its purpose, and the targeted mobile network component. A pivotal discussion in this study is the examination of the current state of AI integration in major telecom standardization entities.

In the realm of privacy and security, the work by Cai et al. \cite{cai2023survey} recognizes the unprecedented potential of GANs, primarily due to their capacity to generate plausible data. The authors lament the absence of comprehensive surveys in this space and embark on a systematic analysis, highlighting the strengths, shortcomings, and future trajectories of GANs. Similarly,
Ayanoglu et. al. \cite{Ayanoglu2023survey} examines the potential of GANs in handling tasks like spectrum sharing, anomaly detection, and security attack mitigation. The study underscores the advantages of GANs, from synthesizing field data to recovering corrupted bits in the spectrum. Cybersecurity, another related domain of interest, has seen the rapid adoption of GANs, especially for handling tasks related to imbalanced datasets. Dunmore \cite{Dunmore2023survey} reviews GANs' role in intrusion detection systems, highlighting their capabilities in generating adversarial examples, manipulating data semantics, and augmenting rare class data. Lastly, adversarial samples in machine learning, particularly those affecting wireless and mobile systems, have garnered significant attention. Liu's survey \cite{liu2022survey} provides a comprehensive review on adversarial ML in these systems, from the physical layer up to the application layer. The study underscores the state-of-the-art adversarial ML techniques, their impact, and the inherent challenges and opportunities in the domain


\begin{table}[]
\caption{Comparison with Related Surveys}
\label{tab:related}
\resizebox{\columnwidth}{!}{%
\begin{tabular}{|c|c|c|l|}
\hline
\textbf{Ref.}                                                                              & \textbf{Style} & \textbf{GMs}                                                                                & \multicolumn{1}{c|}{\textbf{Content and Applications}}                                                                                                                                                               \\ \hline
\cite{yang2019generative}                                                 & Magazine       & GAN                                                                                         & Channel Modeling                                                                                                                                                                                         \\ \hline
\cite{liu2023deep}                                                        & Tutorial       & \begin{tabular}[c]{@{}c@{}}GAN\\ VAE\\ DGM\end{tabular}                                     & AI-Generated Content Services                                                                                                                                                                            \\ \hline
\cite{xu2023unleashing}                                                   & Survey         & \begin{tabular}[c]{@{}c@{}}GAN\\ VAE\\ DGM\\ FGM\end{tabular}                               & AI-Generated Content Services                                                                                                                                                                            \\ \hline
\cite{NAVIDAN2021108149}                                                  & Survey         & GAN                                                                                         & Networking, Cybersecurity, IoT                                                                                                                                                                           \\ \hline
\cite{Suparna2022survey}                                                  & Survey         & GAM                                                                                         & Industrial IoT                                                                                                                                                                                           \\ \hline
\cite{Karapantelakis2023survey}                                           & Survey         & GAN                                                                                         & Mobile Networks                                                                                                                                                                                          \\ \hline
\cite{cai2023survey}                                                      & Survey         & GAN                                                                                         & Security and Privacy                                                                                                                                                                                     \\ \hline
\cite{Ayanoglu2023survey}                                                 & Survey         & GAN                                                                                         & Cognitive Networks                                                                                                                                                                                       \\ \hline
\cite{Dunmore2023survey}                                                  & Survey         & GAN                                                                                         & Cybersecurity and Intrusion Detection                                                                                                                                                                    \\ \hline
\cite{liu2022survey}                                                      & Survey         & GAN                                                                                         & Adversarial ML                                                                                                                                                                                           \\ \hline
\multicolumn{1}{|l|}{\multirow{2}{*}{\begin{tabular}[c]{@{}l@{}}This\\ Work\end{tabular}}} & Tutorial       & \begin{tabular}[c]{@{}c@{}}GAN\\ VAE\\ FGM\\ DGM\\ GTM\\ LLM\\ GAM\\ RBM\\ DBN\end{tabular} & \begin{tabular}[c]{@{}l@{}}The tutorial covers a wide range of GMs \\ to dissect the essentials, concise mathematical \\ underpinnings, virtues, drawbacks, and variants \\ of each model. Essential performance metrics are \\ provided for benchmarking GMs and the \\ ways how GMs can complement DAI techniques \\ are presented.

\end{tabular}  \\ \cline{2-4} 
\multicolumn{1}{|l|}{}                                                                     & Survey         & \begin{tabular}[c]{@{}c@{}}GAN\\ VAE\\ DGM\\ DBN\end{tabular}                               & \begin{tabular}[c]{@{}l@{}}Physical Layer Design\\ Network Optimization, Organization, Management\\ Network Traffic Analytics\\ Cross-Layer Network Security\\ Localization and Positioning\end{tabular} \\ \hline
\end{tabular}%
}
\end{table}

\begin{figure*}[t]
\hspace{-13.5cm}   
\centering
    \resizebox{3.5\columnwidth}{!}{\input{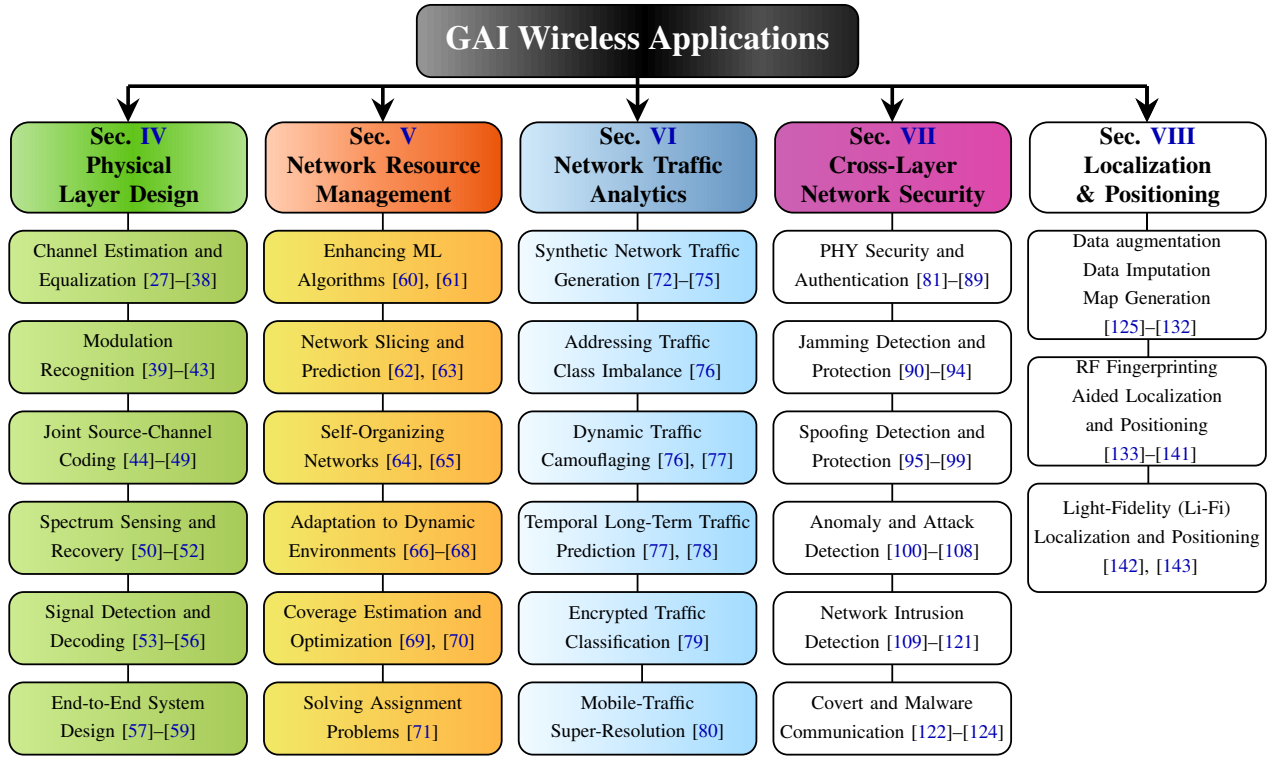}}
    \caption{GenAI applications in wireless communications and networks.}
    \label{fig:GenAI_apps}
\end{figure*}
\subsection{{SURVEY CONTRIBUTIONS, CONTENT, ORGANIZATION, AND READING GUIDELINE}}

{As shown in Table \ref{tab:related}, a recurring limitation of previous surveys is their narrow focus on specific GM types and GenAI application domains in the realm of wireless communications and networks. This tutorial-cum-survey is distinct due to its comprehensiveness, self-containment, rich-content, and key insights, as evidenced by the structure and content presentation in the sequel.}

For the sake of self-containment and providing readers with the necessary backgrounds, we start with preliminaries on 6G networks and wireless intelligence in Sec. \ref{sec:6G_prelims}. {A concise overview of emerging trends in 6G communications and networking technologies, along with their diverse range of applications and services is presented in Sec. \ref{sec:6G_prelims}-\ref{sec:6G_brief}. Then, we step in wireless intelligence domain in Sec.  \ref{sec:6G_prelims}-\ref{sec:DAI} by providing a taxonomy of DAI models and briefly discussing their notable enhancements, especially in complex network environments, by addressing gaps in analytical methods in Sec.  \ref{sec:6G_prelims}-\ref{sec:DAI_usecases}. Finally, Sec.  \ref{sec:6G_prelims}-\ref{sec:GM_ways} discusses how GenAI acts as a perfect counterbalance to augment the inherent challenges of DAI and elucidates ways GMs can boost DAI performance.}

In Sec. \ref{sec:GAI_Models}, we present a brief tutorial on GMs by first providing a taxonomy of GMs in Sec. \ref{sec:GAI_Models}-\ref{sec:GM_taxo}, where GMs are categorized based on how they model the underlying distribution of data. In the order of prominence and wide use, following prime examples of GMs are presented: 
generative adversarial networks in Sec. \ref{sec:GAI_Models}-\ref{sec:GAN},
variational autoencoders in Sec. \ref{sec:GAI_Models}-\ref{sec:VAE},
flow-based GMs in Sec. \ref{sec:GAI_Models}-\ref{sec:flow},
diffusion-based GMs in Sec. \ref{sec:GAI_Models}-\ref{sec:diffusion},
generative transformers models and large language models in Sec. \ref{sec:GAI_Models}-\ref{sec:GTM}, generative autoregressive models  in Sec. \ref{sec:GAI_Models}-\ref{sec:GAM},
restricted Boltzmann machines in \ref{sec:GAI_Models}-Sec. \ref{sec:RBM},
and deep belief networks in Sec. \ref{sec:GAI_Models}-\ref{sec:DBN}. Throughout these subsections, we dissect the essentials, concise mathematical underpinnings, virtues, drawbacks, and variants of each model. Concluding this GM tutorial, we suggest essential performance metrics for benchmarking these GMs. Even though it is non-trivial to provide a quantitative or qualitative comparison among the wide variety of GMs, we conclude this GM tutorial by providing a set of key performance metrics that should be taken into account for comparison purposes.  

Given that we set the stage with all the necessary background information, Sec. \ref{sec:PHY} - Sec. \ref{sec:localization} elucidate potential GM applications in various topics in wireless communication and networks domain. As illustrated in Fig. \ref{fig:GenAI_apps}, we delve into prominent topics, such as physical layer design in Sec. \ref{sec:PHY}; network optimization, organization, and management in Sec. \ref{sec:network}; network traffic analytics in Sec. \ref{sec:traffic}; cross-layer network security in Sec. \ref{sec:security}; and localization and positioning in Sec. \ref{sec:localization}. For readers' convenience, each section is supplemented with visuals displaying sub-topics, pertinent references, and utilized GMs for enhanced comprehension. All sections are also concluded with summary, insights, and future research directions in the corresponding domains. Sec. \ref{sec:PHY} - Sec. \ref{sec:localization} review around 120 technical papers; 91 of which were published in last 5 years with the following distribution 15 in 2018, 21 in 2019, 15 in 2020, 17 in 2021, 16 in 2022, and 8 in 2023, respectively.

{Accordingly, Sec. \ref{sec:6GMfrontiers} delineates the strategic importance of GMs for new frontiers of 6G network research such as semantic communications in Sec. \ref{sec:6GMfrontiers}-\ref{sec:semantic_front}; integrated sensing and communications in Sec. \ref{sec:6GMfrontiers}-\ref{sec:isac_front}; next-generation PHY layer design (e.g., THz communications, ELAA, NFC, holographic beamforming) in   Sec. \ref{sec:6GMfrontiers}-\ref{sec:PHY_front}, digital twins in Sec. \ref{sec:6GMfrontiers}-\ref{sec:twin_front}; AI-generated content in Sec. \ref{sec:6GMfrontiers}-\ref{sec:AIGC_front}; 
mobile edge computing and edge AI in Sec. \ref{sec:6GMfrontiers}-\ref{sec:MEC_front}; and adversarial ML and trustworthy AI in Sec. \ref{sec:6GMfrontiers}-\ref{sec:trust_front}.} Nonetheless, the way forward is not without hurdles at all. Thus, Sec. \ref{sec:GM_challenges} discusses multi-faceted open research challenges and point out strategies and candidate technologies as a remedy. In Sec. \ref{sec:GM_challenges}-\ref{sec:complexity}, we present ramifications of computational complexity on 6G KPIs, which is followed by GM-related scalability and flexibility issues in Sec. \ref{sec:GM_challenges}-\ref{sec:scalability}, as 6G networks are expected to accommodate an exponential growth in connected devices, services, and types of data. 
Subsequently, Sec. \ref{sec:GM_challenges}-\ref{sec:robustness} focuses on GMs' robustness against various forms of errors and reliability in their performance at different conditions. Then, Sec.  \ref{sec:GM_challenges}-\ref{sec:compatibility} sheds lights on interoperability, compatibility, and standardization aspects of GMs to make them work smoothly across diverse devices, vendors, and networks. Thereafter, Sec. \ref{sec:GM_challenges}-\ref{sec:regulation} concentrates on rules, regulations, and policies, both existing and those that are yet to be established. At the end, we conclude the paper in Sec. \ref{sec:conclusions} with a few remarks. 

\section{PRELIMINARIES ON 6G WIRELESS INTELLIGENCE}
\label{sec:6G_prelims}
In this section, we first provide a concise overview of emerging trends in 6G communications and networking technologies, then highlight the challenges faced in the journey towards 6G realization in Sec. \ref{sec:6G_prelims}-\ref{sec:6G_brief}. For the sake of self-containment and facilitating discussions, Sec. \ref{sec:6G_prelims}-\ref{sec:DAI} provides fundamentals, taxonomy, and historical evolution of DAI techniques. In Sec. \ref{sec:6G_prelims}-\ref{sec:DAI_usecases}, we briefly explain how DAI techniques have been utilized to solve main challenges in wireless networks. Finally, we connect previous subsections with a high-level discussion to provide insights into how GenAI can replace or complement DAI techniques. Accordingly, Sec. \ref{sec:6G_prelims}-\ref{sec:GM_ways} elucidates how GenAI acts as a perfect counterbalance and explain ways generative models augment the inherent challenges of DAI.

{
\subsection{WHAT SHOULD 6G BE?}
\label{sec:6G_brief}
The evolution of 6G networks is set to revolutionize the technological landscape, offering a plethora of novel applications and services that extend well beyond the capabilities of 5G. These advancements target various sectors and use cases, integrating emerging technologies to create a more connected and efficient world.

A key feature of 6G would be the use of Terahertz (THz) communications, which provide significantly larger bandwidth and higher spatial resolution compared to current technologies \cite{sarieddeen2020next, rappaport2019_wireless}, enabling ultra-high-speed data transmission and advanced sensing applications. 
As operating at higher frequencies allows us packing more antennas into unit aperture size, Extremely Large Antenna Arrays (ELAA) and Near-Field Communications (NFC) promise enhanced spatial multiplexing and reduced interference, though they pose design and performance challenges \cite{larsson2018massive, bjornson2021rethinking, cui2022near}.

Another innovative approach is Integrated Sensing and Communication (ISAC) that combines communication and sensing within a single network infrastructure to enhance wireless communication performance and support applications in autonomous transportation and smart cities  \cite{demirhan2023integrated, zhang20206G}. Semantic communication, a shift from traditional data transmission to the exchange of meaningful information, is also crucial in 6G networks. This approach considers the context, receiver's knowledge, and data's intended purpose, making the network more efficient and intelligent \cite{luo2022semantic, yang2023semantic}. 

Non-Terrestrial Networks (NTNs), encompassing satellite systems and high-altitude platforms, aim to provide global coverage, especially in remote areas \cite{Kaleem2022enhanced,Kaleem2022uav,Misbah2023phase}. However, integrating these platforms with terrestrial networks presents challenges in network architecture, latency management, and energy sustainability \cite{zeng2016wireless, giordani2020toward, ArzykulovCNE22, ZhangCDS22, BushnaqCEAA19, BushnaqKCAA21, liou2018orbital}. Optical Wireless Communications (OWC) offers a spectrum-abundant alternative to RF communications, immune to electromagnetic interference but sensitive to atmospheric conditions and line-of-sight limitations \cite{khalighi2014survey, pathak2015visible, saeed2019optical, CelikSSAA20}.

Quantum communications and computing are expected to bolster 6G networks with secure data transmission and efficient computations. These technologies necessitate quantum-resistant cryptographic algorithms to counteract potential security threats  \cite{pirandola2019_advances, preskill2018_quantum, bernstein2017_postquantum}. In parallel, 6G will leverage advanced AI for intelligent services, enhancing user experiences in content recommendation, smart industry, and healthcare \cite{alimi2020_toward, zhang2017_deep, mobley2002_an, jiang2017_ai}.

The integration of Mobile Edge Computing (MEC) and Edge AI (EAI) is another significant development, enabling real-time, low-latency processing at the network's edge \cite{letaief2019_roadmap, zhang2021toward, wang2019_edge, shi2016_edge}. This supports applications like smart grids, advanced robotics, and immersive virtual reality (VR) and augmented reality (AR) experiences. Likewise, holographic communications, essential for real-time 3D multimedia transmission, will facilitate teleconferencing and entertainment, demanding high data rates and low latency \cite{saad2019vision, zhang2020holographic, akyildiz20195g}.

6G will also support large-scale IoT deployments and digital twins, providing real-time synchronization and improved decision-making across various sectors \cite{dang2020what, lu2021digital, CelikSE22, CelikE22, kivimaki2021edge}. The metaverse, an immersive digital universe, will benefit from 6G's capabilities, offering interactive and responsive experiences through digital twins, EAI, and holographic communications \cite{hossain2021metaverse, lu2021digital, kivimaki2021edge, gao2021holographic}.

Despite these technological advancements, 6G networks face significant challenges. Spectrum management is crucial, as the radio spectrum becomes scarce and exploration of new frequency bands like THz is necessary. Meeting diverse Quality of Service (QoS) requirements demands advanced resource management and traffic techniques. The heterogeneity of devices, technologies, and services in 6G networks requires flexible network architectures and protocols. Scalability is vital to accommodate the growing number of connected devices and services. Energy efficiency is paramount, especially for battery-powered devices in IoT deployments. Lastly, robust security and privacy protection are imperative in the face of increasing network complexity and scale. Addressing these challenges is essential for the successful implementation and operation of 6G networks, where in AI/ML can play a vital role as explained in the remainder.
}

\begin{figure*}
    \centering
\frame{\includegraphics[width=2\columnwidth]{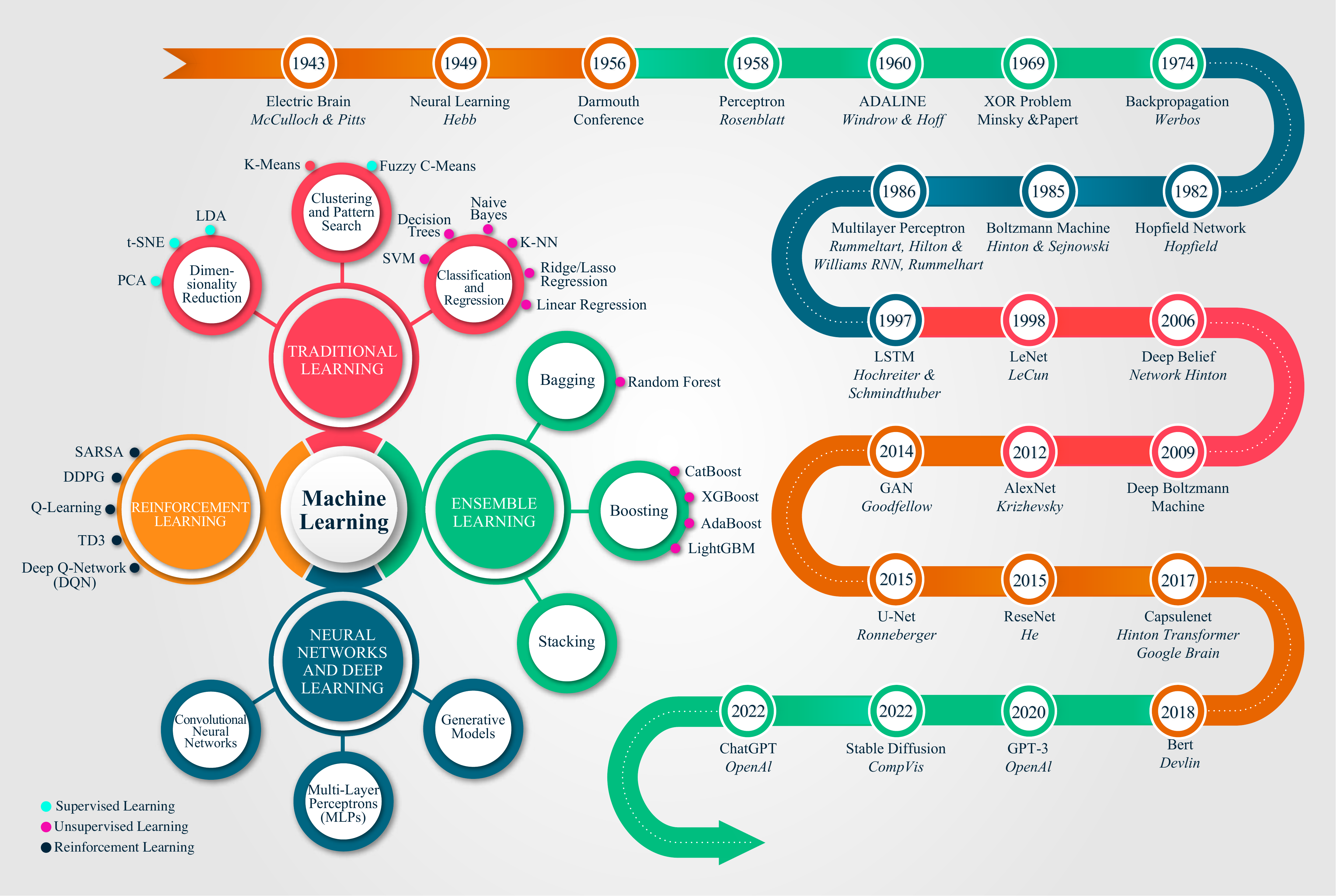}}
    \caption{The Taxonomy and Timeline of ML/AI Techniques}
    \label{fig:taxo_ml}
\end{figure*} 

\begin{table*}[t]
\centering
\caption{Comparison of Supervised, Unsupervised, and Reinforcement Learning.}
\label{tab:ML_comp}
\resizebox{\textwidth}{!}{%
\begin{tabular}{l|l|l|l|}
\cline{2-4}
\multicolumn{1}{c|}{} &
  \multicolumn{1}{c|}{\textbf{Supervised Learning}} &
  \multicolumn{1}{c|}{\textbf{Unsupervised Learning}} &
  \multicolumn{1}{c|}{\textbf{Reinforcement Learning}} \\ \hline
\multicolumn{1}{|l|}{\rotatebox[origin=c]{90}{\textbf{Pros}}} &
  \begin{tabular}[c]{@{}l@{}}- Effective for well-defined tasks with labeled data\\ - Good performance in classification and regression tasks\end{tabular} &
  \begin{tabular}[c]{@{}l@{}}- Does not require labeled data\\ - Can discover hidden patterns and structures in data\\ - Useful for data preprocessing, dimensionality reduction, and clustering\end{tabular} &
  \begin{tabular}[c]{@{}l@{}}- Suitable for sequential decision-making problems\\ - Can learn from delayed feedback\\ - Capable of dealing with dynamic environments\end{tabular} \\ \hline
\multicolumn{1}{|l|}{\rotatebox[origin=c]{90}{\textbf{Cons}}} &
  \begin{tabular}[c]{@{}l@{}}- Requires labeled data\\ - Prone to overfitting if not properly regularized\\ - May not generalize well to new tasks or domains\end{tabular} &
  \begin{tabular}[c]{@{}l@{}}- No explicit target labels, making evaluation challenging\\ - May require more complex models and training techniques\end{tabular} &
  \begin{tabular}[c]{@{}l@{}}- Can be computationally expensive\\ - Requires exploration-exploitation trade-off\\ - Sensitive to reward function design\end{tabular} \\ \hline
\multicolumn{1}{|l|}{\rotatebox[origin=c]{90}{\textbf{Prefs.}}} &
  \begin{tabular}[c]{@{}l@{}}- When labeled data is available\\ - For tasks with clear input-output relationships\\ - When the goal is prediction or classification\end{tabular} &
  \begin{tabular}[c]{@{}l@{}}- When labeled data is scarce or unavailable\\ - For tasks focused on discovering hidden patterns or relationships\\ - When the goal is data analysis or preprocessing\end{tabular} &
  \begin{tabular}[c]{@{}l@{}}- When the problem involves sequential decision-making\\ - In dynamic environments with delayed feedback\\ - When the goal is to learn an optimal policy or strategy\end{tabular} \\ \hline
\end{tabular}%
}
\end{table*}

\subsection{AN OVERVIEW OF DISCRIMINATIVE AI} 
\label{sec:DAI}

6G networks, set to embody an unparalleled complexity, render traditional model-based strategies for wireless network deployment, design, and operation increasingly insufficient. Over recent years, DAI techniques have demonstrated notable enhancements, especially in complex network environments, by addressing gaps in analytical methods. Yet, their deployment hinges on extensive real-life data, often presenting formidable challenges concerning time and cost. Furthermore, DAI is encumbered by a limited grasp on data distribution, leading to potential generalization pitfalls. In contrast, GenAI offers promising solutions, excelling in understanding the innate data distribution and proffering realistic data generation. Integrating this capability can substantially mitigate the limitations of discriminative models, making them invaluable for realizing wireless intelligence in 6G networks.

{To streamline our discourse, Fig. \ref{fig:taxo_ml} visualizes the taxonomy of ML techniques that is divided into four main categories: supervised learning, unsupervised learning, ensemble learning, and deep learning that also encapsulates generative models, which are explained in the sequel. }

\begin{itemize}
    \item[\large \adforn{71}] \textit{Supervised Learning} (SL) trains models based on a dataset containing input-output pairs, where the output (or target) is known and labeled. The goal is to learn a mapping from inputs to outputs so that the model can generalize and make predictions on unseen data. The SL is mainly used for classification and regression tasks such as support vector machines (SVM), decision trees, naive Bayes,  k-nearest neighbors (K-NN), and linear/ridge/lasso regression.   
    
    \item[\large \adforn{71}] \textit{Unsupervised Learning} (USL) models learn from a dataset without explicit target labels. The goal is to uncover hidden patterns, structures, or relationships within the data. Examples of unsupervised learning tasks include 
    \begin{itemize}
        \item  Clustering and Pattern Search: K-means clustering and fuzzy C-means clustering. 
        \item Dimensionality Reduction: principal component analysis (PCA), t-stochastic neighbor embedding, and linear discriminant analysis.
    \end{itemize}
    
    \item[\large \adforn{71}] \textit{Reinforcement Learning} (RL) agents learn to make decisions by interacting with an environment. The agent receives feedback in the form of rewards or penalties, and its goal is to learn a policy that maximizes the cumulative reward over time. RL differs from SL and USL in that it focuses on sequential decision-making problems and operates in an environment where feedback is often delayed. The most common RL algorithms can be summarized as follows: Q-learning, state–action–reward–state–action, Deep Q-Learning (DQN), deep deterministic policy gradient (DDPG), and Twin Delayed DDPG (TD3), which are ordered in the chronological order of development.  
\end{itemize}

In essence, each ML class possesses a distinct collection of techniques and strategies tailored to address specific tasks. In Table \ref{tab:ML_comp}, we compare the advantages and disadvantages of various ML types and explain preferences by describing the appropriate circumstances and objectives for utilizing a particular ML type. Indeed, the unique characteristics of each ML class confer distinct benefits for resolving diverse challenges in wireless networks.

Ensemble learning is a novel ML paradigm where multiple models, a.k.a,  weak learners, are trained to solve the same problem to obtain more robust models and accurate results. The goal of ensemble methods is to judiciously combine the predictions of several base estimators built with a given ML algorithm to improve generalizability and robustness over a single estimator. Different ensemble methods will be effective for different types of problems and data, which are summarized below:

\begin{itemize}
    \item[\large \adforn{14}] Bootstrap aggregating (a.k.a., bagging \cite{breiman1996bagging}) involves manipulating the training set by resampling, and then running a classifier on each version of the training set. The results are then averaged for regression or voted for classification to produce the final prediction. Random Forest is a famous example of bagging.

    \item[\large \adforn{14}] Boosting refers to a group of algorithms that utilize weighted averages to transform weak learners into stronger learners \cite{schapire2003boosting}. Unlike bagging, boosting uses subsets of the original data to produce a series of weak learners. The learners are then weighted according to their accuracy before being combined to form a strong prediction. Examples of boosting algorithms include AdaBoost and Gradient Boosting (e.g., LightGBM, CatBoost, XGBoost).

    \item[\large \adforn{14}] Stacking involves training an ML algorithm to combine the predictions of several others \cite{dvzeroski2004combining}. First, all the individual models are trained based on the complete training set; then, a combiner algorithm is trained to make a final prediction based on all the predictions of the individual models. 
\end{itemize}
 
Recently, neural networks and deep learning (DL) has substantially improved above traditional ML techniques thanks to deep neural networks' (DNNs) ability to learn complex and hierarchical representations of data, leading to better generalization, prediction, and classification performance. The architecture of a DNN typically consists of an input layer, multiple hidden layers, and an output layer. The input layer receives the raw data, while the output layer generates the final prediction or representation. The hidden layers in between perform various transformations on the data, learning increasingly complex features and high level of abstraction as the data passes through the network. The most prominent examples of DNNs include feedforward neural networks (FNNs); multilayer perceptrons (MLPs); convolutional neural networks (CNNs); recurrent neural networks (RNNs) such as long short-term memory (LSTM) and gated recurrent units (GRUs); and autoencoders (AE). GMs and transformers also fall into the class of DNNs/DL and will be discussed in-depth in subsequent sections.  

Above inherent DL features reduces the need for manual feature engineering and improves the performance of models across a wide range of tasks. DL models are also designed to scale well with large datasets and can exploit the computational power of modern hardware. Moreover, DL models often learn end-to-end mappings from input data to the desired output, eliminating the need for intermediate processing steps or handcrafted features, simplifying the overall learning process and allows for a more seamless model integration into various applications. As a result of these key features, DL models has recently become a prominent research tool to solve intricate wireless communication and networking problems. Thus, we subsequently examine the means by which each ML class can enhance wireless networks and identify their most fitting applications.

\subsection{DAI USE CASES IN WIRELESS INTELLIGENCE}
\label{sec:DAI_usecases}
In recent years, DAI models have been utilized to improve performance of wireless networks as exemplified below.
    \subsubsection{CHANNEL ESTIMATION AND EQUALIZATION} Both SL and USL algorithms, particularly DL, can learn to estimate channel state information (CSI) and perform equalization \cite{burse2010channel,AbdallahCME22,AbdallahCME22conf,Abdallah2023RIS}. Accurate CSI estimation is crucial for various tasks, such as beamforming, resource allocation, and link adaptation. Traditional pilot-based approaches may not be efficient for complex and non-linear channels in rapidly changing environments, where DNNs can be trained using historical data to predict CSI accurately. On the other hand, equalization aims to mitigate the detrimental impacts of inter-symbol interference (ISI) and noise. Traditional linear equalizers or decision-feedback equalizers, may not be optimal for highly dispersive channels or non-linear distortions, where DNNs can identify patterns in received signals and adaptively adjust their equalization strategies. 

    \subsubsection{BEAMFORMING \& CODEBOOK DESIGN} By learning from historical and real-time data, DL algorithms may be trained to adaptively optimize beamforming weights based on CSI and other environmental factors \cite{al2022review,asmaa2023drl,abdallah2022madrl}. Unlike traditional codebooks that may not accurately represent the specific environment of a deployment site, DL models can learn the unique features of the environment and generate customized codebooks tailored to the specific site by processing large amounts of data collected from the deployment site. As hybrid architectures has come into prominence to strike a good balance between hardware complexity and MIMO gain, DL algorithms can be used to optimize the analog and digital beamforming weights jointly, yielding improved system performance and energy efficiency. Finally, multi-modal sensory data collected by integrated sensing \& communication systems can be leveraged by DL models to substantially improve beamforming and codebook design by effective beam tracking through localization and movement prediction in mobile environments. 
    
    \subsubsection{COGNITIVE RADIOS \& DYNAMIC SPECTRUM ACCESS} RL agents can learn optimal spectrum allocation strategies in dynamic environments, identifying the best available spectrum bands for communication while minimizing interference and maximizing utilization \cite{zhou2019dynamic}. Cognitive radios can use RL to adapt their operating parameters, such as transmit power, modulation schemes, and coding rates, to the current network conditions, allowing them to optimize communication performance and coexist with other radio systems in a shared spectrum environment. Additionally, RL can be utilized for efficient spectrum handoffs when primary users re-occupy the spectrum, minimizing handoff latency and maintaining seamless communication.
    
    \subsubsection{RADIO RESOURCE MANAGEMENT} 
    ML can play a crucial role in radio resource management and energy efficiency in wireless networks by optimizing various aspects, such as resource allocation, power control, and network configuration \cite{du2020green}. For instance, DL models can predict user demands and traffic patterns  while deep RL (DRL) techniques can help in making intelligent decisions for allocating resources to users. SL techniques can be used to model the relationship between power levels and performance metrics, while RL methods can be employed to determine optimal power control policies in dynamic network environments. These approaches can be amalgamated to optimize efficient use of network resources such that energy consumption is minimized while QoS is maintained. Moreover, USL techniques can identify patterns in network data to design energy-efficient network configurations and topologies. 

    \subsubsection{PHYSICAL LAYER SECURITY \& AUTHENTICATION} 
    SL contributes to the detection and mitigation of eavesdropping and jamming attacks by identifying and classifying malicious activities in the network \cite{kamboj2021machine}. Additionally, SL can help in the fingerprinting of RF signals for user authentication or localization by identifying their unique characteristics. USL is instrumental in detecting network intrusions and anomalies by flagging unusual patterns in network traffic or behavior, thus providing early warning and protection against potential cyber threats, network faults, or performance degradation. By integrating both SL and USL approaches, the confidentiality, integrity, and overall security of wireless communication can be significantly enhanced.

    \subsubsection{HANDOVER MANAGEMENT} ML can significantly improve handover management by providing intelligent decision-making and optimization for maintaining seamless connectivity and QoS for mobile users as they move across different cells or access points \cite{sonmez2020handover}. ML can contribute to different aspects of handover management, such as prediction, decision-making, and optimization, using various types of ML techniques. For instance, DNNs can predict the need for handovers by analyzing historical data, user mobility patterns, and network conditions, thereby enabling the network to proactively initiate the handover process, reducing delays, and improving the user experience. On the other hand, DRL agents can optimize handover decisions by taking into account factors such as user mobility, signal strength, network load, and QoS requirements, which will eventually help determine the best target cell or access point for each user. Finally, SL models can be used to optimize the handover process by dynamically adjusting parameters such as handover thresholds, hysteresis margins, and time-to-trigger values, reducing unnecessary handovers (ping-pong effect) and minimizing the overall handover latency. 

    \subsubsection{TRAFFIC PREDICTION \& CONGESTION CONTROL} While SL algorithms use labeled training data to make predictions about future traffic patterns, USL algorithms can be used to identify underlying patterns and structures in the network traffic data, which can then be used for making predictions \cite{jiang2022cellular, Zhang2020congestion}. Time series forecasting models specifically designed to handle temporal data can also be used for network traffic prediction. By training on historical data with labeled congestion states, SL algorithms can help predict congestion levels and take appropriate actions to alleviate the problem. Moreover, USL approaches (e.g., clustering, dimensionality reduction techniques) can be used to identify patterns, understand the underlying causes of congestion, and design strategies to mitigate it. RL algorithms can also be employed to dynamically adjust network parameters (e.g., routing, bandwidth allocation) based on real-time feedback on congestion levels. 

    \subsubsection{NETWORK ROUTING \& LOAD BALANCING} USL techniques, such as clustering and anomaly detection, can be employed to group network nodes with similar characteristics or traffic patterns and identify unusual traffic patterns or network behaviors \cite{nayak2021routing, gures2022loadbalancing}. This information can help in designing efficient routing strategies, optimizing load distribution, and maintaining network performance in dynamic conditions. On the other hand, RL allows network agents to learn from their interactions with the environment and adapt their routing decisions and load balancing strategies over time. By using RL algorithms, network agents can learn to select the best paths, optimize load distribution, and make better decisions in dynamic network conditions, such as varying traffic loads and changing link capacities.

    \subsubsection{NETWORK FUNCTION VIRTUALIZATION \& NETWORK SLICING} 
    Network function virtualization (NFV) involves deploying network functions as software on virtualized infrastructure, where ML can assist in determining the optimal deployment of these virtual network functions (VNFs) and in orchestrating their instantiation, scaling, and migration based on network conditions \cite{ssengonzi2022survey}. In the case of network slicing, ML can help automate the creation, modification, and termination of network slices, in an adaptive manner based on user requirements and network conditions. Furthermore, ML can be employed to analyze network traffic patterns, identify anomalies, and  predict potential faults within network slices or VNFs. By recognizing patterns that may lead to anamolies and failures, ML can help network operators take preventive measures, such as reallocating resources or migrating VNFs, to minimize service disruptions and maintain network reliability.

\begin{table}[t]
\centering
\caption{The ways GMs complement DAI techniques.}
\label{tab:GMways_comp}
\resizebox{0.5\textwidth}{!}{%
\begin{tabular}{|l|l|l|}
\hline
\textbf{Techniques}                                           & \textbf{Type} & \textbf{Complementary Role}                                               \\ \hline
\multirow{2}{*}{\begin{tabular}[c]{@{}l@{}}Data\\ Augmentation\end{tabular}}        & SL & Generate synthetic data for training                                           \\ \cline{2-3} 
                                                                           & RL            & Generate synthetic experiences                                         \\ \hline
\multirow{3}{*}{\begin{tabular}[c]{@{}l@{}}Data\\ Imputation\end{tabular}} & SL            & Estimate missing feature values                                        \\ \cline{2-3} 
                                                                           & USL           & Fill in missing values for clustering, dimensionality reduction        \\ \cline{2-3} 
                                                                           & RL            & Estimate missing state or reward information                           \\ \hline
\multirow{3}{*}{Disentanglement}                                           & SL            & Learn interpretable features for classification or regression          \\ \cline{2-3} 
                                                                           & USL           & Learn representations for clustering, dimensionality reduction         \\ \cline{2-3} 
                                                                           & RL            & Learn interpretable state representations                              \\ \hline
\multirow{3}{*}{Regularization}                                            & SL            & Encourage robust feature learning for classifiers or regression        \\ \cline{2-3} 
                                                                           & USL           & Impose structure on latent space or learned representations            \\ \cline{2-3} 
                                                                           & RL            & Learn compact state representations, model environment dynamics        \\ \hline
\multirow{3}{*}{\begin{tabular}[c]{@{}l@{}}Dimensionality\\ Reduction\end{tabular}} & SL & Learn lower-dimensional features for classification or regression              \\ \cline{2-3} 
                                                                           & USL           & Mitigate the curse of dimensionality                                   \\ \cline{2-3} 
                                                                           & RL            & Learn compact state representations                                    \\ \hline
\multirow{3}{*}{\begin{tabular}[c]{@{}l@{}}Feature\\ Learning\end{tabular}}         & SL & Learn informative features for classification or regression                    \\ \cline{2-3} 
                                                                           & USL           & Focus on salient aspects of the data                                   \\ \cline{2-3} 
                                                                           & RL            & Learn quickly and generalize better                                    \\ \hline
\multirow{3}{*}{\begin{tabular}[c]{@{}l@{}}Transfer\\ Learning\end{tabular}}        & SL & Transfer knowledge between task-specific classifier/regression models      \\ \cline{2-3} 
                                                                           & USL           & Enable robust representations for clustering, dimensionality reduction \\ \cline{2-3} 
                                                                           & RL            & Facilitate policy or value function transfer                           \\ \hline
Semi-SL                                                                    & SL            & Leverage information in unlabeled data                                 \\ \hline
\multirow{3}{*}{\begin{tabular}[c]{@{}l@{}}Multi-Task \\ Learning\end{tabular}}     & SL & Learn common features for task-specific classifiers or regression models       \\ \cline{2-3} 
& USL           & Learn common representations for clustering, dimensionality reduction  \\ \cline{2-3} 
& RL            & Learn common representations for policy or value function transfer     \\ \hline
\begin{tabular}[c]{@{}l@{}}Exploration\\ Balancing\end{tabular}            & RL            & \begin{tabular}[c]{@{}l@{}}Learn uncertainties and environment dynamics   \end{tabular}
                        \\ \hline
\begin{tabular}[c]{@{}l@{}}Imitation \\ Learning\end{tabular}                       & RL &  \begin{tabular}[c]{@{}l@{}}Learn policies capturing expert behavior \\ and generate synthetic demonstrations\end{tabular}
\\ \hline
\end{tabular}%
}
\end{table}

{
\subsection{THE WAYS GenAI COMPLEMENT DAI} 
\label{sec:GM_ways}
GMs can complement various DAI approaches, enhancing the learning process and improving the overall system's performance. In the remainder, we list the main common ways GMs can help different DAI models. For the sake of a better comprehension of the comparison among above items, Table \ref{tab:GMways_comp} also summarizes complementary GM techniques alongside their beneficial role in each DAI types. {It is worth mentioning that this subsection presents a generic discussion on how GenAI complements DAI without being involved with application specific considerations, which are our main focus in Sec. \ref{sec:PHY}-Sec. \ref{sec:6GMfrontiers}.}

    \subsubsection{DATA AUGMENTATION \cite{bowles2018gan}} GMs can significantly enhance the learning process in both supervised and reinforcement learning by addressing the challenges associated with data acquisition and labeling. Acquiring and labeling a large amount of data can be computationally expensive and time-consuming, especially for versatile data obtained from complex environments. GMs can be employed to generate supplementary synthetic data that can be combined with real data to augment the training dataset. This approach can address the challenges faced by SL models and RL agents by providing faster convergence, improved model performance, better generalization, and reduced overfitting.  

    \subsubsection{MISSING DATA IMPUTATION \cite{zhang2018missing}} GMs can complement various ML techniques by addressing the issue of missing or incomplete data. In SL, GMs can estimate missing values, allowing classifiers or regression models to make use of complete data, potentially improving their performance. Similarly, in USL, GMs can fill in missing values, enabling algorithms (e.g., clustering, dimensionality reduction) to work with complete data and enhance their effectiveness. In RL, GMs can help estimate missing state or reward information, allowing agents to learn more effectively from complete data, improving their decision-making and overall performance. By estimating missing values, GMs can enhance the robustness and reliability of ML techniques when dealing with incomplete data.
    
    \subsubsection{DISENTANGLEMENT \cite{ren2021generative}} Disentanglement is the process of finding a more structured, low-dimensional and interpretable representation of the data, where the different dimensions correspond to distinct factors of variation. GMs can complement ML techniques through disentanglement by learning to separate the underlying factors of variation in the input data for a better generalization. Disentangled representations can help SL and USL models to focus on the relevant factors of variation for a specific task, leading to improved performance and better interpretability of the learned model. By using GMs to learn interpretable and disentangled state representations, RL agents can achieve better generalization and adapt more effectively to new situations or states.

    \subsubsection{REGULARIZATION \cite{goodfellow2014generative, kingma2014semi, xu2015overview}} Regularization is a technique used to improve the generalization by preventing overfitting, which occurs when a model learns the noise or random fluctuations in the training data. That is, regularization helps the model to capture the underlying patterns in the data, rather than memorizing the training data, resulting in better performance on unseen data/environments. At this point, GMs offer unique benefits to regularization due to their ability to learn the underlying data distribution, generate new samples, and their inherent focus on robust feature learning. In SL, incorporating GMs into classifiers or regression models can encourage them to learn more robust and informative features by training the models to reconstruct the input data while also predicting the target labels. This helps the model to capture the underlying structure of the data and avoid fitting to noise or irrelevant details. In USL, GMs can impose a structure on the latent space or learned representations, which can lead to more interpretable and useful representations, encouraging the algorithm to focus more on the most salient aspects of the data. In RL, GMs can act as a regularizer by learning a compact and meaningful state representation that encourages the agent to focus on the essential aspects of the environment, rather than overfitting to specific observations. Additionally, GMs can be used to learn a model of the environment dynamics, which can introduce additional regularization in the form of accurate and efficient environment representation, allowing the agent to perform better planning and make better decisions.

    \subsubsection{DIMENSIONALITY REDUCTION \cite{sorzano2014survey,xu2015overview}} GMs can help dimensionality reduction by learning lower-dimensional representations of high-dimensional data thanks to its innate ability to capture the underlying structure and correlations in the high-dimensional data. In SL tasks (e.g., classification and regression), high-dimensional data can lead to overfitting and increased complexity. The lower-dimensional representation learned by GMs can be used as features for the SL models, thereby reducing the dimensionality of the feature space and leading to improved generalization and better model performance. By learning a compact representation of the data, GMs can help USL algorithms to mitigate the curse of dimensionality by identify meaningful patterns and relationships more effectively. Similarly, GMs can help RL agents dealing with high-dimensional state or observation spaces by learning a compact and meaningful state representation.

    \subsubsection{FEATURE/REPRESENTATION LEARNING \cite{bengio2013representation}} GMs, through their ability to learn meaningful and compact representations of input data, can significantly contribute to the success of feature and representation learning across various learning paradigms.  The learned features and representations can enhance SL by providing more informative and efficient representations of data, leading to improved classification or regression outcomes. In USL, these compact representations allow models to focus on the most salient aspects of the data, thereby improving performance in tasks such as clustering, dimensionality reduction, and density estimation. Additionally, GMs assist RL agents in learning expeditiously and generalizing better to new or unseen situations and states, which is particularly important in high-dimensional or partially observable environments.

    \subsubsection{TRANSFER LEARNING \cite{zhuang2020comprehensive}}  GMs can complement transfer learning across SL, USL, and RL tasks by learning shared features or representations that capture the underlying structure and correlations in the data. This common representation enables the transfer of knowledge between tasks, leading to improved performance and better generalization.  GMs can generate synthetic data that bridges the gap between the source and target domains, assisting models in several ways: 1) adapting more quickly and effectively to the target domain, 2) mitigating the risk of negative transfer, and 3) fine-tuning their learned policies on more representative samples for RL agents in particular.

    \subsubsection{SEMI-SUPERVISED LEARNING \cite{kingma2014semi}} In many real-world scenarios, labeled data may be scarce, while unlabeled data is abundant. GMs can be employed to leverage the information contained in the unlabeled data, thereby improving the performance of supervised learning models. This process involves training a GM on both labeled and unlabeled data and subsequently using the learned representations or likelihood estimates to enhance the supervised learning task. By making the most of the available data, GMs contribute to the overall effectiveness of semi-supervised learning approaches.

    \subsubsection{MULTI-TASK LEARNING \cite{zhao2020generative}} Multitask learning is an approach in which a single model learns to perform multiple related tasks simultaneously. GMs can contribute to multitask learning by learning common features/representations across these tasks, leading to improved performance and better generalization. While GMs can help transfer knowledge between task-specific classifiers or regression SL models, they can enable robust representations for USL tasks (e.g.,  clustering, dimensionality reduction, or density estimation) across related datasets in approaches. Finally, GMs facilitate policy or value function transfer between various RL tasks, promoting efficient learning and better generalization across multiple environments or tasks.

\begin{figure*}[t!]
 \frame{\includegraphics[width=\textwidth]{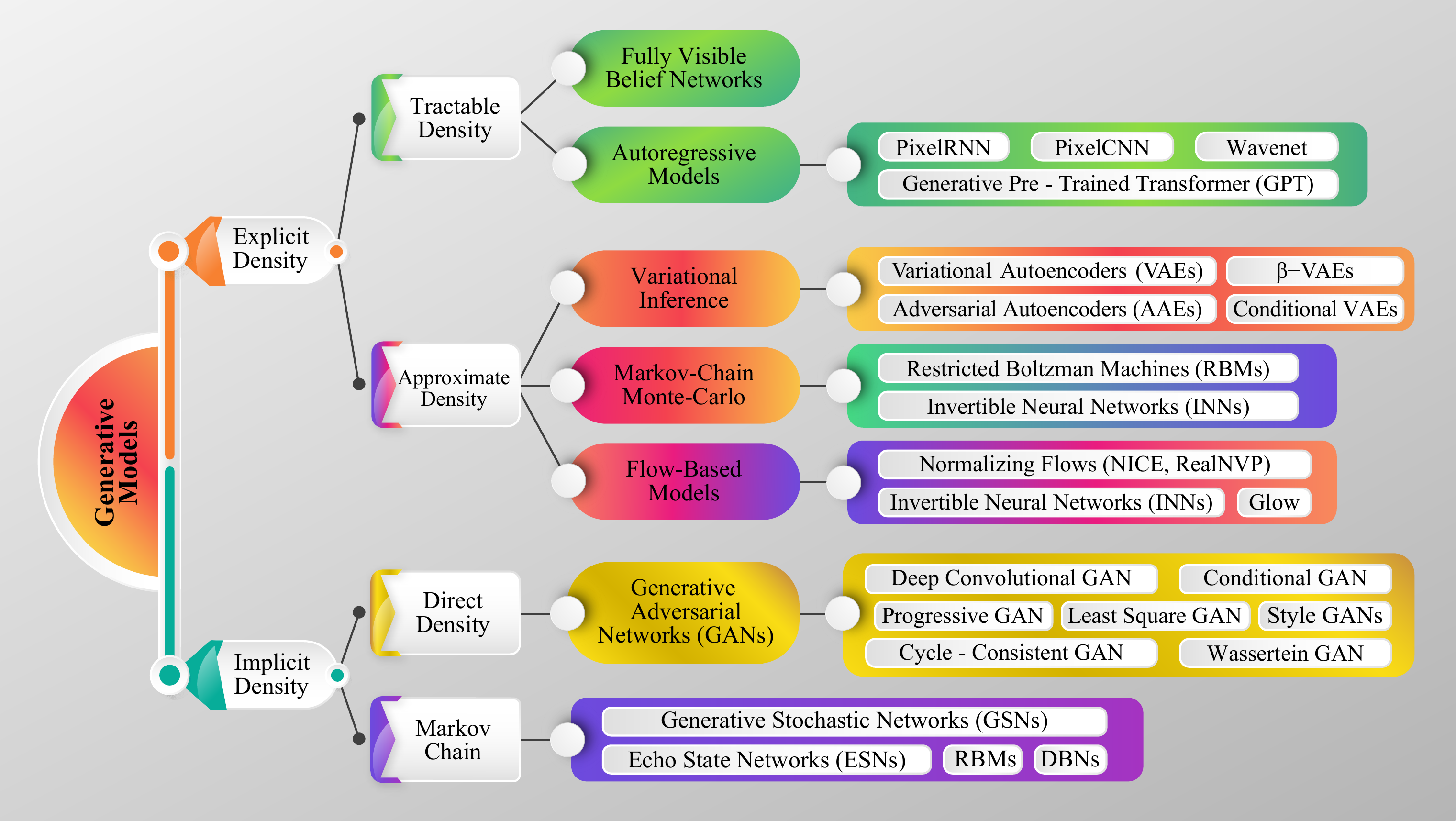}}
    \caption{A taxomony of GMs.}
    \label{fig:GenAI_taxo}
\end{figure*}

    \subsubsection{BALANCING EXPLORATION \& EXPLOITATION \cite{tschantz2020reinforcement,xu2021improving}} GMs can guide RL agents to strike a balance between exploration and exploitation processes by learning the underlying structure and uncertainties in the environment and then generating promising action sequences, state transitions, or estimate the potential value of unexplored states. The learned uncertainties can help RL agents identify areas where their knowledge is incomplete, encouraging them to explore those areas more effectively. Additionally, by simulating the environment dynamics, GMs can help agents evaluate the potential consequences of different actions, enabling them to explore the environment more intelligently and efficiently, while minimizing the risks associated with purely random exploration.

    \subsubsection{IMITATION LEARNING \cite{ho2016generative}} 
    Imitation learning is an RL technique where an agent learns a policy by observing and mimicking the behavior of an expert, which can speed up the learning process through acquisition of knowledge and skills without exploring the environment from scratch. Nonetheless, imitation learning suffers from distribution shift problem causing poor performance over scenarios not observed in the expert demonstrations. GM can learn a policy that captures the underlying structure and decision-making patterns of the expert, generalizing to unseen states. Additionally, they can generate synthetic demonstrations that cover a broader range of states and actions, helping the agent to better understand the task dynamics and improve its performance in unexplored regions of the state space.

}

\section{A TUTORIAL ON GENERATIVE MODELS} 
\label{sec:GAI_Models}
{In this section, we will first provide a taxonomy of GMs, then introduce their most prominent and exploited prime examples. Throughout these subsections, we dissect the essentials, concise mathematical underpinnings, virtues, drawbacks, and variants of these GMs. At the end, we suggest essential performance metrics for benchmarking these GMs that should be taken into account for comparison purposes. It is worth noting that this brief tutorial does not discuss GM use cases wireless communications and networks; hence, readers with solid GenAI background can skip this section and continue with Sec. \ref{sec:PHY}-Sec. \ref{sec:6GMfrontiers}.}

{
\subsection{A TAXONOMY OF GENERATIVE MODELS} 
\label{sec:GM_taxo}

Following the taxonomy of Goodfellow \cite{goodfellow2016nips} depicted in Fig. \ref{fig:GenAI_taxo}, GMs can be categorized in the way they represent and manipulate the probability density function (PDF) of the data they aim to model: explicit and implicit density models.

\subsubsection{EXPLICIT DENSITY MODELS} This class of models are also referred to as prescribed density models and have an explicit representation of the PDF of the dataset, $p(x)$, where $x$ is a data point from the data distribution. There are two primary types of explicit density models: tractable and approximate.

\paragraph*{Tractable Density Models} 
Tractable Density Models refer to a subclass of explicit density generative models that directly define the PDF of the data in a form that permits efficient computation of $p(x)$. ``Tractable'' in this context implies that the model's mathematical structure allows for the direct evaluation, sampling, or both, of the likelihood without resorting to computationally intensive or approximate methods. Examples include Fully Visible Belief Networks (FVBNs) and Autoregressive Models, such as PixelRNN, PixelCNN, WaveNet, and transformer-based models (e.g., GPT-3). 

\paragraph*{Approximate Density Models} 
When $p(x)$ is elusive to direct or efficient computation, devising approximate density models navigate, estimate, or work around to tackle complexities and challenges of directly modeling intricate data distributions. While they introduce additional layers of complexity, both in terms of architecture and training, they are indispensable in scenarios where sheer model expressivity and the capability to capture nuanced data characteristics are paramount. Examples include Variational Inference (VI) models (e.g, Variational AEs (VAEs), Conditional VAEs, $\beta$-VAEs, Adversarial AEs (AAEs), etc.);  Markov Chain Monte Carlo (MCMC) models (e.g., Restricted Boltzmann Machines (RBMs), Deep Belief Networks (DBNs), etc.); and Flow-based GMs (FGMs) (e.g., NICE, RealNVP, Glow, Invertible Neural Networks (INNs), etc.)

The primary advantages of explicit density models stem from their provision of an explicit form for $p(x)$, enabling direct likelihood-based training and evaluation. Moreover, they frequently grant interpretable insights into data structure. However, explicit density models, especially those approximating the likelihood, can be computationally burdensome during training or inference. Additionally, the imposition of a specific functional form for $p(x)$ might occasionally be overly restrictive, leading to potential suboptimal outcomes when faced with intricate data distributions.

\subsubsection{IMPLICIT DENSITY MODELS}
Implicit models represent a distinct class of GMs that do not explicitly define the PDF of the data they aim to model. Instead of providing a tractable expression for $p(x)$, this class of models learn to generate samples from the data distribution by implicitly capturing the underlying PDF through a stochastic process. Implicit density models can be further divided into direct implicit models and Markov chain-based implicit models.

\paragraph*{Direct Implicit Density Models}
These models directly generate samples from the target distribution without explicitly defining a PDF. Generative adversarial networks (GANs) and its variations fall into this category of GMs. 

\paragraph*{Markov Chain-based Implicit Density Models}
These models generate samples from the target distribution using a Markov chain, which is a stochastic process where the future states depend only on the present state. Markov chain-based models often involve a random walk through the state space, where each step is based on the current state and some transition probabilities. Examples include generative stochastic networks (GSNs), Echo State Networks (ESNs), RBMs, and DBNs. 

In summary, implicit models prioritize the generation of high-quality samples over explicit density estimation or evaluation. Their design and training methodologies are crafted around this objective, making them particularly suitable for tasks where the quality of generated samples is the primary concern. However, their implicit nature requires alternative approaches to model evaluation, stability, and interpretability compared to their explicit counterparts.
}

\subsection{GENERATIVE ADVERSARIAL NETWORKS (GANs)}
\label{sec:GAN}
\begin{figure}[t]
    \centering
    \includegraphics[width=0.99\columnwidth]{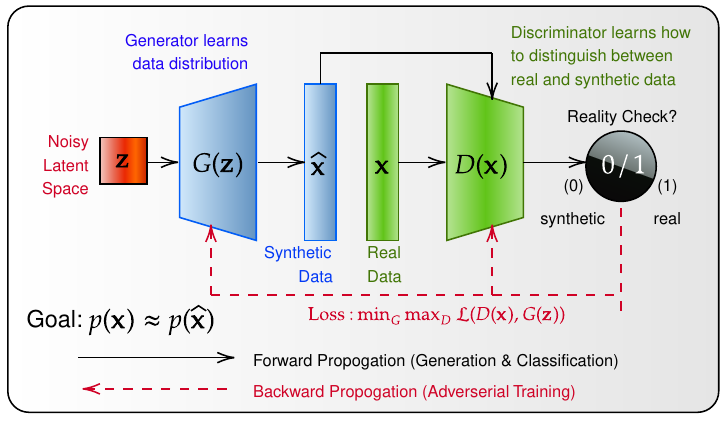}
    \caption{Schematic illustration of GAN architectures: a) direct training, and b) indirect training.}
    \label{fig:GAN}
\end{figure}

GANs were first introduced by the seminal paper of Ian Goodfellow et al. \cite{goodfellow2014generative}, wherein they define GAN as a set of two neural networks, a generator and a discriminator, which compete with each other in a zero-sum game, as shown in Fig. \ref{fig:GAN}. The generator aims to generate realistic samples, $G(\boldsymbol{z})$, from random input noise, $\boldsymbol{z}$. That is, it learns to map the noise distribution, $p(\boldsymbol{z})$, to the data distribution, $p(\boldsymbol{x})$, effectively producing samples that resemble the original data. On the other hand, the discriminator, $D(\boldsymbol{x})$, takes samples, $\boldsymbol{x}$, as input and tries to maximize the likelihood of correctly classifying them as real or fake as an output. Following this architecture, the training process of GANs involves an adversarial game between the generator and the discriminator, with the following objective function,
\begin{align}
\nonumber \min_{G(\boldsymbol{z})} \max_{D(\boldsymbol{x})} \mathcal{L}(D, G) &= \mathbb{E}_{\boldsymbol{x}}[\log D(\boldsymbol{x})]  + \mathbb{E}_{\boldsymbol{z}}[\log (1 - D(G(\boldsymbol{z})))].
\end{align}
In each iteration, the generator creates samples from random noise, and the discriminator evaluates the generated samples alongside real data samples. The generator's goal is to create samples that can deceive the discriminator, while the discriminator's goal is to correctly classify samples as real or fake. The training process alternates between updating the generator and discriminator weights, ensuring that the generator learning to produce better samples and the discriminator learning to better distinguish between real and fake samples.

Unlike the above direct training approach, indirect training is a two-step approach involving training the generator and discriminator networks separately: the generator is trained using a pre-trained auxiliary network that provides guidance on how to create realistic samples. This auxiliary network can be a different discriminator or another pre-trained model (e.g., an autoencoder) that captures the structure of the data distribution and ensures that generator comply with its criteria for realistic samples. The generator's weights are updated based on the feedback from the auxiliary network, without directly updating the auxiliary network's weights during this process.

Indirect training can help alleviate some of the challenges associated with direct training, such as mode collapse and unstable training dynamics. By decoupling the generator and discriminator training, indirect training can lead to more stable training and better convergence properties \cite{ho2016generative, metz2016unrolled}. In Table \ref{tab:GAN_training}, we provide a comparison between two approaches and suggest preferences depending on the specific application, the desired trade-off between training stability and computational complexity, and the characteristics of the dataset.

\begin{table}[t]
\centering
\caption{Comparison of direct and indirect GAN training. }
\label{tab:GAN_training}
\resizebox{\columnwidth}{!}{%
\begin{tabular}{l|l|l|}
\cline{2-3}
& \multicolumn{1}{c|}{\textbf{Direct Training}} & \multicolumn{1}{c|}{\textbf{Indirect Training}} \\ \hline
\multicolumn{1}{|l|}{\multirow{2}{*}{\rotatebox[origin=c]{90}{\textbf{Pros}}}} &
  \multirow{2}{*}{- Simple training procedure} &
  \multirow{2}{*}{\begin{tabular}[c]{@{}l@{}} - More stable training dynamics\\ - Better convergence properties\end{tabular}} \\
\multicolumn{1}{|l|}{} &                                               &                                                 \\ \hline
\multicolumn{1}{|l|}{\multirow{2}{*}{\rotatebox[origin=c]{90}{\textbf{Cons}}}} &
  \multirow{2}{*}{\begin{tabular}[c]{@{}l@{}}- Potential for mode collapse\\ - Training instability\end{tabular}} &
  \multirow{2}{*}{\begin{tabular}[c]{@{}l@{}}- Requires additional pre-training steps\\ - Increased computational complexity\end{tabular}} \\
\multicolumn{1}{|l|}{} &                                               &                                                 \\ \hline
\multicolumn{1}{|l|}{\multirow{3}{*}{\rotatebox[origin=c]{90}{\textbf{Prefs.}}}} &
  \multirow{3}{*}{\begin{tabular}[c]{@{}l@{}}- Dataset structure is simple\\ - Limited computational resources\end{tabular}} &
  \multirow{3}{*}{\begin{tabular}[c]{@{}l@{}}- Dataset has complex structure\\ - Pre-trained auxiliary network available\\ - Focus on training stability\end{tabular}} \\
\multicolumn{1}{|l|}{} &                                               &                                                 \\
\multicolumn{1}{|l|}{} &                                               &                                                 \\ \hline
\end{tabular}%
}
\end{table}

\subsubsection{GAN VARIATIONS}
\label{sec:GAN_variations}
Since its introduction, there have been significant developments and enhancements to address GANs' various challenges and broadening their applicability, which are listed and briefly explained below:
\begin{itemize}
    \item [\large \adforn{5}] \textit{Conditional GANs} (cGANs) extend the original GAN framework, a.k.a. Vanilla GANs \cite{goodfellow2014generative}, by conditioning the generation process on additional information, such as class labels or other attributes \cite{mirza2014conditional}. This allows for more controlled sample generation, as the generated samples exhibit specific characteristics determined by the conditioning information. 

    \item [\large \adforn{5}] \textit{Cycle-consistent GANs} (CycleGANs) are another variant designed for unpaired image-to-image translation tasks, where there is no direct correspondence between the source and target domain images \cite{zhu2017unpaired}. CycleGANs employ a cycle-consistency loss to ensure that the translation process is reversible, leading to more coherent and meaningful translations between the domains. 

    \item [\large \adforn{5}] While\textit{ deep convolutional GANs} (DCGANs) introduced convolutional layers to GANs for a better performance and stability \cite{radford2015unsupervised}, \textit{Wasserstein GANs} (WGANs) proposed a new loss function based on the Wasserstein distance, resulting in improved training stability and convergence \cite{arjovsky2017wasserstein}.

    \item [\large \adforn{5}] \textit{Progressive GANs} (ProGANs) developed a training scheme that progressively increases the resolution of generated images, enabling the generation of high-quality, high-resolution samples \cite{karras2017progressive}.

    \item [\large \adforn{5}]  \textit{Bidirectional GANs} (BiGANs) introduce an encoder alongside the generator and the discriminator, which aims to learn a bijective mapping between the data space and the latent space \cite{donahue2016adversarial}. This framework is particularly useful for tasks like semi-supervised learning and representation learning.

    \item [\large \adforn{5}] \textit{Information Maximizing GANs} (InfoGANs) extend the standard GAN architecture by maximizing the mutual information between a small subset of the latent variables and the observation \cite{chen2016infogan}. This results in a disentangled representation of the latent variables, making the model capable of generating samples with specific attributes easily controllable.

    \item [\large \adforn{5}] \textit{StarGAN} allows for multi-domain image-to-image translation with a single model \cite{choi2018stargan}. This eliminates the need to train a unique GAN model for every pair of domains, thereby making the framework more scalable and efficient.

    \item [\large \adforn{5}] \textit{Stacked GANs} (SGANs) use a hierarchical approach where multiple GANs are stacked together to refine the output progressively \cite{huang2017stacked}. Each GAN in the stack focuses on different aspects of the image, allowing for more complex and high-quality image generation.

    \item [\large \adforn{5}]  \textit{Categorical GANs} (CatGANs) extend GANs to improve semi-supervised learning. The model uses a classifier in the discriminator to predict the probability distribution over the classes for each sample \cite{springenberg2015unsupervised}. This allows the model to learn better feature representations, thereby improving classification performance.

    \item [\large \adforn{5}]  \textit{Least Squares GANs} (LS-GANs) modify the loss function used in the original GAN framework to use the least squares loss function \cite{mao2017least}. This change aims to address issues such as mode collapse and improve the stability during the training process. LS-GANs have been shown to generate higher-quality samples compared to standard GANs.

    \item [\large \adforn{5}]  \textit{StyleGAN} introduces a style-based generator architecture that allows more flexible and controllable design of the generated images \cite{karras2019style}. StyleGAN utilizes adaptive instance normalization (AdaIN) layers to modulate the style/appearance of the generated image at different levels of the generator's network. The architecture allows for incredibly detailed and high-quality image synthesis and has been a groundbreaking development in the field of GMs. A later version, StyleGAN2, further refines the architecture for even better performance \cite{karras2019analyzing}.

   \item [\large \adforn{5}] \textit{Brainstorming GAN} (BGAN) is an architecture enabling multiple agents to collaboratively learn data distributions in a distributed manner without sharing their actual datasets \cite{ferdowsi2020brainstorming}. Instead, agents``brainstorm'' by exchanging generated data samples, thus overcoming the challenges of limited datasets and avoiding the need for a centralized controller. This scalable solution accommodates diverse neural network architectures among agents and has been demonstrated to produce higher quality data samples with lower divergence metrics compared to other distributed GAN models.

\end{itemize}

\subsection{VARIATIONAL AUTOENCODERS (VAEs)}
\label{sec:VAE}

\begin{figure}[t]
    \centering
    \includegraphics[width=0.49\textwidth]{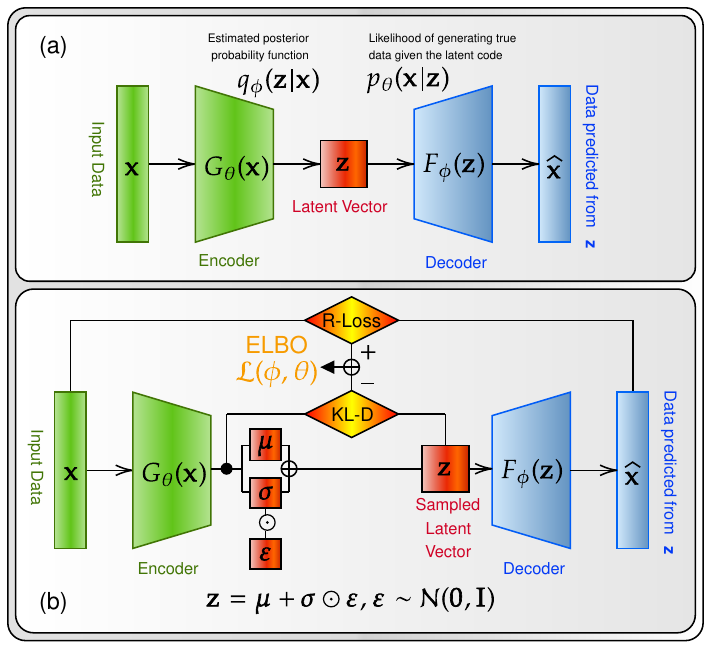}
    \caption{Schematic illustration of a) AEs and b) VAEs.}
    \label{fig:VAE}
\end{figure}

The traditional autoencoders (AEs) aims to learn the best encoding-decoding scheme using an iterative optimisation process, as shown in Fig. \ref{fig:VAE}.a. The encoder with parameter $\theta$, $G_\theta(\boldsymbol{x})$, learns to map input data, $\boldsymbol{x}$, to a latent space representation, $\boldsymbol{z}$, while the decoder with parameter $\phi$, $F_\phi(\boldsymbol{z})$, learns to reconstruct the input data from the latent space\footnote{Latent/feature space is a low-dimensional representation of high-dimensional data and captures the essential features/patterns in the data that help in generating new samples, reconstructing inputs, or performing other tasks.} representation. In this way, AEs learn a mapping between a high-dimensional data space and a lower-dimensional latent space. 

Being first introduced by Kingma and Welling in their seminal work \cite{kingma2013auto}, VAEs are developed to address the limitations of traditional AEs, which often struggled with generating realistic samples from the learned latent space. By incorporating variational inference and a probabilistic framework, as shown in Fig. \ref{fig:VAE}.b, VAEs are able to generate new data samples that more closely resemble the training data distribution.VAEs are trained by optimizing the evidence lower bound (ELBO) of the marginal likelihood of the data, which involves learning a continuous and smooth representation of the data in the latent space and can be formulated as follows:
\begin{equation}
\mathcal{L}(\theta, \phi) = \mathbb{E}_{q_\phi(\boldsymbol{z} \mid \boldsymbol{x})}[\log p_\theta(\boldsymbol{x} \mid \boldsymbol{z})] - \text{KL}(q_\phi(\boldsymbol{z} \mid \boldsymbol{x}) || p(\boldsymbol{z})),
\end{equation}
where $p(\boldsymbol{z})$ is the prior distribution of the latent variables, which is usually chosen as a standard Gaussian distribution, $p(\boldsymbol{z}) \sim \mathcal{N}(\boldsymbol{z}; \boldsymbol{0}, \boldsymbol{I})$. The first term of the ELBO represents the reconstruction loss, which measures the similarity between the input data and its reconstruction, while the second term represents the Kullback-Leibler divergence (KL-D)  between the approximate posterior distribution, $q_\phi(\boldsymbol{z} \mid \boldsymbol{x})$, and the prior distribution, $p(\boldsymbol{z})$, which acts as a regularizer that encourages the learned latent space representation to be close to the prior. In simpler terms, the VAE's training is regularised to avoid overfitting and ensure good enough latent space properties to enable generative process.

VAEs have gained significant attention in the DL community for their ability to generate high-quality samples while still maintaining a tractable and interpretable latent space. 
Since their introduction, numerous improvements and variants of VAEs have been proposed, as listed and briefly summarized below 

\begin{itemize}
    \item [\large \adforn{8}]  \textit{$\beta$-VAE} model introduces a hyperparameter, $\beta$, that controls the trade-off between the reconstruction loss and the regularization term in the VAE objective function \cite{higgins2016beta}. The purpose of this modification is to learn a more disentangled representation in the latent space. By adjusting the value of $\beta$, one can control the level of disentanglement in the latent variables, making the model more interpretable and useful for downstream tasks.

    \item [\large \adforn{8}]  \textit{Wasserstein Autoencoders} (WAEs) are designed to offer a more stable and robust training process by using the Wasserstein distance as the objective function instead of the KL-D used in standard VAEs \cite{tolstikhin2018wasserstein}. This results in a more balanced trade-off between data fidelity and regularization, potentially leading to better quality of generated samples.

    \item [\large \adforn{8}]  \textit{VAEs with Arbitrary Conditioning} (VAEAC) allows for the model to be conditioned on arbitrary subsets of observed variables \cite{ivanov2018variational}. This feature enables VAEAC to be applied to tasks that involve incomplete or partially observed data, making it versatile for various data imputation and generation scenarios.

    \item [\large \adforn{8}] \textit{Vector Quantized VAEs} (VQ-VAEs) incorporate vector quantization into the encoding process to create a discrete latent representation \cite{van2017neural}. This framework is particularly useful for tasks requiring a discrete latent space, like text-to-speech or symbolic music generation.

    \item [\large \adforn{8}] \textit{Ladder-VAEs }introduce a hierarchical latent variable model, aiming for a more interpretable and disentangled latent space \cite{sonderby2016ladder}. This approach is useful for cases where a multi-scale representation can provide added benefits.

    \item [\large \adforn{8}] \textit{Conditional VAEs} (cVAEs) extend the VAE framework by allowing the model to condition on additional variables, such as labels or attributes, similar to Conditional GANs \cite{sohn2015learning}. This allows for more targeted data generation.

    \item [\large \adforn{8}] Although not a VAE variant per se, normalizing flows can be combined with VAEs to model more complex posterior distributions \cite{rezende2015variational}. This can enhance the VAE's ability to capture more intricate data manifolds. Sec. \ref{sec:flow} will cover flow-based GMs more in detail. 

    \item [\large \adforn{8}] VAE-GAN is a hybrid model aiming to combine the strengths of both VAEs and GANs, focusing on balancing the quality of generated samples with the interpretability of the latent space \cite{larsen2015autoencoding}. In VAE-GAN, the VAE is used as the auxiliary network for the generator, while the discriminator is trained directly using the standard GAN training procedure. The VAE-GAN model aims to provide more stable training dynamics and better convergence properties while still generating high-quality samples.
    
\end{itemize}
Above advancements have led to the development of VAEs by addressing various issues related to mode collapse, disentanglement, and scalability; thereby, enabling VAEs to generate even more realistic samples and better disentangled latent space representations, further expanding their applicability to a variety of tasks. While VAEs are generally easier to train and provide a more interpretable latent space, they may generate less sharp samples compared to GANs due to the pixel-wise reconstruction loss. Moreover, GANs have been shown to generate high-quality samples at the cost of training difficulty and aforementioned issues such as mode collapse and unstable dynamic. 


\begin{figure}[t]
    \centering
    \includegraphics[width=0.49\textwidth]{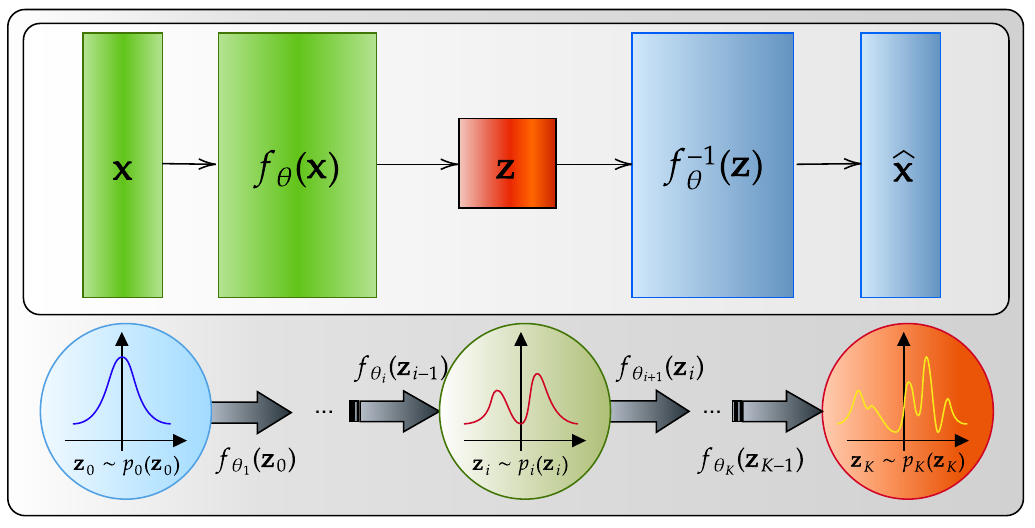}
    \caption{Schematic illustration of flow-based GMs.}
    \label{fig:flow}
\end{figure}

\subsection{FLOW-BASED GMs (FGMs)} 
\label{sec:flow}

FGMs are a class of deep GMs that learn to transform simple, easy-to-sample distributions into complex, high-dimensional data distributions \cite{rezende2015variational, kingma2018glow}. They accomplish this by learning a series of invertible transformations (or flows) that map points between the simple base distribution and the target data distribution. FGMs are advantageous due to their exact likelihood computation and efficient sampling and inference procedures.

As illustrated in Fig. \ref{fig:flow}, let $\mathbf{z} \sim p(\mathbf{z} )$ denote a random variable sampled from a simple base distribution (e.g., isotropic Gaussian), and $\mathbf{x}$ denote a data point from the complex target distribution $p(\mathbf{x})$. The goal of a flow-based model is to learn an invertible transformation $f_\theta$ parameterized by $\theta$ that maps $\mathbf{z} $ to $\mathbf{x}$, i.e., $\mathbf{z} = f_{\theta}(\mathbf{x} )$. The inverse transformation, which maps $\mathbf{x}$ back to $z$, is given by $\mathbf{x}  = f_{\theta}^{-1}(\mathbf{z})$. The change of variables formula relates the densities of $\mathbf{z} $ and $\mathbf{x}$:
\begin{equation}
\nonumber p(\mathbf{x}) = p(\mathbf{z} ) \left| \det \frac{d f_{\theta}^{-1}(\mathbf{x})}{d \mathbf{x}} \right|,
\end{equation}
where the determinant of the Jacobian, $\det \frac{d f_{\theta}^{-1}(\mathbf{x})}{d \mathbf{x}}$, captures how the density of $\mathbf{z} $ is transformed when mapping to $\mathbf{x}$.
FGMs typically employ a sequence of invertible transformations, or layers, to form a normalizing flow:
\begin{equation}
\nonumber \mathbf{x}  = f_{\theta_1} \circ f_{\theta_2} \circ \cdots \circ f_{\theta_K}(\mathbf{z} )
\end{equation}
Training an FGM involves maximizing the likelihood of the observed data, which can be computed exactly due to the change of variables formula. The prominent FGMs are briefly explained below:

\begin{itemize}
    \item[\large \adforn{5}] \textit{Real-valued Non-Volume Preserving} (RealNVP) RealNVP \cite{dinh2016density} introduces a series of coupling layers that alternate between modifying different parts of the input data. In each coupling layer, one part of the input data is left unchanged, while the other part is transformed using a scale and translation function that depends on the unchanged part. The scale and translation functions are implemented using neural networks. This design enables RealNVP to be both invertible and have an easily computable Jacobian determinant, which is essential for computing the likelihood of the data. Applications of RealNVP include image generation, unsupervised representation learning, and domain adaptation tasks.

    \item[\large \adforn{5}] \textit{Glow} \cite{kingma2018glow} extends RealNVP by incorporating several improvements, such as 1x1 invertible convolutions and actnorm, a layer that normalizes activations per channel. These modifications improve the expressive power of the model and allow for more efficient training. Additionally, Glow introduces multi-scale architecture, where the input data is split into different scales during the forward pass, making it easier to capture both local and global structures in the data. Glow has been applied to various tasks, including image synthesis, image-to-image translation, and unsupervised representation learning.

    \item[\large \adforn{5}] \textit{Nonlinear Independent Components Estimation} (NICE) \cite{dinh2014nice} is an earlier FGM that utilizes coupling layers similar to RealNVP but without the scale function. NICE models the data distribution as a composition of simple, invertible transformations that are applied to the input data. The absence of the scale function makes NICE less expressive than RealNVP or Glow but provides a simpler architecture. NICE has been used for tasks like image generation and unsupervised representation learning.
\end{itemize}

On the other hand, autoregressive flows combine the principles of both autoregressive and FGMs. In an autoregressive flow, the flow transformation is framed as an autoregressive model, where each dimension of the transformed variable is conditioned on the previous dimensions. This enables the model to capture complex dependencies between variables in the data distribution and generate samples more effectively. In this way, autoregressive flows can benefit from the strengths of both model types to capture complex dependencies between variables and generate realistic samples, as mentioned for GAMs in Sec. \ref{sec:GAM}, while maintaining the exact likelihood estimation and efficient sampling of FGMs. Examples of autoregressive flows include models like Masked Autoregressive Flow (MAF) \cite{papamakarios2017masked} and Inverse Autoregressive Flow (IAF) \cite{kingma2016improved}. MAF and IAF both employ autoregressive structures in their flow transformations but differ in how the transformations are applied, with MAF using a forward autoregressive structure and IAF using an inverse autoregressive structure.

\begin{figure}[t!]
    \centering
    \includegraphics[width=0.49\textwidth]{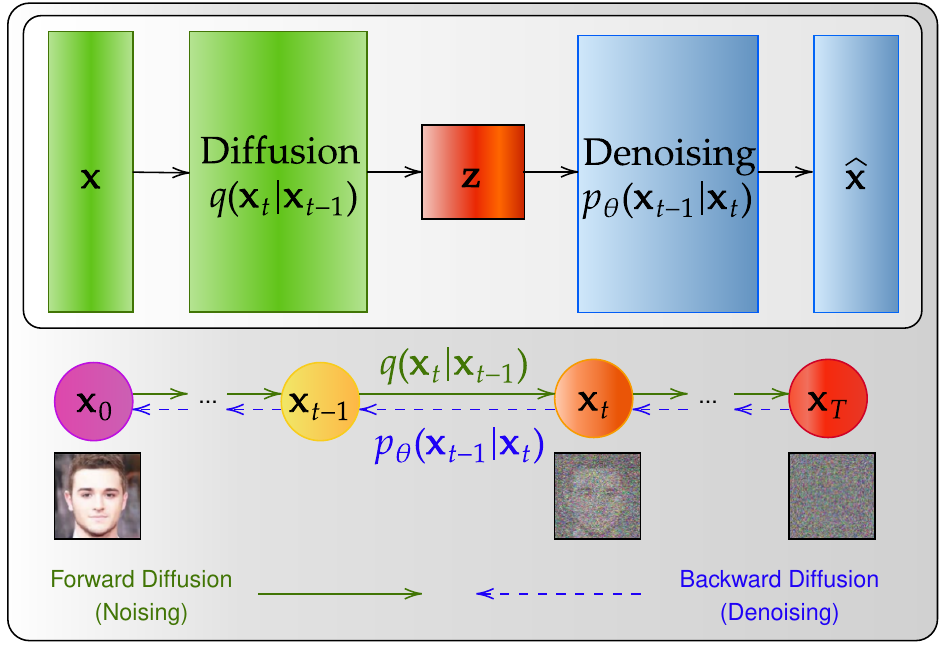}
    \caption{Schematic illustration of diffusion-based GMs.}
    \label{fig:diffusion}
\end{figure}

\subsection{DISFFUSION-BASED GMs (DGMs)} 
 \label{sec:diffusion}
Being inspired by non-equilibrium thermodynamics, DGMs leverage the concept of diffusion processes to learn the underlying data distribution and generate new samples. As illustrated in Fig. \ref{fig:diffusion}, DGMs exploit a continuous-time stochastic process of $T$ steps (e.g., Markov chain) that slowly transforms input data into a simple noise-like distribution (e.g., Gaussian) by adding noise at each step, then learn to reverse the diffusion process (i.e., denoising) to construct desired data samples from the noise.

To be more specific, starting from a data-point 
$\mathbf{x}_0$ sampled from the real data distribution $q(\mathbf{x})$, a Gaussion noise with variance $\beta_t$ is added to $\mathbf{x}_{t-1}$, producing a new latent variable $\mathbf{x}_t$ with distribution $q(\mathbf{x}_t \vert \mathbf{x}_{t-1})$, which can be formulated as $$q(\mathbf{x}_{1:T} \vert \mathbf{x}_0) = \prod^T_{t=1} q(\mathbf{x}_t \vert \mathbf{x}_{t-1}),$$
where $q(\mathbf{x}_t \vert \mathbf{x}_{t-1}) \sim \mathcal{N}(\mathbf{x}_t;\mathbf{\mu}_t= \sqrt{1 - \beta_t} \mathbf{x}_{t-1}, \boldsymbol{\Sigma}_t=\beta_t\mathbf{I}), \forall t$. Noting that  the latent variable $\mathbf{x}_T$
becomes nearly an isotropic Gaussian distribution as $T \rightarrow \infty$, one can generate a novel data point from the original data distribution from a noise input, $\mathbf{x}_T \sim \mathcal{N}(\mathbf{0}, \mathbf{I})$, only if the reverse process can be run to acquire a sample from reverse distributions, $q(\mathbf{x}_{t-1} \vert \mathbf{x}_{t})$.

Unfortunately, $q(\mathbf{x}_{t-1} \vert \mathbf{x}_{t})$ cannot be easily estimate  due to the intractability caused by extensive computations involving the entire dataset for statistical estimates. Alternatively, one can learn a model to approximate these conditional probabilities in order to run the reverse diffusion process given below
$$p_\theta(\mathbf{x}_{0:T}) = p(\mathbf{x}_T) \prod^T_{t=1} p_\theta(\mathbf{x}_{t-1} \vert \mathbf{x}_t),$$
where $p_\theta(\mathbf{x}_{t-1} \vert \mathbf{x}_t) \sim  \mathcal{N}(\mathbf{x}_{t-1}; \boldsymbol{\mu}_\theta(\mathbf{x}_t, t), \boldsymbol{\Sigma}_\theta(\mathbf{x}_t, t))$. The approximation of conditioned probability distributions of the reverse diffusion process, $p_\theta(\mathbf{x}_{t-1} \vert \mathbf{x}_t)$, can be learned through DNNs. At this point, it is worth noting that the combination of $q(\cdot)$ and $p_\theta(\cdot)$ is very similar to a VAE. Thus, DGMs can be trained to maximize the ELBO with respect to the model's parameters, which can be written as
\begin{equation}
\nonumber \mathcal{L}(\theta; \mathbf{x}, \mathbf{z}) = \mathbb{E}{q(\mathbf{z} \mid \mathbf{x})}\left[\log p_{\theta}(\mathbf{x} \mid \mathbf{z})\right] - \operatorname{KL}\left(q(\mathbf{z} \mid \mathbf{x}) | p(\mathbf{z})\right).
\end{equation}
By being able to model the reverse process, we can generate new data by sampling from the learned distribution. There are several variations of DGMs, which mainly differ in the choice of noise schedule, the structure of the reverse model, and the type of data they are applied to. Some notable variations include:
\begin{itemize}
    \item[\large \adforn{10}] \textit{Denoising Diffusion Probabilistic Models }(DDPM) \cite{sohl2015deep} uses an explicit denoising function to model the reverse diffusion process, where the denoising function is parameterized by a DNN. The noise schedule in DDPM is a fixed, linear schedule.
    \item[\large \adforn{10}] \textit{Score-based models} \cite{song2020score} estimate the score function of the data distribution, which is the gradient of the log-density with respect to the data. Score-based models use the score function to guide the reverse diffusion process. They can be used in combination with different noise schedules and can be applied to various types of data.
    \item[\large \adforn{10}] \textit{Denoising Score Matching} (DSM) \cite{vincent2011connection} is a method to train energy-based models by matching the score function of the model to that of the data distribution. This approach has been combined with DGMs to learn expressive GMs.
\end{itemize}

\begin{figure}[t]
    \centering
    \includegraphics[width=0.45\textwidth]{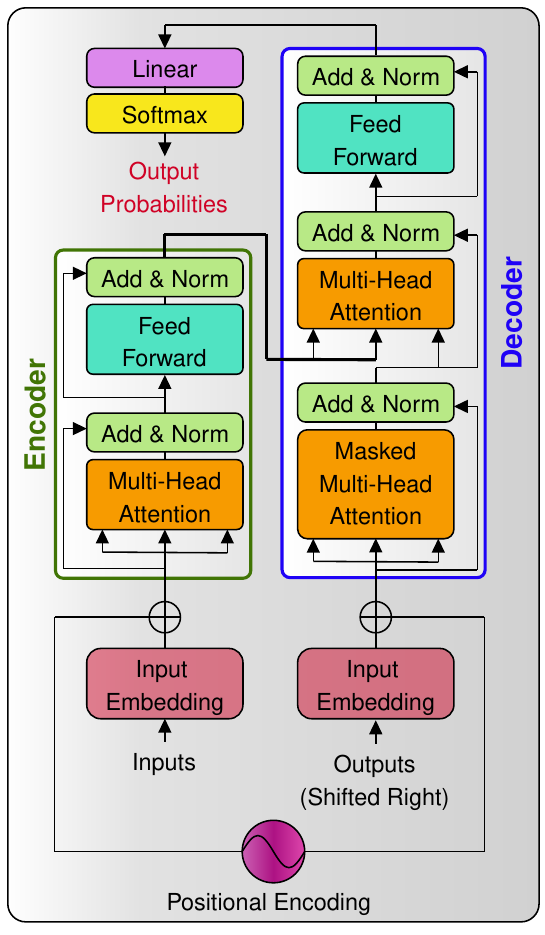}
    \caption{Schematic illustration of GTMs \cite{vaswani2017attention}.}
    \label{fig:GTM}
\end{figure}

\subsection{GENERATIVE TRANSFORMERS AND LARGE LANGUAGE MODELS}
\label{sec:GTM}
Unlike the above GMs learning the underlying probability distribution of the data and/or generate new samples from learned distributions, generative transformer models (GTMs) learn a mapping from input sequences to output sequences, which is then utilized to generate output sequences given new input sequences. GTMs are initially introduced in the seminal work ``Attention is All You Need'' by Vaswani et al.~\cite{vaswani2017attention}, wherein the key innovation is the self-attention mechanism that weighs the importance/relevance of a word in a sequence relative to all other words, thereby capturing the context more effectively. These inherent features allow GTMs to be particularly successful in natural language processing tasks by handling long sequences more effectively than previous architectures, e.g., RNNs and LSTMs.

\subsubsection{GTM ARCHITECTURE}

A GTMs consist of an encoder and a decoder, each composed of a stack of identical layers, as schematized in Fig. \ref{fig:GTM} reproduced from ~\cite{vaswani2017attention}. The encoder processes the input sequence and compresses the information into a sequence of continuous representations, which is then exploited by the decoder to generate the output sequence. The self-attention mechanism in both the encoder and decoder enable GTMs to consider the entire input sequence when encoding a word, or the entire previously generated output sequence when predicting the next word. In the sequel, we dig a bit deeper into the algorithmic and mathematical details of the GTM architecture in simple terms.

\begin{itemize}
    \item[\adforn{74}] \textit{Self-Attention Mechanism:}
    Consider an input sequence $X = [x_1, x_2, ..., x_n]$, where $x_i \in \mathbb{R}^d$ for $i = 1, ..., n$, $d$ is the dimensionality of the input, and $n$ is the sequence length. Moreover, $W_Q, W_K, W_V$,  $W_O$ are learnable query, key, and value, output weight matrices of the attention mechanism, respectively. For each input $x_i$, we compute a query $q_i = x_i W_Q$, a key $k_i = x_i W_K$, and a value $v_i = x_i W_V$. Next, the attention score $s_{ij}$ between each pair of inputs $x_i$ and $x_j$ is calculated as the dot product of their queries and keys, scaled by $1/\sqrt{d}$, and then softmaxed, i.e., $s_{ij} = \text{softmax}\left(q_i \cdot k_j^T / \sqrt{d}\right)$. After that, the output $z_i$ for each input $x_i$ is computed as a weighted sum of all values $v_j$, where the weights are the attention scores, i.e., $z_i = \sum_{j=1}^{n} s_{ij} v_j$. Finally, each $z_i$ is linearly transformed using the output weight matrix $W_O$ to obtain the final output of the self-attention mechanism for $x_i$, i.e., $o_i = z_i W_O$.

    \item[\adforn{74}] \textit{Input Embedding and Position Encoding:}
    The positional encoding and input embedding layer is the first stage of GTM architecture where the input tokens (e.g., words or subwords in the case of NLP tasks) are transformed into continuous vectors. The embedding operation is often a simple lookup operation in an embedding matrix, which is learned during training, thereby enabling the model to understand and utilize inputs in a meaningful way. Mathematically, if we have a vocabulary of size $V$ and we decide to use $d$-dimensional embeddings, the embedding matrix $E$ will be of size $V \times d$. If a particular input sequence consists of token indices $[i_1, i_2, ..., i_n]$, 
    these input embeddings are then combined with positional encodings to give the model information about the order of the tokens in the sequence. The result is a sequence of $d$-dimensional input representations $X = [x_1, x_2, ..., x_n]$, which are then passed through the encoder network.

    \item[\adforn{74}] \textit{Encoder:}
    The encoder consists of a stack of identical layers, each with two sub-layers: a multi-head self-attention mechanism (MHSAM), and a position-wise fully connected FNN. Denoting $X_\ell$ as the input to the $l$-th encoder layer, $\ell \leq L_E$,then the output of the MHSAM and FNN are given by $A_\ell = \text{MHSAM}(X_\ell)$ and $X_{\ell+1} = \text{FFN}(A_\ell)$, respectively. While the MHSAM applies the self-attention mechanism multiple times in parallel (with different weight matrices) and concatenates the results, $\text{FFN}$ applies a two-layer linear transformation followed by a ReLU activation, i.e., $\text{FFN}(x) = \text{max}(0, xW_1 + b_1)W_2 + b_2$.

    \item[\adforn{74}] \textit{Decoder:}
    Similar to the encoder, the decoder takes positional encoding and output embedding layer as inputs and is composed of a stack of identical layers; however, each of which has an additional MHSAM that attends to the output of the encoder. Denoting $Y_\ell$ as the input to the $\ell$-th, $\ell \leq L_D$, decoder layer, then the output of the first multi-head attention mechanism is $A^1_\ell = \text{MHSAM}(Y_\ell)$. The output of the second MHSAM, which attends to the encoder output $X_\ell$, is $A^2_\ell = \text{MHSAM}(A^1_\ell, X_\ell)$, whereas the output of the feed-forward network is $Y_{\ell+1} = \text{FFN}(A^2_\ell)$. 
    
    The decoder is also equipped with a masked MHSAM, a critical component designed to enforce an auto-regressive property by preventing positions from attending to subsequent positions in the output sequence during training phase. In this way, masked MHSAM ensures that the prediction for position $i$ can only depend on the known outputs at positions less than $i$, which is important since the output sequence is generated one token at a time from left to right during inference. and the model has no knowledge of future tokens at each step as they haven't been generated yet. It is worth noting that there are also layer normalization and residual connections involved in both the encoder and the decoder, which help with training DNNs.

    \item[\adforn{74}] \textit{Teacher Enforcing Training:}
    The GTM is trained using a process called teacher forcing, where the model is provided with the correct previous outputs as input to the decoder during training. Teacher enforcing is an efficient and stable method, because the model doesn't have to correct its own past mistakes before it can get the current output right. During training process below steps are repeated for each batch of examples in the training data, often for multiple passes over the data (epochs), until the model's performance on a validation set stops improving: 1) Input embedding and position encoding; 2) Encoding; 3) Decoding with teacher forcing; 4) Calculating loss function (e.g., cross-entropy loss) by comparing the prediction to the actual output; and 5)  Backpropagation and optimization to update hyperparameters to minimize the loss using an algorithm, typically a variant of stochastic gradient descent.
\end{itemize}

\subsubsection{GTM-BASED LARGE LANGUAGE MODELS (LLMs)}
GTMs has pave the way for the advent of LLMs, a significant breakthrough in the realm of NLP such as translation, summarization, and question answering. LLMs are trained on vast text data to learn/emulate patterns in human language (including aspects like grammar, syntax, world facts), eventually generating contextually coherent and relevant human-like text.

\begin{itemize}
    \item[\large \adforn{7}] \textit{ChatGPT by OpenAI~\cite{radford2019language, brown2020language}:} 
    The Generative Pretrained Transformer (GPT)  series (GPT-2/3/4) utilize the decoder part of the original transformer model explained above. ChatGPT's primary attribute is the pretraining on a large corpus of text data in an unsupervised manner, which allows them to predict the next word in a sentence, hence learning a representation of language. These models can then be fine-tuned to adapt to specific tasks like translation, summarization, or sentiment analysis.

    \item[\large \adforn{7}] \textit{Google BARD~\cite{devlin2018bert} and T5~\cite{raffel2019exploring}:}
    Google's Bidirectional Encoder Representations from Transformers (BERT) is also a modification of the above original model and only uses the encoder part. Unlike the GPT series, BERT is pretrained in a bidirectional manner to have a better understanding of context. The BERT utilizes a "masked language model" objective where a percentage of the input tokens are masked, and the model predicts these tokens based on the context provided by the non-masked tokens. This bidirectional pretraining allows BERT to have a better understanding of context.
    
    Google's another unique variant is the Text-to-Text Transfer Transformer (T5) model, which approaches all NLP task as a text-to-text problem, where both the input and output are sequences of text. Being pretrained on a large corpus of text data using a denoising autoencoder objective, T5 allows for its application on a wide range of tasks with minimal task-specific modifications.

    \item[\large \adforn{7}] \textit{Microsoft's Turing NLP~\cite{smith2022using}:}
    The initial version of Turing-NLG is a LLM trained on a diverse and extremely large corpus of data with 17 billion parameters. Similar to OpenAI's and Google's LLMs, Turing-NLG also utilizes the above GMT architecture and is pretrained in a generative manner to generate human-like text and has demonstrated remarkable performance in tasks like writing essays, summaries, and answering questions. Turing-NLG has been taken to another level by Microsoft and NVIDIA's Megatron-Turing Natural Language Generation (NLG), which is trained using DeepSpeed and Megatron, advanced tools designed to streamline the training of exceptionally large neural networks, over 530 billion of parameters, pushing the boundaries of what is possible in LLM and NLP.
\end{itemize}

\begin{figure}[t]
    \centering
    \includegraphics[width=0.49\textwidth]{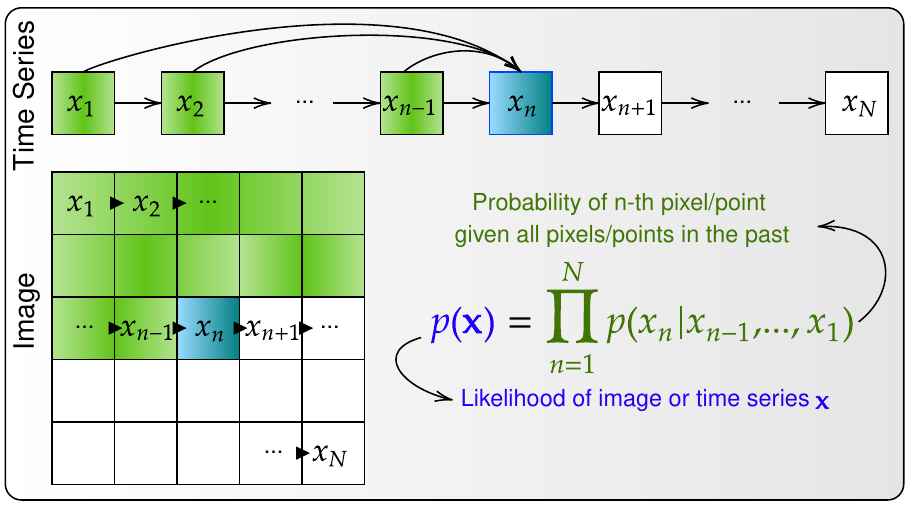}
    \caption{Schematic illustration of GAMs.}
    \label{fig:GAM}
\end{figure}

\begin{table*}[t]
\caption{Comparison of deep generative autoregressive models.}
\label{tab:GAM}
\centering
\resizebox{\textwidth}{!}{%
\begin{tabular}{|l|l|l|l|}
\hline
\textbf{Model} & \textbf{Strengths} & \textbf{Weaknesses} & \textbf{Applications} \\ \hline
FVSBN & \begin{tabular}[c]{@{}l@{}}Simple architecture, easy to train\end{tabular} & \begin{tabular}[c]{@{}l@{}}Limited capacity, not scalable\end{tabular} & General-purpose \\ \hline
NADE & \begin{tabular}[c]{@{}l@{}}Efficient training,
good generalization\end{tabular} & \begin{tabular}[c]{@{}l@{}}More complex than FVSBN,  less efficient than MADE\end{tabular} & General-purpose \\ \hline
MADE & \begin{tabular}[c]{@{}l@{}}Efficient training, good performance\end{tabular} & \begin{tabular}[c]{@{}l@{}}Autoencoder-based,  less interpretable\end{tabular} & General-purpose \\ \hline
PixelRNN/CNN & \begin{tabular}[c]{@{}l@{}}High-quality images,  captures spatial relationships\end{tabular} & \begin{tabular}[c]{@{}l@{}}High computational cost,  sequential generation\end{tabular} & Image generation \\ \hline
WaveNet & \begin{tabular}[c]{@{}l@{}}High-fidelity audio, long-range dependencies\end{tabular} & \begin{tabular}[c]{@{}l@{}}High computational cost,  slow generation\end{tabular} & Audio generation \\ \hline
\end{tabular}
}
\end{table*}

\subsection{GENERATIVE AUTOREGRESSIVE MODELS (GAMs)}
\label{sec:GAM}
{
GAMs have their roots in traditional autoregressive models, which were originally developed for time series analysis \cite{yule1927method, box2015time}. They are designed to generate data by learning the conditional probability distribution of a sequence of data points. These models leverage the autoregressive property, where each data point in the sequence is assumed to be dependent on its previous data points.  To be more specific, consider a sequence of data points represented by the vector $\mathbf{x} = [x_1, x_2, \ldots, x_N]$, as shown in Fig. \ref{fig:GAM}. The GAMs aim to factorize the joint probability distribution of the sequence as a product of conditional probability distributions $p(\mathbf{x}) = \prod_{n=1}^{N} p(x_n \mid x_{<n})$, where data point $x_n, \: \forall n,$ is conditioned on  all its preceding data points in the sequence, $x_{<n}$. By modeling these dependencies, GAMs can generate realistic and diverse new data points by sampling from these conditional probability distributions sequentially, leading to the development of powerful GMs capable of modeling complex data distributions. Training a generative GAMs typically involves using maximum likelihood estimation (MLE), which is given by
\begin{equation}
\nonumber \mathcal{L}(\theta) = \sum_{n=1}^{N} \log p(x_n \mid \theta),
\end{equation}
where $\theta$ represents the model parameters. During the generation/sampling process, GAMs generate new samples sequentially by sampling from the learned conditional probability distributions. 

By employing DL techniques, deep GAMs have pushed the boundaries of GenAI thanks to the following key characteristics: 1) capable of capturing complex dependencies between variables in the data distribution, resulting in more accurate and realistic samples; 2) scalable to model large datasets and high-dimensional data samples, 3) flexible to be adapted to different data types and domains by tailoring DNNs to suit the specific characteristics of the data; and 4) able to capture long-range dependencies in data, which was previously challenging for traditional autoregressive models, through incorporation of techniques such as attention mechanisms and dilated convolutions. The most common deep GAMs are explained below:

\begin{itemize}
    \item[\large \adforn{12}] \textit{Fully Visible Sigmoid Belief Networks} (FVSBNs)  are a type of GAM that uses a feedforward neural network (FNN) to model the conditional probability distribution of each variable given its preceding variables \cite{neal1992connectionist}. The network consists of a single hidden layer with sigmoid activation functions and is trained using maximum likelihood estimation.
    \item[\large \adforn{12}] \textit{Neural Autoregressive Distribution Estimator} (NADE) extends FVSBNs by employing a DNN to model the conditional probability distributions \cite{larochelle2011neural}. NADE shares parameters across different dimensions of the input, resulting in more efficient training and better generalization.
    \item[\large \adforn{12}] \textit{Masked Autoencoder for Distribution Estimation} (MADE) is a deep GAM that uses an AE architecture \cite{germain2015made}. MADE applies masking to the weights of the AE to enforce the autoregressive property, ensuring that each output dimension only depends on the preceding input dimensions. This masking strategy leads to more efficient training and better performance compared to NADE.
    \item[\large \adforn{12}] \textit{PixelRNN} \cite{oord2016pixel} and \textit{PixelCNN} \cite{van2016conditional} are deep GAMs specifically designed for generating images. PixelRNN uses a RNN architecture, while PixelCNN employs a CNN. Both models are capable of capturing long-range dependencies and complex spatial relationships in images, leading to high-quality image generation.
    \item[\large \adforn{12}] \textit{WaveNet} \cite{oord2016wavenet} is a deep GAM designed for generating raw audio waveforms. It employs dilated convolutions to capture long-range dependencies in the audio data efficiently. WaveNet has been successful in generating high-fidelity, realistic-sounding speech and music.
\end{itemize}

\begin{figure}[t]
    \centering
    \includegraphics[width=0.49\textwidth]{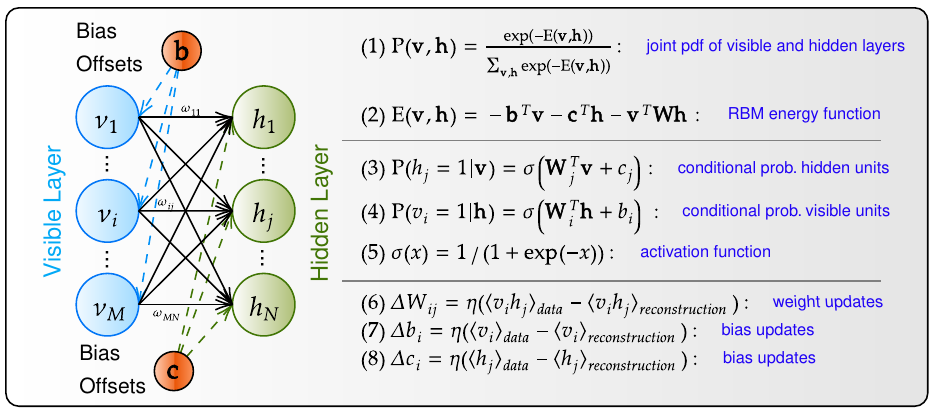}
    \caption{Schematic illustration of RBMs.}
    \label{fig:RBM}
\end{figure}

\subsection{RESTRICTED BOLTZMANN MACHINES (RBMs)} 
\label{sec:RBM}

RBMs are a type of GM that can learn probability distributions over binary-valued input data. 
In 2002, Geoffrey Hinton introduced the RBM, which is a simplified version of the Boltzmann Machine that removes intra-layer connections, making the model more computationally tractable and suitable for training with gradient-based methods like contrastive divergence \cite{hinton2002training}. As shown in Fig. \ref{fig:RBM}, RBMs consist of a visible layer and a hidden layer, with connections only between the visible and hidden nodes. This structure is known as a symmetric bipartite graph, and the "restricted" in the name arises from the absence of intra-layer connections.

RBMs model the joint probability distribution between the visible units $\mathbf{v}$ and hidden units $\mathbf{h}$, $P(\mathbf{v}, \mathbf{h})$ in Fig. \ref{fig:RBM}.(1), where $E(\mathbf{v}, \mathbf{h})$ is the energy function as given in Fig. \ref{fig:RBM}.(2), $\mathbf{W}$ is the weight matrix, and $\mathbf{b}$/$\mathbf{c}$ is the bias offset vector for the visible/hidden units. 
The RBM training process is typically performed using an unsupervised learning algorithm called Contrastive Divergence (CD) which follows below steps: 
\begin{enumerate}
    \item \textit{Initialization}: Random initialization of the weight matrix and bias offset vectors.
    
    \item \textit{Positive phase (forward propagation)}: Given an input data sample, compute the hidden layer activations using the current weights and biases, i.e., $P(h_j = 1 \mid \mathbf{v})$ as given in Fig. \ref{fig:RBM}.(3). 
    
     \item \textit{Negative phase (backward propagation)}: Reconstruct the visible layer activations by sampling from the distribution using the current weights and biases, i.e. $P(v_i = 1 \mid \mathbf{h})$ as given in Fig. \ref{fig:RBM}.(4). 
     
     \item \textit{Update}: Update the weights and biases based on the difference between the outer products of the input data and the hidden activations from the positive phase, and the outer products of the reconstructed visible layer activations and the hidden activations from the negative phase, i.e., Fig. \ref{fig:RBM}.(6)-Fig. \ref{fig:RBM}.(8).
     
     \item \textit{Repeat}: Perform steps 2-4 for a fixed number of iterations or until convergence, using different input data samples for each iteration.
\end{enumerate}

Once an RBM has been trained, it can sample the hidden layer activations from the prior distribution over the hidden layer, then use the sampled hidden activations to generate visible layer activations (input data) through the learned conditional probability distributions. This generative process modeling the joint probability distribution between the visible and hidden layers allows RBMs to produce new samples that share similar characteristics with the input data they were trained on. 

Even though RBMs were originally designed for binary data, this limitation has been overcome with the development of variations of RBMs that can handle different types of data, such as real-valued, categorical, or even multi-modal data. One such variation is the Gaussian-Bernoulli RBM, which is designed to handle continuous real-valued data \cite{welling2004exponential}. There are other types of RBMs as well, such as the softmax RBM \cite{nair2009object}, which can handle categorical data, or continuous-valued RBM to incorporate additional structure or sparsity constraints (e.g., sparse RBMs) \cite{lee2007sparse}. These variants have further expanded the applicability of RBMs to a broader range of ML problems such as feature learning, dimensionality reduction, and collaborative filtering \cite{hinton2006fast}.

\begin{figure}[t]
    \centering
    \includegraphics[width=0.49\textwidth]{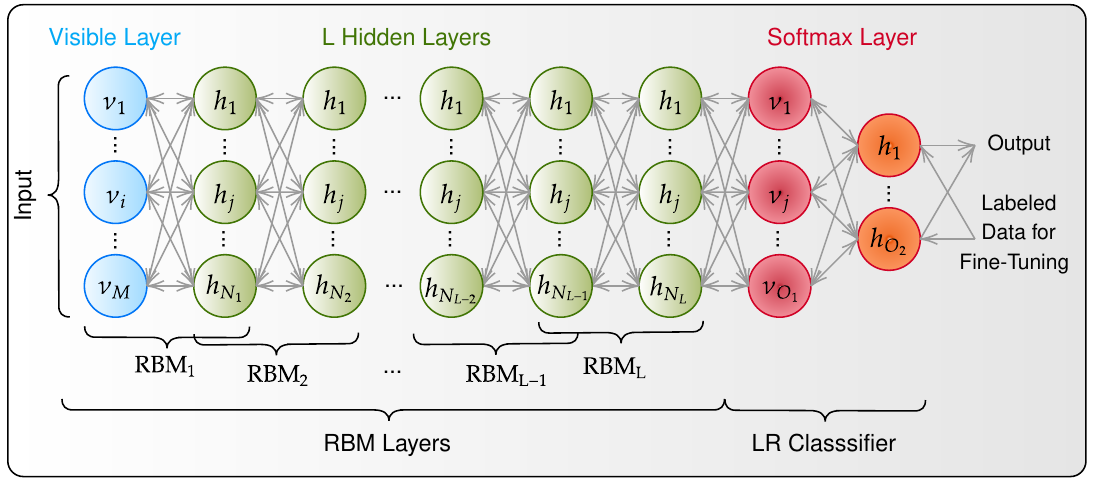}
    \caption{Schematic illustration of DBNs.}
    \label{fig:DBN}
\end{figure}

\subsection{DEEP BELIEF NETWORKS (DBNs)} 
\label{sec:DBN}

Before the introduction of DBNs, training DNNs was challenging due to the vanishing gradient problem and the difficulty of optimizing the objective functions. Hinton et. al. proposed a new method for training deep architectures by stacking multiple RBMs on top of one another to form a DBN \cite{hinton2006fast}. As shown in Fig. \ref{fig:DBN}, a DBN with $L$ hidden layers can learn the joint probability distribution
\begin{equation}
\nonumber p(\mathbf{v}, \mathbf{h}_1, \mathbf{h}_2, \dots, \mathbf{h}_L) = p(\mathbf{v}|\mathbf{h}_1) \prod_{l=1}^{L-1} p(\mathbf{h}_l|\mathbf{h}_{l+1}),
\end{equation}
where each term $p(\mathbf{h}_{l}|\mathbf{h}_{l+1})$ corresponds to an RBM, representing the conditional probability distribution between adjacent layers of hidden variables. In this manner, DBNs can efficiently learn  hierarchical representations from data by a greedy layer-wise unsupervised pre-training method to initialize the weights of the network, followed by a supervised fine-tuning step using backpropagation.

In the unsupervised pre-training phase, the goal is to initialize the weights of the network by learning a hierarchical representation of the input data. The DBN is trained layer by layer, where each layer is an individual RBM. Train the first layer RBM using the input data. The input data is treated as visible units, and the RBM learns to reconstruct it through the hidden units by leveraging CD algoritm as explained in the previous section. The hidden unit activations of the first layer RBM are used as input (data) for the second layer RBM to train the second layer RBM, which is repeated for all layers in the DBN. After completing the unsupervised pre-training, the DBN has been initialized with weights that can capture a hierarchical representation of the input data.

In the supervised fine-tuning phase, the aim is to adjust the weights of the DBN using labeled data to perform a specific task, such as classification or regression.
A supervised output layer (e.g., softmax) uses the labeled data to perform backpropagation to update the weights of the entire network, including the pre-trained layers and the output layer. This step fine-tunes the weights to optimize the DBN for the specific SL task.
The backpropagation process is repeated until a predefined stopping criterion is met, such as reaching a maximum number of epochs or achieving a desired level of performance on the validation set.

The success of DBNs sparked a resurgence of interest in neural networks and DL, paving the way for the development of more advanced architectures and training techniques, such as CNNs, RNNs, and AEs. As a GM, DBNs have been applied to various tasks, including image generation, feature learning, dimensionality reduction, anomaly detection, collaborative filtering and recommendation, sequence modeling, and natural language processing.

\begin{table}[t!]
\centering
\caption{GMs along with Potential Complementary Roles}
\label{tab:GMways_examples}
\resizebox{0.49\textwidth}{!}{%
\begin{tabular}{|l|l|}
\hline
\textbf{Comp. Role }    & \textbf{Generative Model Examples}                                                  \\ \hline
{Data Augmentation} & DCGAN, CycleGAN, VAEs, RBMs, PixelCNN/RNN, FGMs              \\ \hline
{Data Imputation}   & VAEs, RBMs, DBNs, PixelCNN/RNN, Normalizing FGMs                                 \\ \hline
{Disentanglement}   & $\beta$-VAEs, InfoGANs, FactorVAEs, FGMs             \\ \hline
{Regularization}    & VAEs (as Bayesian prior), RBMs (as pre-training)                                    \\ \hline
{Dim. Reduction}    & VAEs, RBMs, DBNs, FGMs                                               \\ \hline
{Feature Learning} & BiGAN, InfoGAN, VAEs, RBMs, DBNs, PixelCNN/RNN, FGMs          \\ \hline
{Transfer Learning}  & StarGAN, CycleGAN, conditional VAEs, PixelCNN/RNN                     \\ \hline
{Semi-SL}   & SGAN, CatGAN, VAEs, PixelCNN/RNN                                 \\ \hline
{Multi-Task Learning}   & MT-GAN, MT-VAEs, PixelCNN/RNN                                  \\ \hline
{Balancing Exp.}    & VAEs, GANs , PixelCNN/RNN (all with model-based RL) \\ \hline
{Imitation Learning}    & GAIL, VAEs, PixelCNN/RNN (all with behavioral cloning) \\ \hline
\end{tabular}%
}
\end{table}

\subsection{KEY GM PERFORMANCE METRICS}
\label{sec:GM_metrics}
 It is non-trivial to provide a quantitative or qualitative comparison among the wide variety of GMs  presented above because performance may vary depending on the specific GM model variant, architecture, and dataset. Instead, we provide a set of key performance metrics that should be taken into account while comparing GMs: 

\begin{itemize}
    \item[\large \adforn{14}] \textit{Sample quality:} How realistic and diverse are the generated samples? High-quality samples should be visually or statistically indistinguishable from real data.
    \item[\large \adforn{14}] \textit{Training Stability:} How stable is the training process? Some GMs may suffer from instability issues, making it difficult to converge to an optimal solution.

    \item \textit{Model Size and Computational Efficiency:}How large is the model in terms of its parameters, architecture, and memory footprint? The size of a model often correlates with its computational demands. Training large models can be resource-intensive and may require specialized hardware, like powerful GPUs or TPUs. Furthermore, larger models may result in slower inference times, which is not ideal for real-time applications.

    \item \textit{Memory Consumption:}Larger models demand more memory, both during training and inference. This can be a challenge when deploying models in resource-constrained environments like mobile devices or edge devices. When evaluating GMs, it's beneficial to consider the trade-offs between model size and performance.

    \item[\large \adforn{14}] \textit{Inference Speed: }How quickly can new samples be generated? Faster inference speeds are desirable for real-time or large-scale applications.
    \item[\large \adforn{14}] \textit{Scalability:} How well does the model handle large datasets and high-dimensional data? GMs should be able to scale efficiently to model complex data distributions.
    \item[\large \adforn{14}] \textit{Expressiveness:} How flexible is the model in capturing a wide range of data distributions? A more expressive model can adapt to different types of data and learn more complex relationships.
    \item[\large \adforn{14}] \textit{Disentanglement: }How well does the model separate the underlying factors of variation in the data? Disentangled representations can improve interpretability and facilitate downstream tasks.
    \item[\large \adforn{14}] \textit{Ease of Training:} How easy is it to train the model, considering factors like hyperparameter tuning and the availability of training resources?
    \item[\large \adforn{14}] \textit{Interpretability:} How interpretable are the model's learned representations and the generative process itself?
    \item[\large \adforn{14}] \textit{Conditional generation:} How well can the model generate samples conditioned on specific attributes, such as labels or latent variables?
\end{itemize}

Following from the discussion in Section \ref{sec:GM_ways}, we provide a comprehensive list of potential GMs relevant to complementary roles across various ML techniques in Table \ref{tab:GMways_examples}.  Please also note that this table is not exhaustive, and there might be other GMs and their corresponding roles that are not listed here. Additionally, many GMs can be used in combination or can be extended to address various complementary roles in different ML techniques.


\begin{figure}[t]
    \centering
\input{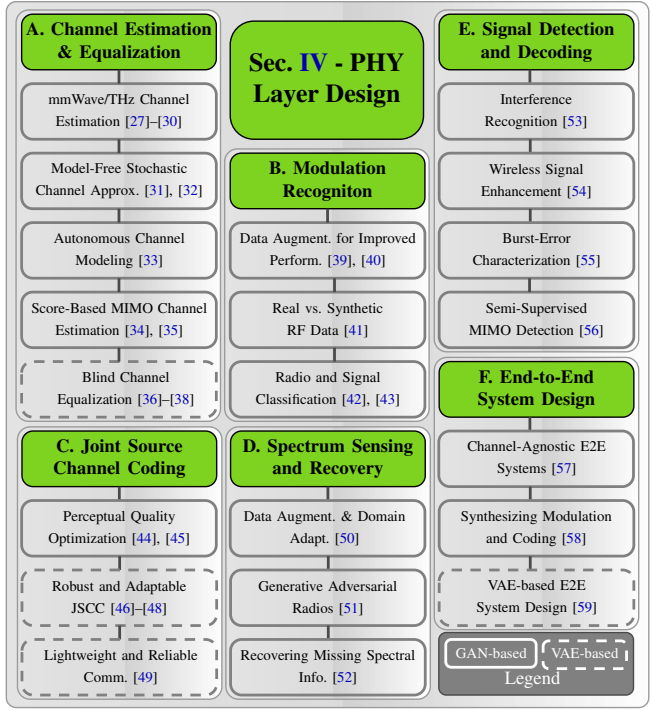}
    \caption{GenAI applications in PHY Layer Design.}
    \label{fig:apps-phy}
\end{figure}

\section{GenAI FOR PHYSICAL LAYER DESIGN}
\label{sec:PHY} 
GMs have shown significant promise in addressing various challenges associated with the physical layer aspects of wireless networks. By learning the underlying structure and patterns in data, GM can contribute to improving the efficiency and reliability of wireless communication in several ways, which are the main focus of this section.

\subsection{CHANNEL MODELING, ESTIMATION, AND EQUALIZATION }

\subsubsection{CHANNEL MODELING AND ESTIMATION}

GMs provide a powerful means for learning complex channel characteristics in wireless communication systems, including those with mathematically intractable or unknown channel models. They are capable of capturing non-linear distortions and time-varying channel properties, thereby enabling accurate channel estimation and modeling. In the remainder, we will cover state-of-the-art GM-based channel modeling and estimation methods studied in the literature:

\subsubsection{mmWave and THz Channel Estimation:}
Due to distinct signal propagation characteristics and harsh channel impediments at high carrier frequencies, mmWave and THz communications present unique challenges in channel estimation and modeling, especially with large number of antenna elements. GMs have been successfully applied in this domain to improve the physical layer aspects of wireless networks. Li et al. \cite{Li2018channel} proposed a novel solution for highly-mobile mmWave systems, leveraging DL tools to efficiently learn and predict mmWave channel covariance matrices by treating them as 2D images. A cGAN is leveraged to learn the implicit mapping function between the received training signals, which act as RF signatures of the environment, and a sparse representation of the large-dimensional channel covariance matrix. In a separate study, Zhang et al. \cite{Zhang2021channel,Zhang2022channel} developed a data-driven air-to-ground channel estimation framework for mmWave communications for UAV networks using a cGAN. To broaden the applicability of the trained channel model, a cooperative framework based on a distributed cGAN architecture is proposed, allowing UAVs to collaboratively learn the mmWave channel distribution in a fully-distributed manner. Meanwhile, Balevi et al. \cite{Balevi2021channel} proposed a GAN-based channel estimation method to accurately estimate frequency selective mmWave and THz channels, even with reduced number of pilots at low SNR conditions. GANs are trained to produce channel samples from the true but unknown channel distribution, which is then used as a prior to estimate the current channel, without need for retraining even if the number of clusters and rays change considerably.

\subsubsection{Model-Free Stochastic Channel Approximations:}
Recent works on training GMs to approximate wireless channel responses has focused on accurately reflecting the PDFs of stochastic channel behaviors \cite{Timothy2019channel,Smith2019channel}. Timothy et al. \cite{Timothy2019channel} introduced the use of variational GANs to provide appropriate architecture and loss functions that accurately capture these stochastic behaviors. Their model-free approach enables the optimization of communication system performance for specific over-the-air channel effects or scenarios exhibiting complex combinations of channel effects. Similarly, Smith et al. \cite{Smith2019channel} demonstrated the versatility of GAN architecture in handling complex communication environments with non-linear amplifier distortion, pulse shape filtering, ISI, frequency-dependent group delay, multipath, and non-Gaussian statistics.

\subsubsection{Autonomous Channel Modeling:}
Yang et al. \cite{Yang2019channel} proposed a novel GAN framework for autonomous wireless channel modeling, eliminating the need for complex theoretical analysis or data processing typically required in traditional methods like ray-tracing and geometry-based stochastic channel models. The GAN is trained with raw measurement data to approximate the distribution of a typical AWGN channel, showcasing its effectiveness and potential to improve the physical layer of wireless networks.

\subsubsection{MIMO Channel Estimation with Score-Based GMs:}
Arvinte et al. \cite{Arvinte2022channel,Arvinte2022channel_journal} presented a novel approach for MIMO channel estimation using score-based GMs, which estimate the gradient of the log-prior of channels in high-dimensional space by leveraging the model for channel estimation through posterior sampling. The score-based model is trained on channel realizations from the clustered delay line (CDL)-D model and demonstrates competitive in- and out-of-distribution performance compared to GAN and compressed sensing methods. In tests on CDL-D and CDL-C channels, the proposed approach achieves significant gains in channel estimation error and end-to-end coded performance, highlighting the robustness and effectiveness of the method for MIMO channel estimation in wireless networks.

\subsubsection{BLIND CHANNEL EQUALIZATION}
Channel equalization is essential to compensate for distortions introduced by communication channels. As the received signals are typically distorted versions of the transmitted ones, channel equalization aims to mitigate these distortions to enhance the quality of the received signals. Recent advancements in channel equalization involve the use of VAEs to develop blind equalization techniques, as outlined below:

In \cite{song2023equalization}, the authors propose an innovative approach for blind channel equalization and estimation using the vector quantized (VQ) VAE framework. This VQ-VAE-based blind channel equalizer addresses the limitations of previous VAE-based equalizers by broadening their applicability to nonlinear systems employing high-order modulation formats. Simulation results reveal that the proposed method achieves performance comparable to a data-aided equalizer using the minimum-mean-squared-error criterion, while surpassing blind constant modulus algorithm (CMA) and VAE-based channel equalizers.

In \cite{Avi2018equalization}, the authors introduce a novel maximum likelihood estimation approach for blind channel equalization using VAEs. The proposed VAE-based blind channel equalizer demonstrates significant improvements in the error rate of reconstructed symbols compared to the CMA. This new equalization method enables lower latency acquisition of unknown channel responses and extends the application of DL techniques to blind channel equalization for noisy ISI channels with unknown impulse responses.

In \cite{Lauinger2022equalization}, the authors explore the potential of adaptive blind equalizers based on variational inference for carrier recovery in optical communications. The proposed generalized VAE equalizer accommodates higher-order modulation formats, probabilistic constellation shaping (PCS), receiver oversampling, and dual-polarization transmission. The VAE equalizer provides reliable channel estimation and outperforms the state-of-the-art CMA for PCS in both fixed and time-varying channels. This research underscores the potential of GMs to enhance the physical layer design, particularly in coherent optical communication systems.

\subsection{MODULATION RECOGNITION}
GMs, particularly GANs, have demonstrated remarkable potential in recognizing and classifying modulation schemes used in wireless communication systems. By learning the unique features and patterns of different modulation schemes, these models accurately identify the modulation type of an incoming signal, paving the way for adaptive modulation techniques and promoting efficient spectrum utilization.

\noindent {\large\adforn{3}} \textit{Data Augmentation and Improved Performance:} In \cite{patel2020modulation} and \cite{Tang2018modulation}, the authors propose data augmentation methods using cGAN and auxiliary classifier GANs (ACGANs), respectively, to tackle the challenge of limited high-quality labeled data in DL-based automatic modulation classification (AMC) for wireless communications. Both approaches generate high-quality, labeled wireless modulation data from a small set of real data while preserving the high-level features of the original dataset, improving the performance of CNN-based AMC systems. These cGAN-based and ACGAN-based data augmentation methods yield significant improvements in modulation classification accuracy, demonstrating the potential of GMs in enhancing the physical layer design.
\noindent {\large\adforn{3}} \textit{Real vs. Synthetic RF Data:} In \cite{Clark2021rf}, the authors examine the impact of data types and data augmentation on the performance of RF-ML systems, particularly in the context of AMC. The authors explore three main questions: the effectiveness of synthetically trained systems in real-world scenarios, the role of augmentation in the RF-ML domain, and how knowledge of signal degradations caused by the transmission channel contributes to the system performance. The paper reveals the value of augmentation compared to extended data capture and whether understanding the distributions of degradation sources affects the network's ability to achieve peak performance. 

\noindent {\large\adforn{3}} \textit{Radio and Signal Classifiers:} In \cite{Mingxuan2018modulation} and \cite{Li2018modulation}, the authors present Radio Classify GAN (RC-GANs) and Signal Classifier GANs (SC-GANs), respectively. These are scalable, end-to-end frameworks that introduce GANs into the radio ML domain for modulation recognition. RCGANs are trained on a synthetic RF dataset and learn features through self-optimization during an extensive data-driven training process. On the other hand, SC-GANs improve GANs by effectively avoiding nonconvergence and mode collapse problems caused by the complexity of RF signals. Both RC-GANs and SC-GAN demonstrate improvements in classification accuracy compared to traditional DL methods and exhibit robustness against noise. By employing GANs for modulation recognition, these methods offer a promising approach to improve the physical layer of wireless networks by enabling AMC without relying on manual experience.

\subsection{JOINT SOURCE-CHANNEL CODING (JSCC)} GMs have been effectively utilized in JSCC schemes, where source and channel codes are designed together to achieve efficient and robust communication. By learning the statistical properties of the source data and the characteristics of the channel, these models facilitate the development of optimized joint source-channel codes, leading to improved end-to-end system performance. In the remainder, we outline recent developments in this area, focusing on the application of GMs to various domains and their ability to address the challenges faced in modern communication systems.

\noindent {\large\adforn{3}} \textit{Image Transmission \& Perceptual Quality Optimization:} Perceptual quality has become an essential aspect of modern communication systems. In \cite{erdemir2022jointsource}, the authors propose the use StyleGAN for two novel JSCC schemes, InverseJSCC and GenerativeJSCC, for enhancing the perceptual quality of wireless image transmission in extreme physical conditions. These schemes leverage deep GMs to optimize a weighted sum of mean squared error and learn perceptual image patch similarity losses. Moreover, \cite{estiri2020jscc} investigates the performance of VAEs compared to standard AEs for the transmission of image data over Gaussian point-to-point wireless channels, demonstrating that VAEs are more robust to channel degradation than AEs. Additionally, the authors employ the structural similarity, a perception-based metric, to optimize neural networks for improved human perceptual quality of the reconstructed images at the receiver.

\noindent {\large\adforn{3}} \textit{Robust and Adaptable JSCC with VAEs:} JSCC schemes must maintain robust performance under varying channel conditions to be effective in real-world communication systems. In a series of studies, \cite{Malur2019jscc}, \cite{Malur2020jscc}, and \cite{Malur2021jscc}, the authors use VAE-based approaches that exhibit robustness to channel condition variations, offering more practical solutions for wireless networks. These approaches demonstrate adaptability in handling different types of data sources and communication scenarios, such as Gaussian sources over multiple AWGN channels, distributed Gaussian sources over a multiple access AWGN channel, and analog independent additive noise channels.

\noindent {\large\adforn{3}} \textit{Lightweight Communication with VQ-VAEs:} In \cite{nemati2022jscc}, the authors investigate the use of VQ-VAEs for point-to-point wireless communication. They modify the VQ-VAE's training process to design a JSCC that is robust against noisy wireless channels, aiming to make the physical/link layer lighter while preserving system reliability. This approach offers an alternative to conventional separated source-channel coding schemes and demonstrates its potential for lighter physical/link layers in IoT networks.

\subsection{SPECTRUM SENSING AND RECOVERY}
GMs have been effectively employed in spectrum sensing techniques, which are crucial for efficient spectrum management and dynamic spectrum access in cognitive radio networks. By learning the characteristics of signal occupancy and interference patterns in the spectrum, these models can accurately detect the presence or absence of primary users, enabling secondary users to access the spectrum without causing harmful interference. 

\noindent {\large\adforn{3}} \textit{Data Augmentation and Domain Adaptation:} In \cite{kemal2019saga}, the authors address the challenges of training data augmentation and domain adaptation in cognitive radio applications by proposing the spectrum augmentation/adaptation with GANs (SAGA) approach. This novel method leverages a deep GAN to generate additional synthetic training data and adapt the training data to spectrum dynamics. The results demonstrate that the SAGA approach significantly increases classifier accuracy, which is sustained with domain adaptation as spectrum conditions change. By augmenting and adapting training data on the fly, SAGA enables cognitive radios to use ML effectively and seamlessly transition between different spectrum conditions.

\noindent {\large\adforn{3}} \textit{Generative Adversarial Radio Spectrum Networks:} In \cite{Roy2019spectrum}, the authors explore the use of GANs for generative replay to improve the physical layer of wireless networks. This approach allows for the creation of signals with similar structure and properties to arbitrary signals, without being exact replicas. The study extends the use of GANs to full-band spectral generation, validating the feasibility of the approach, refining the algorithm, and quantifying its capabilities. The authors aim to push the boundaries of GANs in the radio domain by examining their ability to generate large portions of radio signals, accurately produce samples from random generators, and simulate large radio environments using a naive learning approach.

\noindent {\large\adforn{3}} \textit{Recovering Missing Spectral Information:} In \cite{Dung2018missing}, the authors propose GANs called SARGAN to recover missing spectral information in ultra-wideband radar systems. The proposed SARGAN learns the relationship between original and missing-frequency-band signals by observing numerous possible training pairs. The initial results indicate that the recovered signals achieve more than an 18 dB gain in the SNR, even when up to 90\% of the operating spectrum is missing, without requiring prior knowledge of the missing frequency band locations. This approach offers advantages over traditional spectral recovery techniques, as it shifts the computational complexity to the training phase and does not require information on the missing band locations.

\subsection{ENHANCED DETECTION AND DECODING}
GMs have been employed in the development of advanced signal detection and decoding algorithms, enabling efficient signal detection in complex scenarios where traditional methods may fail. By learning the underlying relationships between transmitted and received signals, these models can adapt to varying channel conditions, reducing the need for explicit channel estimation and minimizing the impact of channel estimation errors on the performance of signal detection algorithms. 

\noindent {\large\adforn{3}} \textit{Interference Recognition with Quadruple GAN:} In \cite{Xiaodong2021interference}, the authors propose an improved Quadruple GAN (QGAN) approachh for wireless interference signal recognition, which is crucial for mitigating cross-technology interference in wireless local area networks (WLANs). The improved QGAN enhances the performance of the original QGAN by substituting it with an auxiliary classifier GAN (ACGAN) and optimizing loss functions for generative, representation, and classification sub-networks. Furthermore, a lightweight model based on knowledge distillation is introduced to reduce memory consumption and computational complexity during the inference phase. 

\noindent {\large\adforn{3}} \textit{Wireless Signal Enhancement:} \cite{Zhou2020signal} presents a wireless Signal Enhancement Generative Adversarial Network (WSE-GAN) for adaptively learning signal characteristics and enhancing wireless communication signals in time-varying systems. The WSE-GAN incorporates the raw time-domain signal as a condition to achieve state-of-the-art enhancement effects while preserving symbol information. The proposed network demonstrates robust learning ability for dynamic channel effects and excellent adversarial capabilities for signal jitter and skews. By focusing on original signals, WSE-GAN maintains exceptional enhancement performance even when the statistical characteristics of channel effects differ between training and testing stages.

\noindent {\large\adforn{3}} \textit{Burst-Error Characterization:} In \cite{Wang2007bursterror}, the authors present deterministic-process-based GMs (DPBGMs) for simulating wireless channels with hard and soft decision outputs, which are crucial for designing and evaluating wireless communication protocols and error-control schemes. The proposed DPBGMs are applied to uncoded enhanced general packet radio service systems with typical urban and rural area channels, demonstrating an excellent approximation to the burst-error statistics of the target hard and soft error sequences. This research showcases the use of GMs to accurately and efficiently simulate digital wireless channels, enabling improved design and performance evaluation of wireless communication systems at the physical layer.

\noindent {\large\adforn{3}} \textit{Semi-Supervised MIMO Detection with CycleGAN:} \cite{zhu2023semisupervised} proposes a novel semi-supervised deep MIMO detection approach using CycleGAN for wireless communication systems without prior knowledge of underlying channel distributions. The proposed CycleGAN detector consists of a bidirectional loop of two modified LS-GAN, where the forward LS-GAN models the transmission process, and the backward LS-GAN detects the received signals. By optimizing the cycle-consistency of transmitted and received signals, the method can be trained online and semi-supervisedly using both pilots and received payload data, effectively reducing overhead. The CycleGAN detector does not require explicit channel estimation and performs well under unknown channel effects. The proposed method outperforms existing semi-blind DL detection techniques and conventional linear detectors in terms of BER and achievable rate, especially in scenarios with nonlinear power amplifier distortion. 

\subsection{END-TO-END (E2E) SYSTEM DESIGN}
GMs offer promising advancements in E2E system design for wireless communications, optimizing the entire communication system from transmitter to receiver. By learning relationships between system parameters and performance metrics, these models can predict optimal configurations and settings for various components, resulting in improved network efficiency and reliability.

\noindent {\large\adforn{3}} \textit{Channel-Agnostic E2E Systems:} In \cite{Ye2018e2e}, the authors propose a channel-agnostic E2E wireless communication system using DNNs for all signal-related functionalities, including encoding, decoding, modulation, and equalization. They address the challenge of not having prior CSI by employing a cGAN to represent the channel effects. Through iterative training, the E2E loss can be optimized in a supervised manner, and the proposed method demonstrates effectiveness on AWGN and Rayleigh fading channels. This approach showcases the potential of building data-driven communication systems that do not rely on prior information about the channel.

\noindent {\large\adforn{3}} \textit{Learning-Based Modulation and Coding Schemes:} \cite{Timothy2018physical} presents a novel learning-based method for synthesizing new physical layer modulation and coding schemes for communication systems without requiring an analytic model of channel impairments. By adopting an adversarial approach, the authors jointly learn a solution to channel response approximation and information encoding applicable over a wide range of channel environments. The proposed adversarial system offers the potential to learn radio communication schemes directly from the sampled response of real physical hardware devices and channels, rather than relying on simplified models that may not fully capture real-world complexities. This method can enable the development of more efficient and adaptive modulation and demodulation functions that optimize system performance for global metrics or loss functions.

\noindent {\large\adforn{3}} \textit{E2E Systems with VAEs:} In \cite{Alawad2022e2e}, the authors propose a novel E2E wireless communication system using VAEs to reconstruct transmitted symbols without sending data bits from the transmitter. The key innovation lies in representing the symbol as an image hot vector containing features such as spikes, closed squared frame, pixel index location, and pixel grayscale colors. These features are inferred by latent random variables (LRVs), which are transmitted through the physical wireless channel instead of the original bits or hot vectors. The proposed VAE architecture can reconstruct the transmitted symbol from the received LRVs, and the results demonstrate that using a VAE with a simple classifier provides a better SER compared to both autoencoder baseline and classical Hamming code with hard decision decoding, particularly at high $E_b/N_0$.

\subsection{SUMMARY, INSIGHTS, AND FUTURE DIRECTIONS}

The utilization of GMs in the physical layer design of wireless communication systems has shown remarkable promise. This paper discussed the key applications of GMs in various aspects of the physical layer, ranging from channel estimation to blind channel equalization, modulation recognition, and JSCC. In the rest, we provide key insights and prospective research directions: 

\begin{itemize}
    \item [\large \adforn{72}]  GMs can accurately model and estimate complex channel characteristics, even in high-frequency mmWave and THz communications; offering state-of-the-art data-driven channel estimation techniques that can outperform traditional approaches, especially in highly dynamic environments. Future work could focus on optimizing these models for real-time deployment and studying their robustness under different environmental conditions. Additionally, investigating the performance trade-offs between model complexity and channel estimation accuracy can be valuable.

    \item [\large \adforn{72}]GANs have shown to be capable of autonomously identifying various modulation schemes, thus enabling adaptive modulation techniques and promoting efficient spectrum utilization. Data augmentation methods improve classification accuracy, and studies have also confirmed the robustness of synthetically trained models. As adaptive modulation becomes more essential for dynamic spectrum management, real-time GM-based techniques will likely be a hot topic for research. Thus, the domain of modulation recognition can benefit from future GM research that focuses on real-time adaptive mechanisms and lower-latency operations.

    \item [\large \adforn{72}]GMs have also been used effectively in JSCC schemes to improve the efficiency and reliability of end-to-end system performance. Novel approaches have been proposed to optimize perceptual quality in image transmission, demonstrating robustness against varying channel conditions. There is room for further investigation into optimization techniques that can handle different types of data transmission, such as video and text, in addition to images. Research can also look into the applicability of GMs in multi-user scenarios and their potential in optimizing system-level performance.

    \item [\large \adforn{72}]In spectrum sensing and recovery, GANs have played a special role in addressing key challenges such as data augmentation, domain adaptation, and spectral recovery. The ability to generate synthetic data and adapt to spectrum dynamics provides a robust framework for dynamic spectrum management. Furthermore, the incorporation of GANs in spectral information recovery illustrates a significant advantage over traditional methods by providing a data-driven approach that moves the computational complexity to the training phase. Future research directions could focus on real-world applications, especially in CRNs with varying degrees of user density and spectrum usage. Exploring how GMs react to real-world radio frequency interference can also provide valuable insights.

    \item [\large \adforn{72}]In the area of signal detection and decoding, GMs have enabled advanced algorithms for efficient signal recognition, especially in complex scenarios where traditional methods may struggle. They offer adaptive capabilities in varying channel conditions, mitigating the impact of channel estimation errors. Applications such as interference signal recognition and MIMO detection have particularly benefited from these advancements, displaying improved performance metrics. Potential areas for future investigation include GMs' adaptability to more complex interference conditions and their ability to learn and adapt without requiring an explicit training phase. The integration of GMs with other ML techniques (e.g., RL) could provide more adaptive solutions for varying channel conditions.

    \item [\large \adforn{72}]GM-based E2E system design has been another milestone, representing a shift from component-wise optimization to holistic system design. Learning-based modulation, channel-agnostic systems, and innovative applications of VAEs exemplify the extensive applicability of GMs in this domain. E2E designs offer a data-driven approach to wireless communication systems, optimizing the overall performance and ensuring robustness against a variety of channel conditions. Future E2E system design research can concentrate on the development of GM-based frameworks that are capable of self-optimization in response to changes in network traffic and conditions. The incorporation of federated learning could be explored for edge-based E2E systems that require low latency and high reliability.
\end{itemize}

Overall, GMs have proven to be invaluable tools for tackling the challenges associated with the physical layer in wireless communication systems. Their ability to learn complex data distributions and generalize to new, unseen data sets them apart as powerful methods for advancing the state-of-the-art in physical layer design. In addition to domain specific directions presented above, future work could potentially explore applicability of GM in solving other pressing issues such as energy efficiency, low-latency communication, and integration with other layers of the communication protocol stack.

\begin{figure}[t]
    \centering
    \input{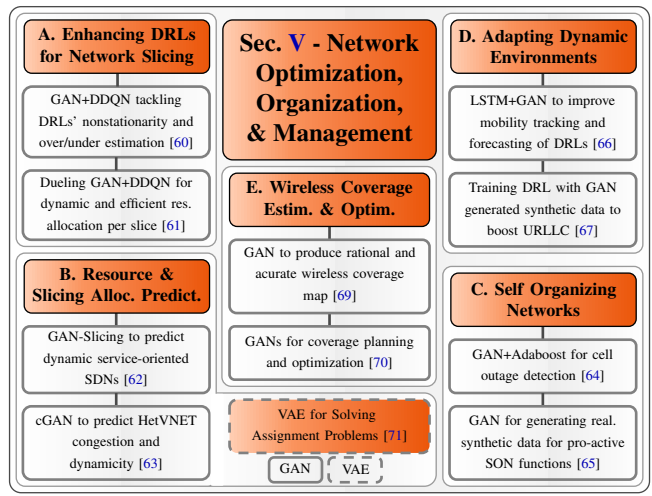}
    \caption{GAN applications for network optimization, organization, and management}
    \label{fig:rrm-phy}
\end{figure}

\section{GenAI FOR NETWORK OPTIMIZATION, ORGANIZATION, AND MANAGEMENT}
\label{sec:network}
GMs have shown great potential for addressing various challenges in network optimization, organization, and resource management. By generating synthetic data and improving learning algorithms, GANs and other GMs can enhance the performance of wireless networks, mobile network slicing, and self-organization features. In the remainder, we present a range of studies that demonstrate the efficacy of GMs in tackling problems related to resource allocation, network organization and management management, and performance optimization in diverse communication scenarios.

\subsection{ENHANCING ML ALGORITHMS} GMs can improve the learning capabilities of RL/DRL agents by generating samples that represent the distribution of state-action values \cite{hua2019gan, hua2019ganJSAC}. This mitigates the effects of learning from nonstationary environments, avoiding action-value overestimation or underestimation issues, and ultimately leading to better network optimization. Hua et al. \cite{hua2019gan} propose a GAN-based Deep Distributional Q Network (GAN-DDQN) to address the challenge of demand-aware resource allocation in 5G network slicing. The GAN-DDQN model generates samples for each action, representing the distribution of state-action values, and mitigates issues stemming from learning in a nonstationary environment. The authors also introduce a reward-clipping mechanism to handle destabilizing effects during training. In a subsequent paper \cite{hua2019ganJSAC}, they develop the Dueling GAN-DDQN, which separates state-value distribution from action-value distribution and combines the action advantage function to obtain action values. Extensive simulations demonstrate the superiority of GAN-DDQN and Dueling GAN-DDQN algorithms over classical DQN in dynamic and efficient resource allocation per slice, leading to better spectrum efficiency and service level agreement satisfaction.

\subsection{NETWORK SLICING AND RESOURCE ALLOCATION} 
GMs can be used to predict user requirements for different resources, enabling efficient and flexible management of network, computing, and storage devices. These predictions 
can be employed in dynamic service-oriented network slicing schemes for a better timely and flexible resource provisioning, and improved quality-of-user-experience (QoE). In this regard, Gu et al. \cite{gu2019ganslicing} propose GANSlicing, a dynamic service-oriented software-defined mobile network slicing scheme for resource allocation prediction in IoT applications. This scheme aims to improve users' QoE by allocating resources in a timely and flexible manner. GANSlicing uses GANs to predict user requirements for different resources, allowing for efficient and flexible management of network, computing, and storage devices. GANSlicing outperforms current network slicing schemes, accepting 16\% more requests with 12\% fewer resources for the same service request batch. Likewise, the work in \cite{falahatraftar2021cGAN} employs a cGAN to augment information related to successful network scenarios, which can be used in the creation of Heterogeneous Vehicular Networks (HetVNETs) slices to address network congestion and dynamicity. The generated data from the cGAN is similar to real data and contributes to enhancing the flexibility and adaptability of the HetVNET. By applying cGAN for the first time in this context, the proposed architecture can intelligently and reliably generate valuable information to create network slices, ultimately avoiding congestion in HetVNETs.

{There has been a recent interest in using DL methods to solve assignment problems
\cite{Lee2018assignment,Kaushik2021assignment}. Similarly, GMs, specifically VAE variants, can be exploited to solve linear sum assignment problems, which are extensively studied in wireless resource allocation problems. In \cite{Zaky2020spectrumsharingGAN}, the authors propose exploiting a VAE variant to solve linear sum assignment problems, commonly exploited in wireless resource allocation problems \cite{celik1,celik4,celik2,celik3,celik5,celik6}. Simulation results demonstrate that the proposed approach can replace the conventional Hungarian algorithm and quickly and accurately find solutions for large sizes of cost matrices.}

\subsection{SELF-ORGANIZING NETWORKS (SONs)} GMs can address data scarcity and imbalanced data issues in SONs by generating realistic synthetic data for training ML algorithms powering SON functions \cite{Zhang2020selforg, Hughes2019selforg}. This can improve the performance of SONs in detecting cell outages and other network management tasks. In \cite{Zhang2020selforg}, the authors present a novel method for cell outage detection in SONs by combining GAN and Adaboost to address the issue of imbalanced data. GAN is utilized to preprocess the imbalanced data, generating more synthetic samples for the minority class, while Adaboost is employed to classify the balanced dataset. This approach effectively detects cell outages in cellular networks and overcomes the limitations of traditional classification algorithms in dealing with imbalanced data. The proposed method demonstrates significant improvements in classification performance based on various metrics such as receiver operating characteristic, precision, recall rate, and F-value. In \cite{Hughes2019selforg}, the authors propose a method for addressing the data scarcity issue in SON for future mobile 5G networks by utilizing GANs to generate large amounts of realistic synthetic data. GANs are used to create synthetic Call Data Records (CDRs), which can be used as training data for ML algorithms that power SON functions. The study demonstrates that GAN-generated synthetic CDRs can be effectively combined with real data to improve the prediction accuracy of ML algorithms, driving proactive SON functions in future 5G networks. This approach is particularly useful when obtaining more real labeled data is not feasible or efficient, and it represents the first time that GANs have been used to generate synthetic CDRs.

\subsection{ADAPTATION TO DYNAMIC ENVIRONMENTS} GMs can enhance the adaptability of algorithms to changes in network conditions, such as UAV mobility patterns \cite{xu2021generativeLSTM} or rare events in URLLC \cite{kasgari2021experienced}, by generating synthetic data that captures the underlying structure of these patterns. Xu et al. \cite{xu2021generativeLSTM} propose a RL approach to maximize the sum-rate of multiple-UAV-served MTC by jointly optimizing transmission power, mode, frequency spectrum, relay selection, and trajectory. They combine LSTM and GAN to improve RL's UAV mobility tracking and forecasting efficacy, and overall reward performance. The improved synthetic data generated by the GAN augments the training set for the LSTM network, leading to a more robust model that better captures UAV mobility patterns. Kasgari et al. \cite{kasgari2021experienced} propose an experienced DRL framework for model-free resource allocation for URLLC in 6G networks without user traffic assumptions. By exploiting GANs to pre-train DRL agents with a mix of real and synthetic data, the framework achieves near-optimal performance within the rate-reliability-latency region under extreme network conditions. The GAN-generated dataset also allows the framework to recover faster during unpredicted rare events. GMs can also be integrated into vehicular networks to address challenges such as navigation optimization, traffic prediction, data generation, and evaluation. Zhang et al. \cite{zhang2023generative} investigate integrating GenAI technologies into vehicular networks, focusing on applications and challenges such as navigation optimization, traffic prediction, data generation, and evaluation. They propose a multi-modality semantic-aware framework to enhance QoS by utilizing multi-modal and semantic communication technologies. A DRL-based approach for resource allocation in GenAI-enabled V2V communication is also proposed, aiming to maximize the system's QoE within constraints.

\subsection{COVERAGE ESTIMATION AND OPTIMIZATION} 
In \cite{Zhuo2018coverage}, the authors propose a weakly-supervised GAN (WS-GAN) with auxiliary information to estimate wireless coverage based on randomly distributed samples of received signal strength. Unlike traditional methods such as kNN or matrix completion, WS-GANs approximate the real distribution of observations by incorporating auxiliary information such as terrain and building data, which significantly impacts the variation of received signal strength, to improve estimation performance. Experiments on a real long-term evolution (LTE) dataset show that WS-GAN outperforms baseline methods in estimation accuracy and produces more rational and practical wireless coverage maps. Moreover, the work in \cite{taras2018coverage} proposes a novel approach for small cell coverage planning and performance optimization in HetNets using GAN to derive knowledge and delivers it to each local SDN controller in a simplified form. The key aspect of this approach is effective training of GANs for various topologies, even when there is limited real data about network behavior and performance. In this way, the authors aim to improve network management and self-organizing functionality, ensuring high QoE across different cell radii.

\subsection{SUMMARY, INSIGHTS, AND FUTURE DIRECTIONS}
Our review reveals that GMs, especially GANs and VAEs, hold considerable promise for various aspects of network management and optimization. From enhanced learning algorithms to solving assignment problems, GMs offer a flexible and effective toolkit for network researchers and practitioners. GMs have demonstrated significant advantages in tackling non-stationary environments, predicting user requirements for resources, dealing with imbalanced and scarce data, and adapting to dynamic network conditions. Techniques such as GAN-DDQN, Dueling GAN-DDQN, GANSlicing, and WS-GAN are among the various innovative approaches that have emerged, providing superior performance over traditional algorithms in tasks like resource allocation, network slicing, and wireless coverage estimation.

One of the crucial insights to draw from this section is that GMs offer a way to make networks more adaptive, efficient, and robust. Through synthetic data generation, GMs can augment real data to create richer datasets for training ML algorithms, thereby improving their performance in tasks like resource allocation and outage detection. Moreover, GMs enhance the capabilities of RL/DRL algorithms, particularly in non-stationary environments, making them more reliable for real-world deployments. The fusion of GMs with existing technologies appears to be not just beneficial but potentially transformative for next-generation networks. Despite the promising advancements, several challenges and opportunities lie ahead:
\begin{itemize}
  \item [\large \adforn{72}] \textit{Scalability}: As networks grow more complex and large-scale, the scalability of GM-based solutions will be an area requiring rigorous investigation.
  \item [\large \adforn{72}] \textit{Security Concerns}: The use of synthetic data and automated decision-making could pose new security and privacy challenges that need to be addressed.
  \item [\large \adforn{72}]  \textit{Real-world Validation}: While many studies rely on simulations, there's a pressing need for more empirical studies to validate the effectiveness of GMs in real-world network environments.
  \item [\large \adforn{72}]  \textit{Interdisciplinary Research}: Combining GMs with other emerging technologies like edge computing or quantum computing could lead to breakthroughs that address some of the most pressing challenges in network optimization.
  \item [\large \adforn{72}]  \textit{Ethical and Regulatory Aspects}: As AI and GMs become more integrated into network management, ethical and regulatory considerations such as data privacy and fairness will become increasingly important.
\end{itemize}

Noting that above works have mainly focused on using GANs and VAEs, without delving into exploring other GMs presented in Section \ref{sec:GAI_Models}. Future research should continue to explore and refine these techniques, as well as investigate the integration of various GMs with other emerging technologies to further advance network optimization and resource management. 


\begin{figure*}[t]
    \centering
    \input{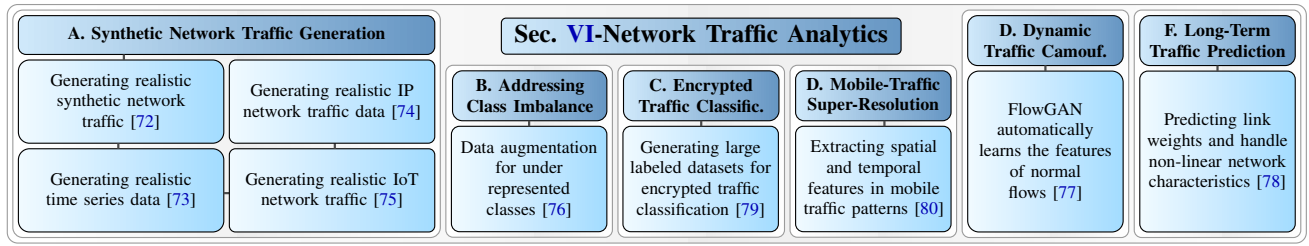}
     \caption{GAN applications for network traffic analytics.}
    \label{fig:apps-traffic}
\end{figure*}
 
\section{GenAI FOR NETWORK TRAFFIC ANALYTICS}
\label{sec:traffic}
GMs have demonstrated significant potential to improve various aspects of network traffic analytics, including generating synthetic network traffic, addressing class imbalance in network traffic classification, enabling dynamic traffic camouflaging, improving temporal link prediction in dynamic networks, generating realistic network traffic data at the IP packet layer, enhancing mobile traffic super-resolution, and assisting in long-term traffic flow prediction, which are all discussed in the sequel. 

\subsection{SYNTHETIC TRAFFIC DATA GENERATION} 
GANs can create statistically similar network traffic samples to a reference dataset, enhancing the performance of ML algorithms and generating live traffic representative of real traffic in various network settings \cite{Aho2018netanalytics}, wherein a GAN-based approach is proposed for generating synthetic network traffic that is statistically similar to real traffic samples. By evaluating different GAN variants, they demonstrate the broader applicability of the proposed technique in generating realistic traffic for various network settings, with the primary goal of enhancing ML performance. Moreover, GANs and DL-based auto-regressive models can generate realistic time-series data, with GANs performing better than the auto-regressive approach in some cases. Forecasting models trained on data generated by GANs yield error rates comparable to those trained on real data \cite{Naveed2022timeseries}. On the other hand, GAN methods can be adapted to create realistic IP packet layer network traffic data, crucial for developing and testing cyber and network security techniques \cite{Cheng2019pacGAN}. Specifically, CNN-GANs successfully generate and transmit real traffic flows (e.g., ICMP Pings, DNS queries, and HTTP web requests), demonstrating the feasibility of GAN-based traffic generators at scale. In \cite{Shahid2020generative}, Shahid et. al. proposes generation of realistic IoT network traffic using a combination AEs and GANs for enhancing network-based intrusion detection systems (NIDS) and evaluating their performance. The authors train an autoencoder to learn the latent representation of real sequences of packet sizes, then train a WGAN on the latent space to generate latent vectors that can be decoded into realistic sequences. The synthetic bidirectional flows generated by this method closely resemble the real traffic produced by a Google Home Mini and can fool anomaly detectors, making them appear as legitimate traffic. The approach can be used for both benign purposes, like data augmentation and NIDS evaluation, and malicious purposes, such as evading NIDS through mimicry attacks.

\subsection{ADDRESSING TRAFFIC CLASS IMBALANCE} 
As encrypted Internet applications grow in prevalence, accurately identifying traffic types for network management and security monitoring becomes increasingly challenging. FlowGAN \cite{Wang2019flowgan} addresses the class imbalance problem in network traffic classification by leveraging GAN's data augmentation capabilities to generate synthetic traffic data for underrepresented classes. The experimental results show improvements in precision, recall, and F1-score over both unbalanced and balanced datasets.

\subsection{ENCRYPTED TRAFFIC CLASSIFICATION} 
ByteSGAN \cite{wang2021bytesgan}, a GAN-based semi-supervised learning encrypted traffic classification method, addresses the challenge of acquiring large labeled datasets for supervised learning. Utilizing a small number of labeled traffic samples and a large number of unlabeled samples, ByteSGAN modifies the structure and loss function of a regular GAN discriminator network, resulting in improved traffic classification performance over traditional supervised learning methods.

\subsection{IMPROVING TRAFFIC SUPER-RESOLUTION} 
The novel GAN architecture ZipNet-GAN, tailored for mobile traffic super-resolution, extracts spatial and temporal features in mobile traffic patterns. The proposed ZipNet-GAN demonstrates remarkable accuracy in inferring fine-grained mobile traffic distributions, achieving up to 100x higher granularity compared to standard probing and outperforming existing interpolation methods \cite{zhang2017zipnet}.

\subsection{DYNAMIC TRAFFIC CAMOUFLAGING} 
Following the same denomination with \cite{Wang2019flowgan}, Li et. al. proposed another FlowGAN approach that can automatically learn the features of "normal" network flows and dynamically morph ongoing traffic flows based on these features, in order to bypass Internet censorship \cite{Lei2019temporal}. By making censored flows indistinguishable from multiple target flows, FlowGAN overcomes the limitations of existing traffic morphing and protocol tunneling techniques by employing GANs to automatically learn the features of "normal" network flows. The authors introduce the concept of $\varepsilon$-indistinguishability to measure similarity between target and morphed traffic flows, demonstrating FlowGAN's effectiveness and efficiency on a dataset of real-world flows.

\subsection{IMPROVING TEMPORAL TRAFFIC PREDICTION}
A novel DL model that combines the strengths of graph convolutional networks (GCN), LSTM networks, and GANs can effectively predict link weights and handle non-linear characteristics of networks \cite{Lei2019temporal}, outperforming state-of-the-art competitors in temporal link prediction tasks. On the other hand, the residual deconvolution-based deep generative network model effectively predicts long-term traffic flow for elevated highways by transforming the original traffic flow data into spatio-temporal matrices and learning to extract multiscale spatio-temporal features of the traffic flow at three different time scales \cite{zang2019traffic}, demonstrating superior performance over state-of-the-art methods for long-term traffic flow prediction.

\subsection{SUMMARY, INSIGHTS, AND FUTURE DIRECTIONS}
Overall, GMs have shown a transformative impact network management, organization, and resource allocation by addressing various challenges related to network traffic analytics. Their ability to generate synthetic network traffic data, improve network traffic classification by handling the class imbalance, enable dynamic traffic camouflaging, and enhance mobile traffic resolution and prediction capabilities makes them a valuable tool in the field. 

To be more specific, GANs have been effective in generating synthetic network traffic that can fool anomaly detectors, thus serving both benign and potentially malicious purposes. The capacity for GMs to address class imbalance issues in traffic classification opens new avenues for improving network management and security monitoring, especially in the age of encrypted Internet applications. GANs have been successful in morphing ongoing traffic flows to bypass internet censorship, providing a new perspective on traffic camouflaging techniques. DL models combining GMs with other architectures, like graph convolutional networks and LSTMs, have excelled in temporal and long-term traffic flow prediction, marking a significant leap over traditional methods. These promising advancements can be futher extended to the research directions summarized below 
\begin{itemize}
    \item[{\large \adforn{72}}] \textit{Network Performance Prediction:} By learning the relationships between network conditions and performance, GMs can  help network administrators proactively manage network resources and optimize performance by forecasting KPIs (e.g.,  throughput, latency, and packet loss), aiding proactive network management.
    
    \item[{\large \adforn{72}}] \textit{Topology Inference and Modeling:} GMs can synthesize realistic network topologies--that resemble real-world networks--for studying network behavior and performance under different conditions.
    
    \item[{\large \adforn{72}}] \textit{Network Protocol analysis:} GMs can be used to study and analyze network protocols by learning their characteristics and generating synthetic protocol data, allowing researchers understand protocol behavior, identify vulnerabilities, and design more secure protocols.

    \item[{\large \adforn{72}}] \textit{Traffic Load Balancing:} GMs can predict future traffic patterns, enabling administrators to identify potential bottlenecks and implement load balancing strategies to redistribute traffic and ensure optimal network performance.
    
    \item[{\large \adforn{72}}] \textit{Simulation and testing:} GMs can create realistic synthetic network data for simulating environments and testing new equipment, protocols, and algorithms, enabling researchers to validate and optimize their designs in a controlled environment before deployment in real networks.

    \item[{\large \adforn{72}}] \textit{NFV and SDN optimization:} Based on historical network data, GMs can recommend optimal configurations for virtual network functions and SDN controllers, improving efficiency and performance.
\end{itemize}

\begin{figure*}[htbp!]
    \centering
    \input{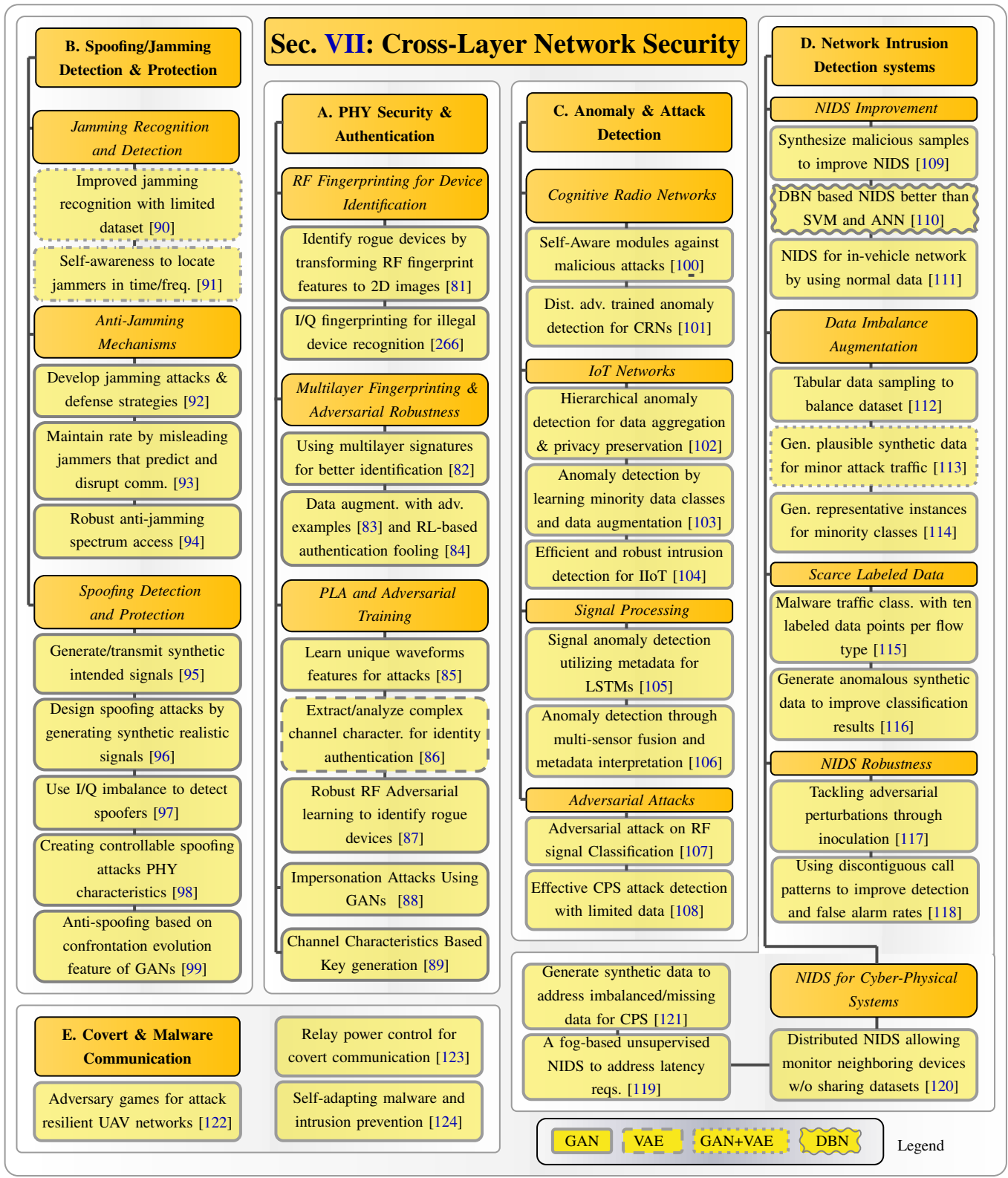}
    \caption{GenAI applications for cross-layer network security}
    \label{fig:apps-security}
\end{figure*}

\section{GenAI FOR CROSS-LAYER NETWORK SECURITY}
\label{sec:security}

This section reviews how GMs can boost various facets of cross-layer network security, encompassing topics such as PHY Layer security and authentication; jamming/spoofing recognition and protection; anomaly and attack detection; network intrusion detection; and covert, malware, and resilient communications.

\subsection{PHY LAYER SECURITY AND AUTHENTICATION}
The rapidly evolving domain of wireless communication and the proliferation of IoT devices have necessitated advanced security measures to counter malicious activities and unauthorized access. RF fingerprinting, which leverages the unique hardware imperfections and wireless channel variations as distinct signatures for device identification, emerges as a compelling solution. Recent advancements have demonstrated the potential of GANs in enhancing the robustness and accuracy of RF fingerprinting methods. However, these same advancements have also unveiled potential vulnerabilities to adversarial attacks.

\noindent {\large\adforn{12}} \textit{RF Fingerprinting for Device Identification:} Several studies have focused on leveraging GANs for accurate identification of rogue or unauthorized devices. Chen et al. \cite{chen2021fingerprint} introduce a novel rogue device identification technique, transforming RF fingerprint features into two-dimensional figures and employing a GAN for unsupervised rogue device discrimination. Based on an experimental verification system built with 54 ZigBee recognized devices regarded, the paper reports a 95\% identification accuracy in a real-world environment, illustrating the efficacy of GANs for device identification without prior information. In a similar vein, Yang et al. \cite{yang2021fingerprint} propose a GAN-based method for RF fingerprint recognition, designed to identify illegal transmitters endangering wireless communication security. They utilize in-phase and quadrature (I/Q) data through wavelet transform data pre-processing to highlight RF fingerprinting characteristics and achieve a remarkable 98.1\% accuracy in distinguishing between legitimate and illegal transmitters, surpassing the performance of traditional CNNs and fully connected DNNs.

\noindent {\large\adforn{12}} \textit{Multi-layer Fingerprinting and Adversarial Robustness:} As RF fingerprinting continues to evolve, more sophisticated approaches that go beyond traditional methods have emerged. Huang et al. \cite{huang2023fingerprint} propose a multi-layer fingerprinting framework that jointly considers multi-layer signatures for improved identification performance. By capitalizing on the multi-view ML paradigm and an information-theoretic approach, they demonstrate superior performance to state-of-the-art baselines in both supervised and unsupervised settings. However, the susceptibility of RF fingerprinting to adversarial attacks remains a pressing concern. Merchant et al. \cite{merchant2019fingerprint} expose a weakness in the training protocol of neural network-based physical-layer wireless security, demonstrating that a GAN can generate signals realistic enough to force classifier errors. They suggest augmenting the training dataset with adversarial examples from a different GAN to reinforce the classifier against this vulnerability. Similarly, Karunaratne et al. \cite{karunaratne2021fingerprint} highlight the potential of adversarial attacks to undermine the security provided by DL-based authenticators. Extensive simulations and experiments on a software-defined radio testbed indicate that developed RL-based attacks capable of fooling the authenticator with a success rate of more than 90\% under appropriate channel conditions.

\noindent {\large\adforn{12}} \textit{Physical Layer Authentication and Adversarial Learning:}
PLA exploits unique hardware and propagation channel features from legitimate users, which are challenging for attackers to replicate. However, GMs can enable intelligent adversaries can impersonate a legitimate user including its signal pattern, hardware impairments and wireless channel characteristics. In \cite{xu2022pla}, authors investigate three waveform candidates with unique signal features (such as orthogonality, non-orthogonality, windowing, etc.) and reveal that GAN attackers can successfully learn unique waveforms features. Meng et al. \cite{meng2023pla} propose a channel impulse response (CIR)-based PLA scheme named "hierarchical variational autoencoder (HVAE)" for the Industrial IoT. HVAE consists of an AE module for CIR characteristics extraction and a VAE module for improving the representation ability of the CIR characteristic and outputting the authentication results. Proposed hierarchical framework can extract and analyze complex channel characteristics for identity authentication without requiring attackers’ prior channel information, demonstrating high performance even in complex environments. Moreover, Roy et al. \cite{roy2020pla} propose the Radio Frequency Adversarial Learning (RFAL) framework, using a GAN to build a robust system that can identify rogue RF transmitters. The RFAL framework uses transmitter-specific signatures (e.g., the I/Q imbalance) and a DNN to learn feature representations and categorize trusted transmitters, demonstrating an impressive 99.9\% accuracy in distinguishing between trusted and counterfeit transmitters.

\noindent {\large\adforn{12}} \textit{Impersonation Attacks through GANs:} Xu et al. \cite{xu2022fingerprint} delve into the largely unexplored territory of impersonation attacks against RF fingerprinting. They propose a colluding impersonation attack framework that employs a GAN to emulate the RF fingerprint of legitimate users. The colluding attacker observes and compares the signal features of the impersonation attacker and the legitimate user, subsequently providing feedback to refine the impersonation technique. This study underscores the necessity of bolstering the security of RF fingerprinting against sophisticated attacks.

\noindent {\large\adforn{12}} \textit{Key Generation via WGANs with Gradient Penalty:}
In \cite{han2022keygen}, the authors propose a key generation method based on WGANs with Gradient Penalty (WGAN-GP) adversarial autoencoder for PLS in wireless channels. The method leverages the channel reciprocity between legitimate nodes to extract symmetric keys using DNNs, outperforming traditional methods in terms of freedom and performance. The proposed method also addresses the issue of unpredictable output from neural network hidden layers and the difficulty of estimating high-dimensional features in advance. The extracted features can be fitted to a Gaussian distribution, and the Wasserstein distance and gradient penalty are used in game training to prevent gradient explosion. Compared to the PCA method, this approach provides a higher security key capacity, lower key error rate, and eliminates interaction in the key generation process, thereby preventing key leakage.

\subsection{JAMMING/SPOOFING RECOGNITION AND PROTECTION}
GMs have shown great potential in addressing signal jamming and spoofing in wireless networks. They can help in developing both attack and defense strategies by learning complex patterns and relationships in the data. Here, we elaborate on how GMs contribute to jamming and spoofing in wireless networks:

\noindent {\large\adforn{12}} \textit{Jamming Recognition:} Tang et al. \cite{tang2020jamming} address the issue of performance deterioration in jamming recognition due to limited sample sizes by proposing a jamming recognition method based on AC-VAEGAN. They modify the structure of the ACGAN network, incorporating the core idea of VAE, resulting in higher correct recognition rates compared to ACGAN and CNN networks under small sample datasets. Similarly, Krayani et al. \cite{krayani2020jamming} explore the integration of CR and UAVs, introducing a self-awareness framework to enhance the physical layer security. They propose a dynamic Bayesian network model for representing the radio environment and a modified Markov Jump Particle Filter for prediction and state estimation. Their novel jammer detection framework allows the UAV to effectively locate the jammer in both time and frequency domains, showcasing the effectiveness of the proposed approach in terms of detection probability and accuracy.

\noindent {\large\adforn{12}} \textit{Anti-Jamming Mechanisms:} Adversarial ML (AML) approaches, such as those proposed by Shi et al. \cite{shi2018jamming} and Erpek et al. \cite{erpek2019jamming}, use GMs to devise defense schemes that deliberately take a small number of wrong actions in spectrum access. Shi et al. \cite{shi2018jamming} present an AML approach to launch jamming attacks in wireless communications and develop a defense strategy against such attacks. By exploiting the sensitivity of DL to training errors, they devise a defense scheme that deliberately takes a small number of wrong actions in spectrum access, aiming to prevent the attacker from building a reliable classifier. In a similar vein, Erpek et al. \cite{erpek2019jamming} introduce an AML method for jamming attacks and defense in wireless communications. The jammer employs a DL classifier to predict and disrupt successful transmissions. To counteract these attacks, the transmitter takes a limited number of incorrect actions, systematically selecting when to do so, in order to mislead the jammer and maintain throughput. In dynamic and unknown environments, Han et al. \cite{han2021jamming} propose a robust intelligent anti-jamming spectrum access scheme which integrates spectrum sensing, completion, learning, and access. They employ a GAN to complete missing spectrum data and utilize DRL for channel selection. Their approach outperforms conventional DRL-based methods in avoiding complex jamming attacks.

\noindent {\large\adforn{12}} \textit{Signal Spoofing:} Shi et al. \cite{shi2019spoofing} introduce a novel spoofing approach using GANs to generate and transmit synthetic wireless signals indistinguishable from intended signals. By employing DL techniques, the adversary pair of transmitter and receiver play a minimax game, training a DL-based spoofing mechanism to potentially fool defense mechanisms like RF fingerprinting. Their results indicate that GAN-based spoofing attacks significantly increase the success probability of wireless signal spoofing. In a similar vein, Shi et al. \cite{shi2021spoofing} present a DL-based spoofing attack to generate synthetic wireless signals that cannot be statistically distinguished from intended transmissions. Their approach captures the waveform, channel, and radio hardware effects inherent to wireless signals under attack, demonstrating the increased probability of misclassifying spoofing signals as intended signals when compared to random or replayed signals.

\noindent {\large\adforn{12}} \textit{Spoofing Detection:} Roy et al. \cite{roy2019spoofingjamming} propose an adversarial learning technique using GANs to identify rogue RF transmitters and classify trusted ones. They exploit the I/Q imbalance of transmitters to learn unique high-dimensional features that serve as fingerprints for identification and classification. Their experiments showcase the discriminator's ability to discriminate between trusted and fake transmitters with 99.9\% accuracy.

\noindent {\large\adforn{12}} \textit{Controllable Spoofing Attacks:} Ma et al. \cite{ma2023spoofing} investigate building a controllable wireless spoofing attack launch framework using fundamental channel modeling and practical wireless datasets. They propose a conditional boundary equilibrium GAN with AAE to generate controllable spoofing signals that can fool legitimate classifiers. Their results show an average attack success probability of over 80\% when mimicking multiple emitters and modulation types, outperforming random, replay, and GAN-based attacks.

\noindent {\large\adforn{12}} \textit{Anti-Spoofing Mechanisms:} GMs have shown promise in the realm of signal spoofing for wireless networks, posing new challenges for NIDS. Li et al. \cite{li2021spoofingjamming} analyze the characteristics of spoofing signals in the acquisition phase of Global Navigation Satellite System (GNSS) spoofing jamming and propose a GNSS anti-spoofing method based on the idea of confrontation evolution in GAN. Their simulations demonstrate a high detection probability of over 98\% when the pseudo-code phase difference exceeds 0.5 chip.


\subsection{ANOMALY AND ATTACK DETECTION}
This section presents an overview of research in using GMs for anomaly detection within cognitive IoT networks, highlighting different methods and applications for detecting abnormalities in wireless signals and networks.

\noindent {\large\adforn{12}} \textit{Cognitive Radio Networks:} Toma et al. \cite{toma2020anomaly} introduced data-driven self-awareness modules in cognitive radio (CR) systems to establish secure networks against malicious attacks. They proposed AI-based abnormality detection techniques at the physical layer in CR, utilizing GMs such as cGAN and DBN. Similarly,  Katzef et al. \cite{Katzef2020anomaly} proposed a distributed anomaly detection scheme based on adversarially-trained data models for cognitive radio networks. Their approach reduces reliance on a central decision-making server, demonstrating performance matching that of state-of-the-art centralized methods.
\noindent {\large\adforn{12}} \textit{IoT Networks:} Zixu et al. \cite{zixu2020anomaly} presented an unsupervised hierarchical approach for anomaly detection through cooperation between GANs and AEs in IoT networks. This approach addresses data aggregation and privacy preservation by reconstructing a sampling pool at a centralized controller using a collection of generators from individual IoT networks. Moreover, Imtiaz et al. \cite{imtiaz2021anomaly} developed a novel framework for detecting anomalies in IoT networks using cGANs. By learning minority data classes and generating augmented data, their proposed models outperformed other anomaly detection models in various performance metrics. Likewise, Benaddi et al. \cite{benaddi2022anomaly} suggested a mechanism to improve the efficiency and robustness of intrusion detection systems in Industrial IoT networks using distributed RL and GANs. Their proposed models outperformed the standard RL models in various performance metrics.

\noindent {\large\adforn{12}} \textit{Wireless Signal Processing:}
Blake et al. \cite{blake2022anomaly} proposed a novel method for signal anomaly detection using GAN output processed by LSTM recurrent neural networks. The authors showed that utilizing metadata for analytics provided robust detection while minimizing computation, bandwidth, and generalizing to numerous effects. Rathinavel et al. \cite{rathinavel2022anomaly} investigated anomaly detection in wireless signals through multi-sensor fusion. They introduced several baselines and generative methods for interpreting metadata into high-level views of the air interface environment, with each of the proposed methods achieving high performance in detecting anomalous activities.

\noindent {\large\adforn{12}} \textit{Adversarial Attacks and Security Concerns:}
Sadeghi et al. \cite{Sadeghi2019anomaly} explored the vulnerability of DL to adversarial attacks in RF signal classification tasks. They demonstrated that these attacks significantly reduce classification performance, raising concerns about the security and robustness of DL-based algorithms in wireless physical layers. Furthermore, Huang et al. \cite{huang2022anomaly} proposed a ST-GAN system to detect attacks in wireless CPSs. The ST-GAN system addresses the issue of limited data in wireless cyber-physical system security, showing effective attack detection capabilities in experiments.

\subsection{NETWORK INTRUSION DETECTION SYSTEMS}
In recent years, GMs have emerged as a promising solution for addressing various challenges in NIDSs for wireless networks. The robustness and effectiveness of DGMs have been explored to develop advanced NIDS models capable of adapting and protecting against evolving threats. In this context, several studies have proposed novel GMs and techniques to improve network threat detection and enhance system security.

\noindent {\large\adforn{12}} \textit{Improved NIDS:} Vu et al. \cite{vu2023nids} propose a solution employing DNNs to enable robust cloud-based NIDSs. They develop two deep GMs to synthesize malicious samples, augmenting training datasets to improve IDS accuracy. Gao et al. \cite{gao2014nids} introduce DBNs for intrusion detection, demonstrating superior performance compared to SVM and ANN. In the context of in-vehicle networks, Seo et al. \cite{seo2018nids} propose a GAN-based NIDS model, capable of detecting unknown attacks with high accuracy using only normal data.

\noindent {\large\adforn{12}} \textit{Addressing Class Imbalance:} To address class imbalance, Ding et al. \cite{ding2022nids} present a tabular data sampling method that balances normal and attack samples through undersampling and oversampling. Park et al. \cite{park2023nids} propose an AI-based NIDS that leverages state-of-the-art GMs to generate synthetic data for minor attack traffic, resolving the data imbalance issue. Huang et al. \cite{huang2020nids} develop an imbalanced GAN to tackle class imbalance, generating representative instances for minority classes and outperforming the state-of-the-art.

\noindent {\large\adforn{12}} \textit{SS-GANs for Scarcely Labeled Data:} In scenarios with scarce labeled data, Jeong et al. \cite{jeong2021nids} propose a semi-supervised GAN model for network intrusion detection, requiring only ten labeled data points per flow type. Salem et al. \cite{salem2018nids} leverage GANs to generate synthetic anomalies, particularly in domains with costly data creation processes and inherently imbalanced datasets. By employing a Cycle-GAN, they generate anomalous data to improve classification results, demonstrating the potential of GANs in anomaly generation.

\noindent {\large\adforn{12}} \textit{Enhancing NIDS Robustness in Adversarial Settings:} The robustness of GMs has also been explored in adversarial settings. Usama et al. \cite{usama2019nids} address the vulnerability of ML-based NIDSs to adversarial perturbations by proposing an AML attack using GANs, which evades ML-based NIDSs while making the detection more robust to adversarial perturbations through inoculation.

\noindent {\large\adforn{12}} \textit{Reducing False Alarm Rates}
Host-based anomaly intrusion detection system design poses challenges due to high false alarm rates. Creech et al. \cite{creech2014nids} introduced a new methodology using discontiguous system call patterns, increasing detection rates while reducing false alarm rates by applying a semantic structure to kernel-level system calls to reflect intrinsic activities hidden in high-level programming languages.

\noindent {\large\adforn{12}} \textit{NIDS for Cyber-Physical Systems:} The convergence of CPS, IoT, and AI has paved the way for novel technological advancements while also increasing the risk of cyberattacks. For CPSs with strict latency requirements, Freitas et al. \cite{freitas2021nids} proposed FID-GAN, a novel fog-based, unsupervised NIDS for CPSs using GANs. The approach addresses latency constraints by training an encoder to accelerate the reconstruction loss computation, achieving higher detection rates and improved speed compared to baseline approaches. To ensure security in large-scale IoT systems, Ferdowsi et al. \cite{ferdowsi2019nids} introduced a distributed GAN-based NIDS that allows IoT devices to monitor their own data and neighboring devices without sharing datasets, suitable for privacy-sensitive applications such as health monitoring and financial services. This distributed GAN-based NIDS exhibited improved accuracy, precision, and false positive rates compared to standalone GAN-based NIDSs. Moreover, Shahriar et al. proposed a GAN-based NIDS that generates synthetic samples to address challenges of imbalanced and missing data in emerging CPS fields, resulting in better attack detection and model stabilization compared to standalone NIDSs \cite{shahriar2020nids}.




\subsection{COVERT, MALWARE, AND RESILIENT COMMUNICATIONS}
The emergence of advanced adversarial methods poses significant challenges to the stability and security of communication networks. In this regard, GANs can bolster network resilience and design covert communication systems while also exploring ways to disguise malware traffic. One such effort by Bo et al. \cite{Bo2019resilient} presented an attack-resilient network connectivity approach for multi-UAV networks. The researchers utilized GANs to formulate a three-agent adversary game, comprising a pair of neighboring UAVs and an attacker. The latter acts as a generator, striving to craft information similar to the UAVs' in order to enhance its jamming capabilities. Conversely, the UAVs act as discriminators, augmenting their capacity to reject false information from the attacker. The proposed framework, integrating a conditional GAN with a least square objective loss function and mean square error, demonstrated improved convergence efficiency, reduced connection latency, and significantly enhanced attack resilience.

Parallelly, Liao et al. \cite{Liao2020covert} investigated a power allocation problem in a cooperative cognitive covert communication system. The team designed a novel GAN-based power allocation algorithm to handle power allocation at the relay secondary transmitter for covert communication. Under the proposed approach, the generator adaptively crafts the power allocation solution, while the discriminator determines the feasibility of covert message transmission. This approach exhibited near-optimal power allocation solution and rapid convergence, enhancing the balance between covert rate and detection error probability.

Lastly, in the context of disguising malicious network activity, Rigaki et al. \cite{rigaki2018malware} proposed using GANs to mimic legitimate network traffic, thereby evading detection. By altering the source code of malware to accept parameters from a GAN, the malware's network behavior was modified to mirror that of a legitimate application, in this case, Facebook chat network traffic. Consequently, the altered malware could bypass detection by modern Intrusion Prevention Systems utilizing ML and behavioral characteristics. This research highlights the potential of self-adapting malware and self-adapting intrusion prevention systems.

\subsection{SUMMARY, INSIGHTS, AND FUTURE DIRECTIONS}
This section has reviewed various facets of cross-layer network security, encompassing topics such as PHY Layer security and authentication, jamming/spoofing Recognition and Protection, anomaly and attack detection, network intrusion detection, and resilient communications. The reviewed works reveal the impressive performance of GANs in RF fingerprinting for device identification. Techniques based on DL architectures demonstrate high accuracy rates, often surpassing traditional methods. However, the susceptibility to adversarial attacks and impersonation remains a critical concern. Interestingly, multi-layer fingerprinting frameworks and PLA schemes using complex ML models have been proposed to combat such vulnerabilities, although with limited success.

GMs have been effective in both recognizing and mitigating jamming and spoofing attacks. Through various methods such as AML, dynamic Bayesian networks, and RL, solutions have been proposed that adaptively respond to complex jamming patterns. Nevertheless, signal spoofing remains a considerable challenge, especially with GMs making it easier for attackers to generate indistinguishable signals. 

Across PLS, PLA, and Jamming/Spoofing topics, a consistent theme emerges: while GMs have been instrumental in fortifying security, they have simultaneously opened new attack vectors, making the adversary more intelligent and adaptable. This duality serves as both a warning and a catalyst for future research endeavors.

GMs, such as cGANs and DBNs, are proving pivotal in detecting anomalies and protecting against malicious attacks at both the network and physical layers. Distributed approaches in CRNs and cooperative methods in IoT networks have shown performance benefits, particularly in scalability and robustness. When it comes to NIDSs, DGMs have emerged as flexible and robust solutions. Various models are employed to address challenges such as class imbalance, scarcity of labeled data, and high false alarm rates.

The versatility of GMs extends to the resilience of communication networks as well. Whether it's ensuring network connectivity in multi-UAV systems or solving power allocation problems in covert communications, GANs have found important applications. However, this power also comes with the threat of being used to disguise malware traffic.

\begin{itemize}
    \item [\large\adforn{72}] \textit{Adversarial Robustness and Explainability:} There is a need for more rigorous studies investigating the robustness of RF fingerprinting and PLA mechanisms, particularly in adversarial settings. Injecting transparency into how GMs make decisions is critical, especially in security-sensitive applications.
  
    \item [\large\adforn{72}] \textit{Counter-Counter Measures:} As attackers adapt, so must defense mechanisms. Research should focus on how to stay ahead in this perpetual arms race, perhaps through dynamic, self-evolving systems.
        
    \item [\large\adforn{72}]\textit{Real-world Deployments:} Most works rely on simulated or controlled environments. Studies in real-world, dynamically changing conditions would offer more practical insights. For CPSs with real-time requirements, further research is needed to balance computational efficiency and detection accuracy. Therefore, research on lightweight models that are effective but computationally less intensive can be particularly useful for CPSs.
\end{itemize}

This synthesis highlights the diverse yet interconnected ways in which GMs contribute to advancing the state-of-the-art in network security and communications. The field is ripe for innovative solutions and interdisciplinary research to fully harness the potential of GMs in creating more secure and efficient networks.

\begin{figure}[t!]
    \centering
    \input{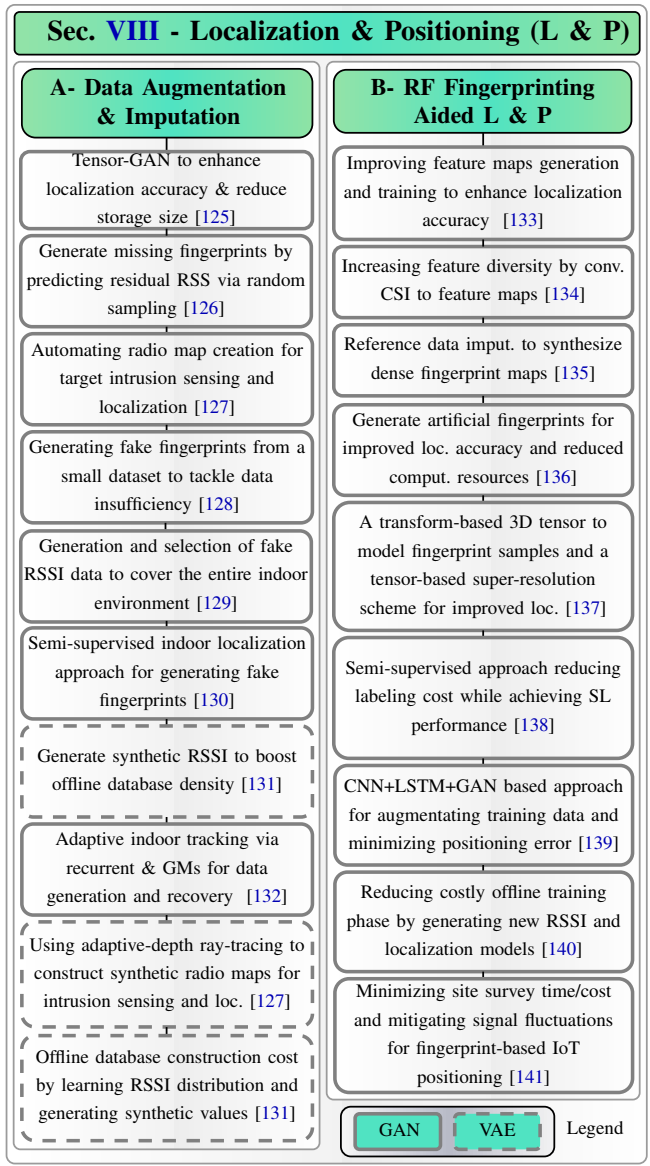}
    \caption{GenAI applications for localization and positioning}
    \label{fig:lp-apps}
\end{figure}

\section{GenAI FOR LOCALIZATION AND POSITIONING}
\label{sec:localization}

GMs have also been applied to a variety of localization and positioning problems, including enhancing indoor localization, addressing WiFi localization issues, automating radio map creation, data augmentation for cost-effective localization, and improving indoor tracking. Furthermore, researchers have employed GMs to address challenges in RF fingerprinting-based localization, such as extending fingerprint databases, generating synthetic fingerprint maps, modelling fingerprint samples, and providing low-cost labeling, which are all discussed throughout this section.

\subsection{DATA AUGMENTATION, IMPUTATION, AND MAP GENERATION}

\noindent {\large\adforn{4}} \textit{Tensor-GAN for Enhancing Indoor Localization:} 
In the realm of mobile computing and social networking, accurate localization in indoor environments is a critical necessity. The complexity of indoor environments, however, poses significant challenges to the traditional RF fingerprint-based localization methods, including limited sample availability, storage constraints, and real-time response requirements. To address these challenges, Liu et al. \cite{liu2021loc} devised a tensor GAN (Tensor-GAN) capable of enhancing localization accuracy and reducing storage consumption. This innovative scheme, incorporating a unique three-player game involving a regressor, a generator, and a discriminator, demonstrated impressive improvements in localization accuracy.

\noindent {\large\adforn{4}} \textit{ACO-GAN for Addressing WiFi Localization Issues:} 
The intensifying demand for location information services underpins the attractiveness of WiFi-based indoor localization. A pressing issue associated with these technologies, however, is the time-consuming task of manually updating the WiFi fingerprint database due to environmental changes affecting signal propagation. To tackle this problem, Ren et al. \cite{ren2021loc} proposed the Adaptive Context GAN model, which efficiently rebuilds missing fingerprints by predicting residual received signal strength indicator (RSSI) values based on random sampling, thereby resolving the remeasurement problem.
\noindent {\large\adforn{4}} \textit{RaGAM for Automating Radio Map Creation:} 
The use of indoor target intrusion sensing techniques is becoming prevalent in various fields, such as smart home management, security monitoring, and disaster relief. In response to the immense labor and time costs associated with traditional radio map construction, Zhou et al. \cite{zhou2020loc} introduced the ray-aided generative adversarial model (RaGAM) to automate radio map creation for WLAN-based target intrusion sensing and localization. The RaGAM model utilizes adaptive-depth ray tree-based quasi three-dimensional ray-tracing, which results in a lower computational cost and increased accuracy.
\noindent {\large\adforn{4}} \textit{Data Augmentation for Cost-effective Localization:} 
Despite their benefits, ML techniques require large amounts of data for training DNN models, which is often a costly endeavor. To address this challenge, Njima et al. \cite{njima2021loc,njima2021loc_journal,njima2022loc} proposed the use of GANs to generate synthetic fingerprints based on limited real data samples. This data augmentation approach, combined with semi-supervised learning, demonstrated notable improvements in localization accuracy. Njima and colleagues addressed the challenge of data insufficiency in indoor localization by utilizing GANs to generate fake fingerprints from a small dataset \cite{njima2021loc},  leading to a mean localization accuracy improvement of 9.66\% in comparison to the conventional semi-supervised localization algorithm. They expanded upon this in subsequent studies, proposing a method for the generation and selection of fake RSSI data to cover the entire indoor environment, which improved localization accuracy  by 21.69\% and 15.36\% for simulated and experimental data, respectively \cite{njima2021loc_journal}. Their latest work introduced a weighted semi-supervised DNN-based indoor localization approach and a system for generating fake fingerprints, called Weighted GAN based indoor localization \cite{njima2022loc}. This system trained a DNN on a mixed dataset of real and fake data samples, avoiding overfitting and enhancing location prediction performance by 8.53-10.11\%. This set of studies collectively demonstrated significant improvements in localization accuracy compared to conventional supervised localization schemes.

\noindent {\large\adforn{4}} \textit{VAEs for Boosting Offline Database Density:} 
DL techniques have been increasingly used for indoor localization, primarily for fingerprinting-based methods. The major drawback of these techniques is the high cost and effort associated with building a large offline database. To alleviate this, Suroso et al. \cite{suroso2022loc} implemented VAEs, a popular deep GM, to learn the distribution of RSSI and generate synthetic RSSI values. This novel approach enhanced the density of the offline database grids, despite some challenges with the accuracy of synthetic RSSI values.

\noindent {\large\adforn{4}} \textit{RecTrack-GAN for Improved Indoor Tracking:} 
The intricate signal propagation effects in indoor environments make accurate indoor positioning and tracking a daunting task. Leveraging advances in ML and DL models, Alberto et al. \cite{alberto2020loc} proposed an innovative framework for adaptive indoor tracking using recurrent models, in conjunction with GANs for data generation and recovery (RecTrack-GAN). This approach improved the accuracy of tracking and reduced error deviation by up to 15\% compared to previous methods.

\noindent {\large\adforn{4}} \textit{Automated Indoor Intrusion Sensing with RaGAM:} Zhou et al. propose the ray-aided generative adversarial model (RaGAM) for efficient WLAN-based indoor intrusion sensing and localization, leveraging adaptive-depth ray tree-based ray-tracing to construct synthetic radio maps and joint synthetic and unsupervised learning to refine these maps using unlabeled RSSI data. The refined map is used to train a probabilistic neural network (PNN) for classifying new RSSI data, leading to accurate intrusion sensing and localization \cite{zhou2020loc}.

\noindent {\large\adforn{4}} \textit{Indoor Localization Improvement with VAEs:} Suroso et al. propose using VAEs to reduce the effort and cost of constructing large offline databases for fingerprinting-based indoor localization. They use VAEs to learn the distribution of the fluctuating RSSI signals and generate synthetic values, which shows potential for enhancing database density. However, the accuracy of the synthetic values is lower with smaller numbers of training epochs, suggesting a need for larger datasets and adjustments in the assumed RSSI-to-image size \cite{suroso2022loc}.

\subsection{RF FINGERPRINTING FOR  LOCALIZATION}

\noindent {\large\adforn{4}} \textit{AF-DCGAN for Fingerprint Database Extension:} 
Li et al. pioneered a method to extend fingerprint databases using a Wavelet Transform-Feature Deep Convolutional Generative Adversarial Network model, which accelerated the training phase and diversified generated feature maps, improving indoor localization accuracy \cite{li2018locfinger}. In a follow-up work, they proposed an Amplitude-Feature Deep Convolutional GGAN  model, which converted collected CSI data into amplitude feature maps, significantly increasing CSI amplitude feature map diversity \cite{li2021locfinger}.

\noindent {\large\adforn{4}} \textit{Generating Synthetic Fingerprint Maps:} 
Lim et al. devised a method to minimize the number of reference points needed for Wi-Fi localization. Utilizing a GAN, they interpolated the values between reference points and synthesized a dense fingerprint map, which led to improved localization accuracy \cite{lim2021locfinger}. Meanwhile, Wei et al. suggested using a single AC-GAN to generate artificial fingerprints, achieving improved localization accuracy and reducing computational resources compared to multi-GAN approaches \cite{wei2021locfinger}.
\noindent {\large\adforn{4}} \textit{3D Tensor for Modeling Fingerprint Samples:} 
Zhu et al. tackle the challenge of indoor localization for smartphones where GPS fails, by proposing a novel DL architecture, namely the tensor-GANs \cite{zhu2018locfinger}, that uses a transform-based 3D tensor to model fingerprint samples and employs DL for network classifiers. Additionally, a tensor-based super-resolution scheme is introduced using a GAN, which improves localization accuracy, response time, and implementation complexity, outperforming existing, energy-intensive methods.

\noindent {\large\adforn{4}} \textit{Device-Free Fingerprinting for Low-Cost Labeling:} 
Chen et al. introduce a semi-supervised, GAN-based device-free fingerprinting system for indoor localization, an essential technology for the IoT \cite{chen2020locfinger}. The proposed system leverages a minimal amount of labeled data and a larger pool of unlabeled data, thereby substantially reducing the need for costly data labeling. Compared to the current supervised schemes, this semi-supervised system offers comparable performance with equal amounts of labeled data, and superior performance with limited labeled data. The paper also provides a mathematical description of the system and discusses the interactions between its GAN components.

\noindent {\large\adforn{4}} \textit{Enhanced Indoor Positioning with CNN, LSTM, and GAN:} Darwin et al. propose a novel architecture combining CNN, LSTM, and GAN to improve indoor positioning accuracy \cite{darwin2022locfinger}. This integrated approach, combining supervised and unsupervised ML models, aims to increase training data and minimize positioning error. Tested across 17 public datasets, the proposed model demonstrated a substantial reduction in positioning error in more than 70\% of the datasets.

\noindent {\large\adforn{4}} \textit{LTE-A Signal-Based Localization with GAN:} Serreli et al. propose a fingerprint-based localization approach leveraging LTE-Advanced Pro Signal and ML techniques \cite{serreli2021locfinger}. The approach analyses the RSSI between the base station and user equipment, using a GAN to understand the correlation between RSSI and the user's position. The authors address the typically costly and time-consuming offline training phase in fingerprinting-based localization by using the GAN to generate new RSSI and localization models. Experimental results indicate a significant improvement in accuracy compared to conventional methods, achieving an impressive 0.4 meters and 10 meters accuracy in indoor and outdoor environments, respectively, while saving time in the measurement campaign.

\noindent {\large\adforn{4}} \textit{Fingerprint-Based IoT Positioning Using DCGANs:} Tarekegn et al. propose a novel method of fingerprint-based positioning for IoT applications, utilizing DCGANs to minimize site survey time and cost, and to mitigate signal fluctuations \cite{tarekegn2020locfinger}. The radio-map designed for receiving signals from WLAN and cellular networks in scalable environments is constructed using a sequential combination of a hybrid support vector machine and long short-term memory algorithms. Experimental results suggest that the proposed method provides promising and reasonable positioning performance for IoT devices operating in scalable wireless environments.

\subsection{OWC LOCALIZATION AND POSITIONING}  
\noindent {\large\adforn{4}} \textit{USL for LiFi Positioning in Indoor Environments:} Hua et al. study the potential of light-fidelity (Li-Fi) indoor positioning systems due to its high accuracy, robustness, and absence of electromagnetic interference \cite{hua2021locvlc}. However, they note that DL models used for Li-Fi positioning are susceptible to device heterogeneity and setup deviations, leading to poor performance during the testing phase. To address this issue, the authors propose an USL method based on adversarial training for visible light network regression. The proposed method employs a feature extractor to align feature distributions from different domains and a domain classifier to differentiate data from the training and test sets. The adversarial network model outperforms directly applying a trained model to the test set, especially under multi-dimensional attack conditions.

\noindent {\large\adforn{4}} \textit{Pedestrian Localization in V2X Networks:} Liu et al. propose a GAN architecture to enhance the accuracy of pedestrian location estimations in Smart City and V2X systems \cite{liu2022locvlc}. Recognizing the challenges posed when multi-modal data is not associated, such as when pedestrians are outside the camera field of view or camera data is missing, their GAN learns the underlying linkage between pedestrians' camera-phone data correspondences during training, and generates refined position estimations based only on pedestrians' phone data during inference. The GAN shows promising results, producing 3D coordinates with a 1-2 meter localization error across various outdoor scenes. The authors also demonstrate that the proposed model supports self-learning, improving localization accuracy by up to 26\% after fine-tuning on an automatically expanded dataset.

\subsection{SUMMARY, INSIGHTS, AND FUTURE DIRECTIONS}

This review has delved into the advancements and challenges in the realm of localization and positioning, particularly focusing on the potential and implications of ML techniques. In particular, GANs and VAEs have emerged as effective tools for enhancing indoor and outdoor localization, tracking, and data augmentation. These methods have significantly addressed the limitations of traditional RF fingerprint-based localization systems, such as storage constraints and limited sample availability.

GAN variations have displayed promising results in indoor localization and radio map automation. VAEs have been instrumental in increasing the density of offline databases. Data augmentation methods have shown substantial improvements in localization accuracy. Advanced models incorporating CNN, LSTM, and GANs offer integrated solutions that minimize positioning errors. In the domain of Li-Fi, adversarial training approaches have also been successful in tackling issues related to device heterogeneity and setup deviations.

While GMs have overcome several hurdles, they also introduce new challenges such as the high cost of data collection and susceptibility to setup deviations. Additionally, there is a need for more rigorous evaluation of these methods in real-world settings to validate their performance and applicability. Several avenues for future research are evident:

\begin{itemize}
    \item[\large \adforn{72}] \textit{Data Efficiency:} Further research could focus on developing models that require fewer training data but still deliver high localization accuracy.
    \item[\large \adforn{72}] \textit{Robustness:} Studies could explore the stability of these models in dynamic indoor and outdoor environments with fluctuating signal conditions.
    \item[\large \adforn{72}] \textit{Scalability:} Investigating the ability of these models to scale across different types of environments and use-cases is crucial for their broader applicability.
    \item[\large \adforn{72}] \textit{Multi-modal Fusion:} Future work may also delve into the fusion of data from multiple sensors or networks to improve localization and tracking.
    \item[\large \adforn{72}] \textit{Cost-Effectiveness:} Addressing the high costs associated with data collection and labeling can open the door for more cost-effective solutions.
    \item[\large \adforn{72}] \textit{Security and Privacy:} As localization technologies become more pervasive, research into preserving user privacy and data security will become increasingly important.
\end{itemize}
In conclusion, advancements in GMs offer promising solutions to long-standing challenges in localization and positioning, setting the stage for more reliable, efficient, and versatile systems in the years to come.

\section{NEW 6G FRONTIERS FOR GENERATIVE MODELS}
\label{sec:6GMfrontiers}
{Based on the GM tutorial presented in Sec. \ref{sec:GAI_Models} and literature on GenAI wireless networking use cases discussed through Sec. \ref{sec:PHY}-Sec. \ref{sec:localization}, this section will share our vision on how GMs can be utilized for the new research frontiers of 6G networks.} 

\begin{table*}[ht]
\centering
\caption{GM Roles in Semantic Communications.}
\label{tab:semantic}
\resizebox{1.45\columnwidth}{!}{%
\begin{tabular}{|l|l|l|}
\hline
\textbf{GMs} & \textbf{Semantic Roles} & \textbf{Description} \\
\hline
\multirow{4}{*}{LLMs} 
& Natural Language Understanding & Advanced interfaces for querying databases. \\
& Contextual Analysis & Ensure transmitted data fits within receiving entity's context. \\
& Semantic Compression & Generate concise summaries or coded representations. \\
& Intent Prediction & Predict future user interactions for proactive data transmission. \\
\hline
\multirow{3}{*}{GANs} 
& Data Synthesis & Generate synthetic data for training. \\
& Image Translation & Convert data into easier-to-interpret forms. \\
& Adaptive Modulation & Adaptively modulate data signals. \\
\hline
\multirow{2}{*}{VAEs} 
& Semantic Compression & Compress high-dimensional data into a latent semantic space. \\
& Context-Aware Messaging & Produce a compressed semantic state for various applications. \\
\hline
\multirow{3}{*}{DGMs} 
& Data Enhancement & Improve or restore signal quality. \\
& Security & Employ for steganography, hiding information. \\
& Variable Rate Communication & Adapt to varying degrees of data complexity. \\
\hline
\multirow{3}{*}{FGMs} 
& Real-Time Adaptability & Provide real-time updates. \\
& Efficient Encoding and Decoding & Maintain semantic integrity in limited bandwidth. \\
& Probabilistic Reasoning & Provide probabilistic estimates for decision-making. \\
\hline
\end{tabular}
}
\end{table*}

\subsection{SEMANTIC COMMUNICATIONS}
\label{sec:semantic_front}
Semantic communication is poised to be a cornerstone technology in the development of 6G networks. Unlike traditional communication paradigms that prioritize the transmission of raw bits with minimal distortion or data loss, semantic communication aims to transmit "meaningful" data and considers various factors such as context, prior knowledge of the receiver, and the purpose of the transmission must be considered. This nuanced approach positions semantic communication as a critical component of future 6G networks, which are expected to leverage GenAI to understand, interpret, and act upon vast heterogeneous data streams. In the rest, we provide insights into how GMs can facilitate semantic communications by fulfilling various functions depending on their inherent capabilities explained in Sec. \ref{sec:GAI_Models}.

\subsubsection{Contextual Analysis, Semantic Compression, and Intent Learning}
LLMs play a pivotal role in revolutionizing the landscape of semantic communications, bringing about an era of sophisticated, intelligent, and context-aware data exchanges. LLMs have also envisioned to be the cornerstone toward realizing self-governed, interactive AI agents that understands telecom language \cite{bariah2023understanding}, which will eventually lead us into collective intelligence rather than previous concept of connected intelligence \cite{zou2023wireless}. Moreover, LLMs can be designed as a multimodal large model trained over Telecom data and fine-tuned to perform several downstream tasks, which can eliminate the need for dedicated models and realized artificial general intelligence (AGI)-empowered wireless networks \cite{bariah2023large}.

Firstly, by harnessing the power of contextual analysis \cite{zhuyun2019contextual}, these models ensure that the data being transmitted is not just a mere collection of bits and bytes but is curated to fit seamlessly within the context of the receiving entity, enhancing the relevance and meaningfulness of the data exchange. Another transformative capability is semantic compression \cite{gilbert2023semantic}, where these models can distill vast amounts of information into concise summaries or coded representations, preserving the essence while optimizing for efficiency. This not only economizes on the resources but also facilitates quicker and more effective comprehension. Furthermore, intent learning and prediction can foresee future user interactions \cite{piantadosi2022meaning}. This proactive approach allows the system to preemptively transmit data based on predicted needs, ensuring timely availability and enhancing user experience. Lastly, with their capabilities in natural language understanding \cite{wang2018glue}, LLMs pave the way for advanced interfaces that can be employed for querying databases. This means rather than using structured queries, users can pose questions in natural, conversational language, and the system would comprehend and fetch the desired data.

\subsubsection{Data Synthesis, Image Translation, and Adaptive Modulation:}
GANs have emerged as powerful tools that are reshaping the fabric of semantic communications, making the process not only more effective but also adaptive and insightful. Inherent data synthesis feature of GANs proves invaluable for training, allowing communication systems to be fine-tuned and optimized without the need for vast amounts of real-world data, which might be difficult to obtain or sensitive in nature. Moreover, GANs can adeptly convert complex or esoteric data forms into more accessible, easier-to-interpret visual representations, ensuring that the transmitted information is not just received, but also easily comprehended by the receiving end. For instance, image translations facilitate better understanding and can bridge the gap between raw data and actionable insights \cite{isola2017image}. And in the dynamic and often unpredictable realm of data transmission \cite{bobrov2021massive}, GANs can adaptively modulate data signals based on the receiving environment, ensuring optimal transmission quality by dynamically adjusting to real-time conditions, ensuring data integrity, minimizes loss, and optimizes resource usage. 

\subsubsection{Context-Aware Messaging and Latent Space Compression:}
VAEs are at the forefront of transforming the dynamics of semantic communications with their unique capabilities tailored to ensure meaningful data exchange. Unlike traditional compression mechanisms, VAEs can efficiently compress high-dimensional data into a latent semantic space, capturing the underlying essence or meaning of the data rather than just its raw form \cite{spinner2018towards}. This compression is not about just reducing size but ensuring that the compressed data carries the most important and semantically rich information.  Moreover, VAEs are adept at encapsulating complex data scenarios into a compressed semantic state, which can be more efficiently transmitted across systems and be beneficial for various applications ranging from real-time analytics to context-sensitive response systems. By understanding the broader context, these models ensure that the transmitted data is not just a series of values, but a meaningful representation tailored for specific application needs \cite{Perdikis2021context}

\subsubsection{Data Enhancement, Steganography, and Variable Rate Communication:}
DGMs can bring a novel approach to semantic communications, addressing various challenges inherent to data transmission and interpretation. Primarily, DGMs possess the capability to refine or restore signal quality, ensuring the meaningful components of transmitted data remain undistorted even under suboptimal conditions \cite{yang2022diffusion}. This attribute proves invaluable in scenarios where signal degradation might obscure the semantic essence of the communicated data. On the security front, DGMs introduce a paradigm shift by enabling steganography—the art of hiding information within other non-secret data \cite{chen2018provably}. This not only ensures the concealment of sensitive data but also its seamless transmission without raising suspicions. Such covert communication modes elevate the security standards of semantic communications, ensuring that the intended message reaches its destination without external interference. Lastly, DGMs cater to the dynamic nature of communication by facilitating variable rate communication. Recognizing the varying complexities of data, DGMs can adapt their transmission rates accordingly, ensuring efficient data flow without overwhelming the communication channels. In this way, DGMs may ensure that the richness of semantics is relayed appropriately, regardless of the intricacy of the underlying data. 

\subsubsection{Real-Time Adaptability, Efficient Encoding and Decoding, and Probabilistic Reasoning:}
FGMs are poised to redefine the landscape of semantic communications, drawing from their unique strengths in modeling complex data distributions. Unlike other GMs, FGMs excel in generating samples and adapting their parameters in real-time, making them suitable for scenarios where information changes rapidly and systems must update their understanding promptly \cite{kim2018flowavenet}. This adaptability ensures that semantic communications remain current and reflective of the most recent data or context. By leveraging the bijective nature of flow models, FGMs can seamlessly transform high-dimensional data into compact representations without significant loss, ensuring that the semantic essence of information remains intact even in environments with constrained bandwidth \cite{zhou2022neural}. This efficient data transformation is crucial for 6G networks, which will handle massive heterogeneous data from various devices. Lastly, by capturing the underlying data distribution, FGMs can offer probabilistic estimates about uncertain scenarios, aiding decision-making processes \cite{papamakarios2021normalizing}. This is pivotal in semantic communications, where understanding the likelihood of various interpretations can greatly enhance the reliability and clarity of transmitted information. The GM roles and their brief description is summarized in Table \ref{tab:semantic}.

\begin{table*}[]
\caption{GM Roles in ISAC}
\label{tab:ISAC}
\footnotesize
\begin{tabular}{|p{2cm}|p{12.5cm}|p{2.5cm}|}
\hline
\textbf{ISAC Roles}        & \textbf{Role Descriptions}                                                                                          & \textbf{GMs} \\ \hline
Data \newline Augmentation &
  Creation of synthetic yet realistic sensory data to augment existing real-world data for robust ML training. &
  GANs, VAEs \\ \hline
Anomaly \newline Detection &
  Identification of data points or patterns that deviate significantly from expected behavior, essential for early warning and reliability. &
  GANs, FGMs, DGMs \\ \hline
Resource \newline Optimization &
  Simulation of various network conditions and prediction of future states for optimal resource allocation. &
  GANs, FGMs \\ \hline
Time-Series \newline Forecasting    & Prediction of future sensor readings based on past and current data, useful for proactive decision-making.          & LSTMs, GMs            \\ \hline
Sensor/Data \newline Fusion   & Combining data from multiple sensor types to create a more comprehensive and accurate environmental representation as well as capturing their complex distributions for advanced analytics and decision-making. & VAEs, FGMs, GTMs      \\ \hline
Decision \newline Making & Assistance in making a series of decisions that depend on a temporal sequence of sensor and communication data.     & LSTMs, GTMs           \\ \hline
Uncertainty Quantification & Estimation of the uncertainty or confidence level associated with sensor data or predictions.                       & VAEs, DGMs            \\ \hline
\end{tabular}%
\end{table*}

\subsection{INTEGRATED SENSING AND COMMUNICATIONS}
\label{sec:isac_front}
ISAC represents an exciting frontier for network technologies as the fusion of sensing and communication capabilities allows a network to adapt and respond dynamically to changes in its environment, thereby enabling more efficient and intelligent operations. GMs can play a pivotal role in realizing ISAC systems in multiple ways as summarized in Table \ref{tab:ISAC} and explained below. 

\subsubsection{Data Augmentation:} This role underscores the generation of synthetic yet convincingly real sensor data to bolster real-world data repositories, ensuring a comprehensive base for robust training. The GANs emerge as front-runners in this realm, renowned for their prowess in crafting high-fidelity and realistic data, rendering them exceptionally suited for such augmentation tasks \cite{sallab2019lidar}. VAEs, with their probabilistic nature, can also be leveraged to create a rich spectrum of data samples, accurately capturing environment nuances.

\subsubsection{Anomaly Detection:} Here, the goal is to pinpoint data patterns or individual data points that starkly deviate from established norms, a vital mechanism for prompt alerts and sustained system reliability. GANs have made strides in this domain, adeptly delineating the expected distribution of sensor data and subsequently singling out anomalies \cite{li2018anomaly}. Furthermore, FGMs, especially their subclass - normalizing flows, may excel in precise density evaluations, becoming foundational for statistical anomaly spotting \cite{gudovskiy2022cflow}. DGMs also enrich this process, infusing uncertainty quantification within their iterative modus operandi, amplifying anomaly detection efficacy \cite{yang2022diffusion}.

\subsubsection{Resource Optimization} This entails the simulation of varying network conditions and the prediction of future states to ensure judicious resource allocation. GANs shine in emulating a plethora of network scenarios, facilitating superior model training geared towards resource maximization \cite{song2021gansim}. Normalizing flows, under the FGM umbrella, with their adeptness in representing intricate network state distributions, emerge as invaluable aids for decision-making processes in this context.

\subsubsection{Time-Series Forecasting} This function is centered on the anticipation of subsequent sensor readings, rooted in historical and present data. This foresight is instrumental for pre-emptive decision-making. DGMs, through their iterative refinement processes, are uniquely positioned to project impending states, drawing inspiration from both past data and current inferences \cite{li2022generative}. LSTMs, architected specifically for time-series data, are naturally inclined to flourish in such forecasting roles \cite{siami2019performance}.

\subsubsection{Sensor/Data Fusion} This duty revolves around amalgamating data stemming from a diverse set of sensors, crafting an exhaustive and precise representation of the environment. VAEs simplify this intricate task, condensing multifarious data into a streamlined latent space \cite{duffhauss2022fusionvae}. Normalizing flows can further bolster this task by adeptly modeling intricate data distributions \cite{kobyzev2020normalizing}. GTMs, known for their sequence-to-sequence capabilities, also play an instrumental role in assimilating lengthy sequences of varied sensor data, underlining their significance in data fusion endeavors \cite{rao2023tgfuse}.

\subsubsection{Decision Making} This role aids in sequential decision-making rooted in a temporal cascade of sensor and communication data. GTMs, primed for processing extensive data sequences, are invaluable here, particularly in sequential decision-making scenarios. Additionally, LSTMs, with their design tailored for data sequences, are indispensable in contexts demanding a robust temporal or sequential backdrop \cite{amiri2020learning}.

\subsubsection{Uncertainty Quantification} Here, the focus is on discerning the confidence or uncertainty level tethered to sensor data or its corresponding predictions. VAEs, with their intrinsic probabilistic nature, facilitate not just precise estimates but also extrapolation of confidence intervals, cementing their place in this realm \cite{cheng2023bi}. DGMs, through their innate ability to integrate uncertainty in their iterative procedures, further augment this estimation process.

\begin{table*}[t!]
\centering
\caption{GM Roles in 6G Digital Twins}
\label{tab:DigitalTwinGM}
\footnotesize
\begin{tabular}{|p{1.5cm}|p{14cm}|p{0.7cm}|}
\hline
\textbf{DT Roles} & \textbf{Role Description} & \textbf{GMs} \\
\hline
Network \newline Simulation& Creating detailed and high-resolution simulations to represent 6G network states, which can be useful for testing scenarios, training other models, or predictive maintenance. & GANs \\
\hline
Uncertainty \newline Quantification  & Accurately capturing and depicting the inherent uncertainties of sensory data to provide confidence bounds of DTs. & VAEs \\
\hline
Temporal \newline Evolution & Addressing the inherently temporal nature of 6G network conditions, tracking their changes over time and predicting future states for more proactive network management and adaptation, understanding patterns, and trends. & GTMs, LSTMs \\
\hline
Distribution  \newline Modeling & Accurately representing intricate data distributions captured by sensors to ensure DTs can represent a wide range of scenarios and conditions. & FGMs \\
\hline
DT \newline Refinement & Continual improvement and refinement of DTs, ensuring evolution and adaptation as more sensory data becomes available and as the real-world scenario it mirrors changes. & DGMs \\
\hline
\end{tabular}
\end{table*}

{
\subsection{NEXT-GENERATION PHYSICAL LAYER DESIGN}
\label{sec:PHY_front}

The core of 6G technological advancement is anchored in the integration of high-frequency spectrums (e.g., mmWave \cite{AbdallahCME22} and THz \cite{sarieddeen2020next}), along with the implementation of massive size antenna arrays, i.e., RIS \cite{Abdallah2023RIS}, ELAA \cite{bjornson2021rethinking}, etc. This combination is pivotal in enhancing the density and effectiveness of connections. Advancements in the spectrum range and the increase of antenna concentration per area eventually lead us into the realm of NFC \cite{cui2022near}, offering precise control over the orientation, form, and structure of beams. In this context, holographic communication comes into play, particularly in the near-field zone, to intricately craft a “hologram” of electromagnetic fields \cite{zhang2020holographic}. This technique uses expansive arrays to create beams with unparalleled spatial resolution, morphing them in sophisticated manners akin to the process of optical holography in shaping light. The convergence of THz, ELAA, NFC, and holographic beamforming strengthens the forefront of advanced wireless communications, though it also intensifies the complexity in channel estimation and beamforming processes. The remainder of this subsection will visit prominent GMs and elaborate their roles in mitigating related challenges of these technologies. 

\subsubsection{THz communications} Channel modeling and signal propagation are paramount for THz communications. GANs can provide a solid foundation in creating realistic simulations of THz channel conditions and analyzing signal propagation, particularly in data-scarce scenarios. VAEs may supplement this by generating synthetic data that encapsulates THz-specific characteristics like atmospheric absorption, aiding in signal reconstruction under challenging environmental conditions. FGMs might offer precision in simulating THz wave propagation, focusing on interactions with environmental elements, crucial for optimizing system design. Additionally, DGMs can play a significant role in simulating the propagation of electromagnetic waves in intricate scenarios, providing valuable insights into optimal configurations and beam paths, enhancing the understanding of NFC and holographic beamforming dynamics in the THz spectrum. In device design and optimization, VAEs could be notably effective, proposing innovative antenna configurations and assisting in material selection for THz devices. FGMs might complement this by determining optimal designs to maximize efficiency and coverage.

\subsubsection{Extremely Large Antenna Arrays} GMs can collectively address the challenges of beamforming and interference management for ELAAs. GANs could be instrumental in generating dynamic beamforming algorithms and simulating various interference patterns. VAEs may contribute by modeling optimal beam patterns for different scenarios, reducing interference by understanding spatial interference distribution. FGMs might optimize the physical layout of ELAA to maximize signal coverage and minimize interference, while developing dynamic beamforming strategies. DGMs could further enhance this by simulating noise and interference patterns in wireless networks, assisting in the development of robust communication strategies that are critical in densely populated network environments. Network planning and resource management in ELAA benefit from GMs; GANs are valuable in simulating deployment scenarios and resource allocation, while VAEs optimize power distribution and signal routing. FGMs can predict the impact of resource allocation strategies on system performance. DGMs, with their ability to simulate complex noise and interference patterns, may inform strategies for efficient resource utilization and network management. Finally, in system performance analysis and predictive maintenance, VAEs might detect deviations for potential system issues, enhancing network health and security. FGMs could offer predictions on system behavior, essential for verifying and validating ELAA designs. DGMs, by simulating a wide range of noise and interference conditions, can also contribute to understanding and enhancing the robustness of communication systems.

\subsubsection{Near-Field and Holographic Beamforming}
The integration of GMs with NFC and holographic beamforming represents a pivotal advancement in these communication technologies. GMs can bring specialized capabilities that collectively address the complex challenges in NFC and holographic beamforming.

In NFC, GANs can play important roles in generating realistic electromagnetic wave propagation scenarios, which is key given NFC's sensitivity to environmental factors. GANs may create intricate simulations that closely mimic real-world conditions, thereby aiding in the optimization of NFC systems for diverse operational scenarios. Concurrently, VAEs can enhance NFC by modeling the unique propagation characteristics of electromagnetic fields at short distances, generating high-fidelity data crucial for system fine-tuning. FGMs in NFC might offer detailed simulations of electromagnetic field behaviors, providing deep insights into the subtle interactions within NFC environments, which is essential for designing more accurate and efficient systems. Additionally, DGMs can bring advanced capabilities in simulating complex wave interactions in NFC, offering valuable perspectives on optimal configurations and enhancing overall system performance.

Holographic beamforming benefits similarly from GMs; GANs are instrumental in modeling diverse beam patterns and operational environments, aiding in the development of beamforming strategies that optimize signal quality and efficiency. VAEs can play a significant role in holographic beamforming by analyzing and reconstructing beam patterns, thereby improving the resolution and focus of holographic beams in complex communication environments. FGMs might contribute to holographic beamforming by simulating the dynamic manipulation of electromagnetic waves, enabling the creation of sophisticated and adaptable beamforming strategies. This enhances the capability to focus and direct signals with high precision. Similarly, DGMs could be highly valuable in creating detailed simulations of electromagnetic field interactions for holographic beamforming. They assist in developing techniques that intricately control and shape beam patterns, which is crucial for high-resolution communication applications.
}

\begin{table*}[]
\centering
\caption{GM Roles in AIGC}
\label{tab:AIGC}
\resizebox{\textwidth}{!}{%
\begin{tabular}{|l|l|l|}
\hline
\textbf{AIGC Roles}                                                                 & \textbf{Description}                                                                               & \textbf{Generative Models}                                        \\ \hline
\begin{tabular}[c]{@{}l@{}}Real-time Content \\ Creation/Customization\end{tabular} & Generating or modifying content in real-time for personalized user experiences.                    & \begin{tabular}[c]{@{}l@{}}LLMs, GPTs, \\ GANs, DGMs\end{tabular} \\ \hline
Enhanced VR/AR                                                                      & Creating realistic textures and environments for VR and AR applications.                           & GANs, DGMs, FGMs                                                  \\ \hline
Content Distribution                                                                & Optimizing content distribution for various devices and network conditions.                        & VAEs, FGMs                                                        \\ \hline
Data Compression                                                                    & Compressing and decompressing data while maintaining quality, crucial for images and videos.       & VAEs, DGMs                                                        \\ \hline
Interactive AI Services                                                             & Generating interactive text-based content like conversational AI and personalized recommendations. & LLMs, GPTs                                                        \\ \hline
\begin{tabular}[c]{@{}l@{}}Smart Cities and \\ IoT Integration\end{tabular}         & Simulating urban environments and generating data for IoT devices.                                 & GANs, FGMs, DGMs                                                  \\ \hline
\begin{tabular}[c]{@{}l@{}}EAI and Localized \\ Content Creation\end{tabular}       & Generating content directly on edge devices, reducing latency.                                     & VAEs, GANs                                                        \\ \hline
\end{tabular}%
}
\end{table*}

\subsection{DIGITAL TWINS OF 6G NETWORKS}
\label{sec:twin_front}
The seamless integration of the digital and physical realms lies at the heart of the 6G vision \cite{bariah2022interplay}. As 6G networks emerge, accommodating a myriad of connected devices and facilitating large-scale IoT deployments, there is a dire need for tools that can create accurate digital representations of the vast and complex real-world scenarios 6G networks inhabit \cite{alkhateeb2023real}. This is where ISAC and DT paradigm handshake, utilizing sensory data to craft intricate virtual replicas of both network conditions and the wireless environment. However, transforming vast, diverse, and dynamic sensory data into precise, actionable DT models presents its challenges. GMs stand out as potent solutions to these challenges, intertwining the promise of 6G with the efficacy of DTs.

\subsubsection{High-resolution Network Simulations:} GANs, known for their aptitude in generating high-resolution, detailed simulations, become pivotal in translating sensory data into vivid digital replicas of 6G network states \cite{njima2021loc_journal}. Given the dynamism of 6G environments, where network conditions can fluctuate based on myriad factors, GANs assist in simulating various potential network states. Their capability to produce realistic representations from sensory data can also enhance predictive maintenance, ensuring networks operate optimally and preemptively addressing potential bottlenecks.

\subsubsection{Uncertainty Quantification:} VAEs, with their probabilistic underpinnings, are particularly adept at capturing and representing the inherent uncertainties of sensory data \cite{cheng2023bi}. As sensors capture the multifaceted nuances of the environment, VAEs can effectively process this data, offering predictions alongside confidence intervals. This becomes essential when representing complex 6G network conditions, ensuring that the digital twin remains a true reflection of the ever-evolving physical state.

\subsubsection{Temporal Modeling and Prediction:} The temporality of network conditions and their evolution finds a match in GTMs \cite{shabaninia2022transformers}, which can excel in processing sequential data, translating it into a coherent narrative of how 6G network conditions might evolve in response to changes in the environment. This time-series interpretation offers network operators foresight, allowing for timely interventions and adjustments.

\subsubsection{Distribution Modeling:} FGMs provide a nuanced approach to modeling intricate data distributions \cite{bond2021deep}, ensuring that the digital representation encapsulates even the most subtle changes in network conditions as captured by sensors. In the fast-paced world of 6G, where real-time synchronization between digital twins and their physical counterparts is crucial, the swift density estimation capabilities of these models become indispensable. 

\subsubsection{Continuous DT Refinement:} Rounding off with DGMs, their iterative approach ensures continual refinement of the digital representation \cite{cao2022survey}. In a realm where sensory data might exhibit noise or where network conditions can change unpredictably, these models' emphasis on uncertainty provides a more layered, refined understanding.

In essence, as 6G networks embark on revolutionizing our digital landscapes, the confluence of sensory data, DTs, and GMs promises an ecosystem where network conditions and environments are not just monitored but truly understood. This understanding, rooted in the dynamic interplay of real-world data and advanced modeling techniques, is set to be the bedrock upon which 6G will thrive.

{

\subsection{AI-GENERATED CONTENT FOR 6G NETWORKS}
\label{sec:AIGC_front}

In the rapidly evolving landscape of telecommunications, the advent of 6G networks is set to revolutionize the way we interact with digital content. This new era is not just about faster speeds and lower latencies; it heralds a transformative shift in the creation, distribution, and consumption of AIGC. The integration of advanced generative models into the fabric of 6G technology is a critical factor in this transformation. GMs offer unprecedented capabilities in generating realistic, interactive, and personalized content. As summarized in Table \ref{tab:AIGC}, this section delves into the synergistic relationship between various GMs and their specific roles in enhancing the AIGC landscape within 6G networks. Through a comprehensive analysis, the remainder will illuminate how GMs are not just facilitating but actively shaping the future of digital media and communications. 

\subsubsection{Real-time Content Creation and Customization}
LLMs and GPTs stand out for their ability to generate personalized text content, including chat responses and customized news articles. Meanwhile, GANs and DGMs are highly effective for generating or modifying images and videos in real-time. These models offer high-quality visual content, making them ideal for dynamic and interactive media applications.

\subsubsection{Enhanced VR and AR}
In the domain of enhanced VR and AR, GANs and DGMs excel in creating realistic textures and environments. Their proficiency in generating high-resolution, detailed images is crucial for the immersive experiences demanded by VR and AR technologies. Additionally, Flow-based Generative Models are instrumental in 3D object generation and transformations due to their ability to model complex distributions, further enhancing the realism and interactivity of virtual spaces.

\subsubsection{Efficient Content Distribution}
The improved efficiency in content distribution is another critical area where VAEs and FGMs play a significant role. They are capable of efficient data compression and reconstruction, which is vital for adaptive content streaming across diverse devices and network conditions. This efficiency is essential in optimizing network resource utilization and ensuring seamless content delivery.

\subsubsection{Advanced Data Compression }

Advanced data compression techniques, particularly relevant for images and videos, are effectively handled by VAEs and DGMs. These models are adept at compressing and decompressing data while maintaining high quality, a key requirement for the bandwidth-intensive media content typical in 6G networks.

\subsubsection{Interactive AI Services}

Interactive AI services, such as conversational AI and personalized recommendations, benefit greatly from the capabilities of LLMs and Generative Pre-trained Transformers. These models are best suited for generating interactive text-based content, enabling more engaging and responsive user experiences.

\subsubsection{Smart cities and IoT Integration}

In the context of smart cities and IoT integration, GANs and FGMs are invaluable for simulating urban environments and generating data for IoT devices. This capability aids in city planning and environmental monitoring. DGMs also contribute significantly, particularly for tasks like environmental modeling or generating realistic simulations of city dynamics, enhancing the efficiency and effectiveness of smart city operations.

\subsubsection{Edge AI and Localized Content Creation}
Lastly, in the sphere of edge AI and localized content creation, VAEs and Lightweight GANs are particularly suitable. Due to their relatively lower computational requirements, these models are ideal for edge devices, enabling them to generate content directly on the device. This reduces latency and is crucial for applications requiring quick response times, such as autonomous vehicles or local media generation.

Each of these GMs brings unique strengths to the table, making them well-suited for specific tasks within the evolving landscape of AIGC in 6G networks. Their integration promises to revolutionize how content is created, distributed, and experienced, paving the way for more immersive, efficient, and personalized digital interactions.

}
\subsection{MOBILE EDGE COMPUTING AND EDGE AI}
\label{sec:MEC_front}
MEC and EAI represent a transformative step for 6G networks, particularly in the way they decentralize data processing. The traditional model, which relied heavily on central servers or cloud-based processing, introduced latency and inefficiencies—particularly for time-sensitive tasks or for devices in remote locations. By integrating computation and AI/ML capabilities directly at the network's edge, devices can process data and make decisions in real-time, right where they are. This shift is crucial for the evolving demands of 6G and its myriad applications. As summarized in Table \ref{table:MEC_EAI}, GMs are a potent tool in this landscape \cite{wang2023overview}, and below details how they fit into the MEC-EAI paradigm for 6G:

\subsubsection{Localized Data Augmentation:} GANs can generate synthetic data samples that mimic real-world data \cite{weng2019data}. In a MEC scenario, GANs can augment the data directly at the edge, useful for scenarios where large datasets are required for training or evaluation but are not available. 

\subsubsection{On-device Uncertainty Modeling:} VAEs are probabilistic models, and their inherent ability to quantify uncertainty can be crucial for edge devices \cite{ahmadi2022variational}. 

\subsubsection{Real-time Sequence Processing:} Many edge applications, such as VR and AR, depend on sequential data. LSTMs and GTMs excel at processing sequences, enabling real-time adjustments in AR experiences based on the user's movements and interactions.

\subsubsection{Complex Distribution Modeling at the Edge:} Edge devices often encounter intricate data distributions, especially in multi-modal or multi-sensor environments. Normalizing Flows can help in modeling these complex distributions, ensuring that the localized AI can understand and act upon the nuanced information it receives.

\subsubsection{Iterative Refinements with DGMs:} In scenarios where the edge data might be noisy or unpredictable, DGMs can play a pivotal role. They iteratively refine their outputs, improving accuracy over time, which can be instrumental for applications like advanced robotics where environmental changes can be erratic.

The blend of MEC and EAI in 6G, supported by GMs, opens up enormous possibilities. It not only decentralizes data processing but also makes it smarter, faster, and more adaptable. The importance of real-time, efficient, and localized decision-making can't be overstated for future networks, especially as the demand for intelligent, autonomous systems continues to rise. GMs, with their varied capabilities, are poised to be pivotal players in the realization of the full potential of MEC in 6G networks.

\begin{table*}[!]
\centering
\caption{GM Roles in MEC and EAI for 6G Networks}
\label{table:MEC_EAI}
\begin{tabular}{|p{1.5cm}|p{15cm}|p{0.7cm}|}
\hline
\textbf{MEC Roles} & \textbf{Role Description} & \textbf{GMs} \\
\hline
Local Data \newline Augmentation &  Generating synthetic data samples mimicking real-world data, aiding in scenarios where on-device datasets for training or evaluation are limited & GANs \\
\hline
Uncertainty \newline Modeling & Predicting uncertain conditions due to their probabilistic nature, aiding edge devices in adjusting operations based on dynamic environments & VAEs \\
\hline
Real-time \newline Processing & For applications like AR that rely on sequential data, processing data sequences in real-time, offering real-time interactions. & LSTMs GTMs \\
\hline
Distribution \newline Modeling & Modeling intricate data distributions to ensure AI models using edge devices interprets nuanced information, especially in multi-sensor setups & FGMs \\
\hline
Iterative \newline Refinements & In environments where data can be noisy, DGMs refine their outputs over iterations, improving accuracy, especially beneficial for unpredictable scenarios. & DGMs \\
\hline
\end{tabular}
\end{table*}


\subsection{ADVERSARIAL ML AND TRUSTWORTHY AI}
\label{sec:trust_front}
AML focuses on understanding how AI models can be susceptible to specially crafted inputs that can deceive the model into making incorrect predictions or classifications. These adversarial examples are designed by introducing small, often imperceptible, perturbations to the input data, such that the model's output is manipulated, which may appear virtually unchanged to humans. The study of AML is crucial not just to understand these vulnerabilities, but also to develop defenses against potential adversarial attacks. In what follows, we explain how GMs can play role in AML domain. 

\subsubsection{Attack Mechanisms using GMs:}
GMs can be used to generate adversarial examples or attack patterns \cite{chen2020mag}. For instance, GANs have been employed to craft adversarial samples that can fool a classifier. The generator in GANs can be trained to produce samples that, although they look genuine, lead a target classifier into making wrong decisions.

\subsubsection{Defense Mechanisms using GMs:}
On the flip side, GMs can also be leveraged to detect and counter adversarial attacks. For instance, a GAN can be trained where the generator creates adversarial examples and the discriminator learns to distinguish between genuine and adversarial samples \cite{deldjoo2021survey}. Once well-trained, the discriminator can act as a filter to detect adversarial examples in real-world deployments. VAEs can be employed as a defense mechanism \cite{ghosh2019resisting}. A VAE can be trained on the original dataset, and when an input (potentially adversarial) is passed through the VAE, the reconstructed output is fed into the classifier. Since VAEs tend to reconstruct what they have learned and ignore tiny perturbations, this can mitigate the impact of adversarial attacks.

\subsubsection{Evolving Threats in 6G:} 6G networks, with enhanced connectivity and distributed intelligence, will provide ample opportunities but will also be vulnerable to sophisticated cyber-attacks. With the convergence of MEC and EAI techniques in 6G, edge devices may be exposed to adversarial attacks aiming to deceive local AI inferences. For instance, a malicious actor could attempt to tamper with an IoT device's sensory data to alter its AI-driven decisions, necessitating robust defense mechanisms. GTMs and LSTMs, given their capacity to handle sequences, can be employed in AML to detect adversarial attacks in time-series data \cite{xu2022tgan}, especially relevant for communications data in 6G networks. FGMs and DGMs can be applied for modeling the complex data distributions of regular traffic and identifying discrepancies in the form of adversarial attacks.

AML directly addresses the challenge of adversarial attacks, making AI systems more resilient and robust. On the other hand, as defenses against adversarial attacks are developed, there's a growing need for transparency to understand how these defenses work and how the model processes information, which falls within the realm of trustworthy AI (TAI) \cite{kaur2022trustworthy}.
Various frameworks have been proposed to guide the development of TAI, with the European Commission's High-Level Expert Group on Artificial Intelligence\footnote{\url{https://digital-strategy.ec.europa.eu/en/policies/expert-group-ai}} being one of the more prominent bodies to have formalized TAI as a concept revolving around seven pillars that encompasses various facets of AI ethics and robustness. In the remainder, we discuss how GMs can contribute to each of these pillars, especially in the context of 6G networks:

\noindent {\large\adforn{38}} \textit{Human Agency and Oversight:} GMs can be used to create interpretable representations of complex data, helping human operators understand and oversee AI-driven processes. Especially in the dense, fast-paced environment of 6G networks, such insights can be invaluable for human administrators.

\noindent {\large\adforn{38}} \textit{Privacy and Data Governance:} GMs can help in data augmentation without breaching privacy. For instance, GMs can be used to create synthetic datasets that resemble the original without carrying sensitive information. Additionally, they can aid in federated learning, where model training occurs at the edge without centralizing data.

\noindent {\large\adforn{38}} \textit{Transparency:} GMs, especially VAEs, can create latent space representations of data that are more understandable. By visualizing these representations, one can glean insights into what the AI model deems significant, fostering transparency.

\noindent {\large\adforn{38}} \textit{Diversity, Non-discrimination, and Fairness:} GANs and other GMs can be trained to ensure that AI models are tested across diverse scenarios and data distributions, ensuring they don't inadvertently discriminate against particular groups or conditions.

\noindent {\large\adforn{38}} \textit{Societal and Environmental Well-being:} GMs can simulate various societal and environmental scenarios, allowing for the predictive assessment of AI's impact. In the context of 6G, this could involve understanding the network's response to different societal needs or environmental conditions.

\noindent {\large\adforn{38}} \textit{Accountability:} By creating simulations and synthetic scenarios, GMs can play a role in "stress-testing" AI systems. In cases where AI-driven decisions in 6G lead to unintended outcomes, these simulations can be revisited to understand the decision-making process, facilitating accountability.

As 6G networks pave the way for more pervasive and intricate AI integrations, the principles of trustworthy AI become even more vital. Generative models, with their versatile capabilities, will play a pivotal role in ensuring that the AI systems of the future are not only intelligent but also ethical, robust, and trustworthy.

\section{OPEN CHALLENGES AND THE WAY FORWARD}
\label{sec:GM_challenges}
In this section, we discuss multi-faceted open research challenges and point out strategies and candidate technologies as a remedy.   
 
\subsection{RAMIFICATIONS OF COMPUTATIONAL COMPLEXITY}
\label{sec:complexity}
The computational overhead associated with GMs presents a major challenge for real-time deployment in 6G networks, which are anticipated to have even more stringent latency requirements than their predecessors. Several dimensions add to this complexity:

    \subsubsection{Energy Efficiency} High computational demands generally translate to increased energy consumption. In the context of 6G, which aims for greener and more sustainable operations, the energy-intensive nature of GMs poses a significant hurdle. Energy-efficient algorithms and hardware acceleration techniques may be needed to make the deployment of GMs viable.

    \subsubsection{Real-time Requirements} 6G networks are expected to support real-time applications like augmented reality, autonomous vehicles, and telemedicine, which have stringent latency requirements. Achieving low-latency operations with computationally demanding GMs remains a formidable challenge.

    \subsubsection{QoS and QoE} Meeting QoS and QoE expectations in a 6G environment laden with computational complexities is another challenge. Network operators will need to balance these quality metrics against the limitations and costs imposed by the computational complexity of GMs.

    \subsubsection{Trade-off Between Speed and Accuracy} Strategies to reduce computational complexity often involve simplifying models or using approximation techniques. However, this can come at the cost of accuracy or reliability, making it a challenging trade-off to navigate.
    
    \subsubsection{Distributed Computing} One potential solution could be the use of distributed computing or edge computing to offload some of the computational tasks. While this can mitigate latency, it introduces new complexities in data synchronization, resource allocation, and network architecture.

    \subsubsection{Hardware Considerations} Specialized hardware accelerators like GPUs or TPUs may be necessary to meet performance criteria. However, these come with their own set of challenges, including increased cost, power consumption, and potential issues with hardware-software compatibility.

Given the expected heterogeneity and dynamism of 6G networks, GMs may need to dynamically adapt to varying computational resources and latency requirements. This calls for the development of GMs and algorithms that can self-optimize based on the operating conditions. Addressing the challenges of computational complexity and latency will likely require a multi-disciplinary approach, involving advances in algorithm design, hardware acceleration, network architecture, and possibly even new paradigms in computing. The cost, energy efficiency, and user experience implications of these challenges make them central to the successful deployment of GMs in 6G networks. 

\subsection{SCALABILITY AND FLEXIBILITY}
\label{sec:scalability}
As 6G networks are expected to accommodate an exponential growth in connected devices, services, and types of data, generative models must be engineered for high scalability and flexibility, which involves several facets:
        \subsubsection{Horizontal Scaling} To manage increasing computational demands, the ability to distribute tasks across multiple machines or clusters becomes vital. Horizontal scaling is often more desirable than vertical scaling (adding more power to a single machine) because it offers greater flexibility in managing workloads and can adapt more fluidly to varying demand.
        
        \subsubsection{Resource Efficiency} For real-time or near-real-time applications, the GMs must be designed to perform tasks under stringent resource constraints. This could involve optimizing algorithms for speed and low memory usage, perhaps by simplifying GMs where possible without sacrificing performance quality.
        
        \subsubsection{Adaptive Architectures} GMs should feature architectures that can adapt based on the hardware availability. For example, a model deployed in a high-resource cloud environment may be able to utilize more layers or nodes, whereas the same model running on an edge device with limited computing capabilities might need to operate in a simplified mode.
    
        \subsubsection{Data Stream Adaptability} 6G networks will not only deal with an increasing amount of data but also a more diverse range of data types and sources. GMs will need to handle streaming data effectively, making real-time adjustments to adapt to changes in data features or volume. This is especially important for time-sensitive applications like autonomous vehicles or real-time health monitoring.

    Scalability and flexibility can be improved through following main approaches: 
    \begin{itemize}
        \item [\large \adforn{73}] \textit{Decentralization and Edge Computing} To further enhance scalability, employing decentralized approaches and edge computing can distribute the computational load. Instead of sending all data to a central server for analysis, pre-processing can be performed closer to the source of the data. This reduces latency, bandwidth usage, and the central computational load, but also poses challenges for coordinating and updating GMs across a distributed network.
        
        \item [\large \adforn{73}] \textit{Modularity and Interchangeability} Scalability is not just about handling more devices or more data; it's also about flexibility in incorporating new features or capabilities. Modular design principles can help by allowing different parts of the GMs to be updated independently, without requiring a complete overhaul of the system. This makes it easier to adopt new algorithms, adapt to new data types, or interface with new devices, thereby ensuring that the system remains both scalable and flexible.
    \end{itemize}

\subsection{ROBUSTNESS AND RELIABILITY} 
\label{sec:robustness}
For GMs to be effective in 6G networks, they must be both robust against various forms of errors/conditions and reliable in their performance. Several facets contribute to these challenges:
        \subsubsection{Environmental and Contextual Variability} 6G networks will operate in diverse settings, from densely populated urban areas to remote rural locations. GMs need to be robust against different types of noise and interference, which can vary significantly across these environments.
    
        \subsubsection{Fault Tolerance} GMs should be designed to continue functioning even when certain network nodes or data streams fail. This involves engineering redundancy and self-healing mechanisms.
        
        \subsubsection{Uncertainty Quantification} The ability to estimate the uncertainty associated with GM predictions can improve robustness, as it allows for the quantification of confidence levels and assists in making more informed decisions.
    
        \subsubsection{Data Quality} The reliability of a GM is highly dependent on the quality of the data used for training and validation. Robust mechanisms for data verification and validation are therefore crucial.

Optimizing both the software and hardware aspects can lead to more robust and reliable GMs, but also adds complexity to the design and deployment process. To increase robustness, GMs could benefit from fusing data from multiple types of sensors, where ISAC can play a vital role.

\subsection{INTEROPERABILITY, COMPATIBILITY, AND STANDARDIZATION}
\label{sec:compatibility}
The challenges of making generative models (GMs) work smoothly across diverse devices, vendors, and networks involve numerous complexities:

    \subsubsection{Variability in GM Architectures} Given that GMs can vary widely in terms of architecture, data requirements, and output formats, achieving interoperability is highly challenging. Devices from different vendors may rely on different versions or types of models, adding further complexity to standardization efforts.

    \subsubsection{Legacy Systems and Backward Compatibility} As advanced GMs are deployed in newer devices, there's a pressing need to ensure that these new models can interact seamlessly with older, legacy systems. This involves intricate problems like data format conversion and feature degradation to make them compatible with older systems, often at the cost of limiting advanced functionalities.

    \subsubsection{Middleware and Translation Layers} To address the issues of interoperability and backward compatibility, middleware solutions or translation layers might be necessary. These would serve as intermediaries to convert and interpret data or commands between disparate GM architectures and legacy systems.
    
    \subsubsection{Versioning and Updates} Managing updates to GMs in a way that doesn't disrupt existing services is a delicate balance. Network operators will have to develop robust versioning strategies to roll out improvements without causing system-wide issues.

    \subsubsection{Security and Trust} Interoperability also brings challenges in ensuring that communications between different GMs and systems are secure. Standardized security protocols that work across various generative models will be essential.

    \subsubsection{Standardization Bodies and Consensus} The development of universally accepted standards for GMs in 6G networks is a massive undertaking requiring multi-stakeholder involvement. This includes device manufacturers, network operators, policy makers, and even end users. Given the fast pace of technological advancements, these standards also need to be agile and adaptive.

    \subsubsection{Global Coordination} 6G networks are expected to be global, adding another layer of complexity to standardization efforts. International bodies may need to be involved to ensure that standards are universally applicable and that they meet the regulatory requirements of different countries.

Creating a standardized, interoperable, and backward-compatible ecosystem for GMs in 6G networks will be one of the defining challenges of this technological shift. It is a monumental task that involves not only technical solutions but also coordinated efforts from regulatory bodies and industry stakeholders. Without such standardization, the full potential of GMs in enhancing 6G services may remain unrealized.

\subsection{REGULATION AND POLICY} 
\label{sec:regulation}
As GMs become an integral part of 6G networks, they should be subject to a myriad of rules, regulations, and policies, both existing and those that are yet to be established. Here are some key challenges in this arena:

    \subsubsection{Data Protection, Ethics, and Social Impacts} GMs may  potentially breach privacy norms by inadvertently acquiring sensitive or personally identifiable information, especially when dealing with unstructured data such as text or images. Striking a balance between data privacy and the operational efficiency of semantic communication poses a substantial challenge. GMs are prone to incorporating the biases present in their training data. In a 6G context, these biases could manifest in various forms—such as discriminatory service delivery or unfair resource allocation—which may have direct social implications. With 6G networks enabling more powerful data analytics and AI capabilities, there's potential for mass surveillance and intrusions into personal privacy. Ethical guidelines are needed to define the acceptable limits of such technologies.

    Above issues are generally governed by regulatory bodies which provide protection by restricting certain kinds of data collection or data usage. For instance, General Data Protection Regulation (GDPR) in the European Union requires a compliance checklist that spans over 7 principles: 1) lawfulness, fairness, consent, and transparency for automated decision-making in critical applications (e.g., healthcare, transportation, or law enforcement, etc.) ; 2) purpose limitation; 3) data minimisation; 4) accuracy; 5) storage Limitations; 6) integrity and confidentiality; and 7) accountability, some of which will be covered in the remainder. Ensuring compliance without compromising functionality is a challenge. 

    \subsubsection{National Security} GMs could be exploited for malicious purposes, such as creating deepfakes or other forms of misinformation, which should be recognized by 6G networks up to a certain extend and transferred to a human-in-the-loop system for further analysis and actions. Governments may also regulate the export, import, and use of certain types of GMs for reasons of national security, which may hinder adoption of certain 6G technologies and services in some part of the globe.
    
    \subsubsection{Cross-Border Coordination \& Public/Private Partnerships} 6G networks are expected to be global in nature, raising the challenge of aligning different countries' policies and regulations related to GMs. This could involve mutual recognition of data protection laws, or establishing international standards for model behavior and interoperability. Moreover, implementing regulatory frameworks for 6G networks utilizing GMs is likely to require collaboration between the public and private sectors. This involves agreeing on shared responsibilities, aligning commercial incentives with public good, and negotiating potential conflicts of interest.

Addressing these regulatory and policy challenges would necessitate a multi-stakeholder approach that takes into account the technical complexities as well as the societal implications of using GMs in 6G networks. Industry standards, legal frameworks, and international cooperation will all play a critical role in navigating these challenges effectively.

\section{CONCLUSIONS}
\label{sec:conclusions}
{
In the dawn of the 6G era, the integration of novel communication paradigms and emerging technologies is rapidly transforming wireless communications, with ML and AI emerging as game-changers in addressing complex challenges. While DAI dominates the AI-driven wireless research landscape, the capabilities of GenAI in enhancing and supplementing DAI methods are becoming increasingly evident, especially in scenarios with limited or incomplete data. Our comprehensive tutorial-survey not only provides insights into the fundamentals of 6G and a taxonomy of cutting-edge DAI models but also delves deep into the pivotal role of GMs across various sectors of wireless research. Through our extensive analysis of numerous technical works, we debunk the misconception of GenAI being merely a nascent trend and highlight its profound implications in shaping the future of 6G networks. Furthermore, by pinpointing potential challenges and recommending promising strategies, we believe our paper equips researchers and professionals with valuable insights, positioning itself as a benchmark resource in this rapidly evolving field.
}
\section{ACKNOWLEDGEMENTS}
Fig. 1, Fig.3, and Fig.4 were produced by Aysenur Kucuksari, a freelancer scientific illustrator (e-mail: aysenurkucuksari22@gmail.com). 

\bibliographystyle{IEEEtran}
\bibliography{IEEEabrv,References}

\newpage
\begin{IEEEbiography}[\frame{\includegraphics[width=1in,height=1.25in,clip,keepaspectratio]{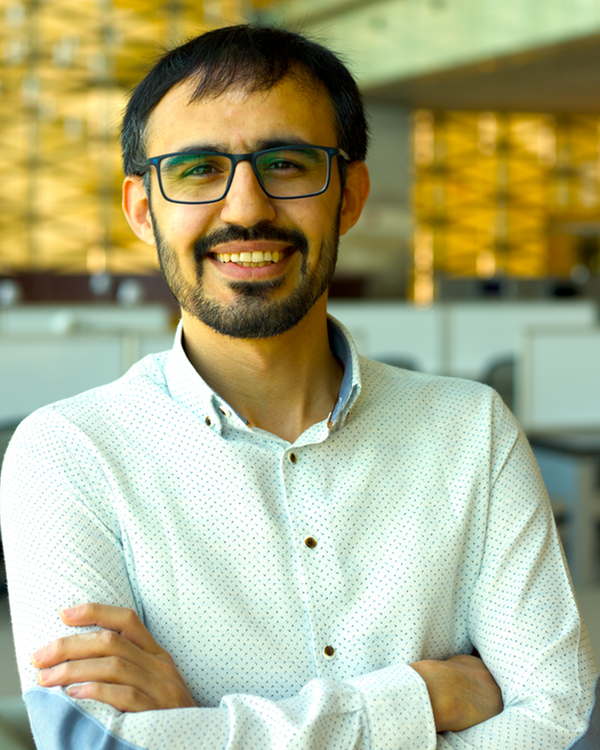}}]{ABDULKADIR CELIK} (Senior Member, IEEE) received the first M.S. degree in electrical engineering in 2013, the second M.S. degree in computer engineering in 2015, and the Ph.D. degree in co-majors of electrical engineering and computer engineering from Iowa State University, Ames, IA, USA, in 2016. He was a Postdoctoral Fellow with the King Abdullah University of Science and Technology, Thuwal, KSA, from 2016 to 2020, where he is currently a Senior Research Scientist with the Communications and Computing Systems Laboratory. Dr. Celik is the recipient of \textsc{IEEE} Communications Society's 2023 Outstanding Young Researcher Award for Europe, Middle East, and Africa (EMEA) region. He currently serves as an editor for \textsc{IEEE} Communications Letters, \textsc{IEEE} Wireless Communication Letters, and Frontiers in Communications and  Networks. His research interests are in the broad areas of next-generation wireless communication systems and networks. 
\end{IEEEbiography}

\begin{IEEEbiography}[\frame{\includegraphics[width=1in,height=1.25in,clip,keepaspectratio]{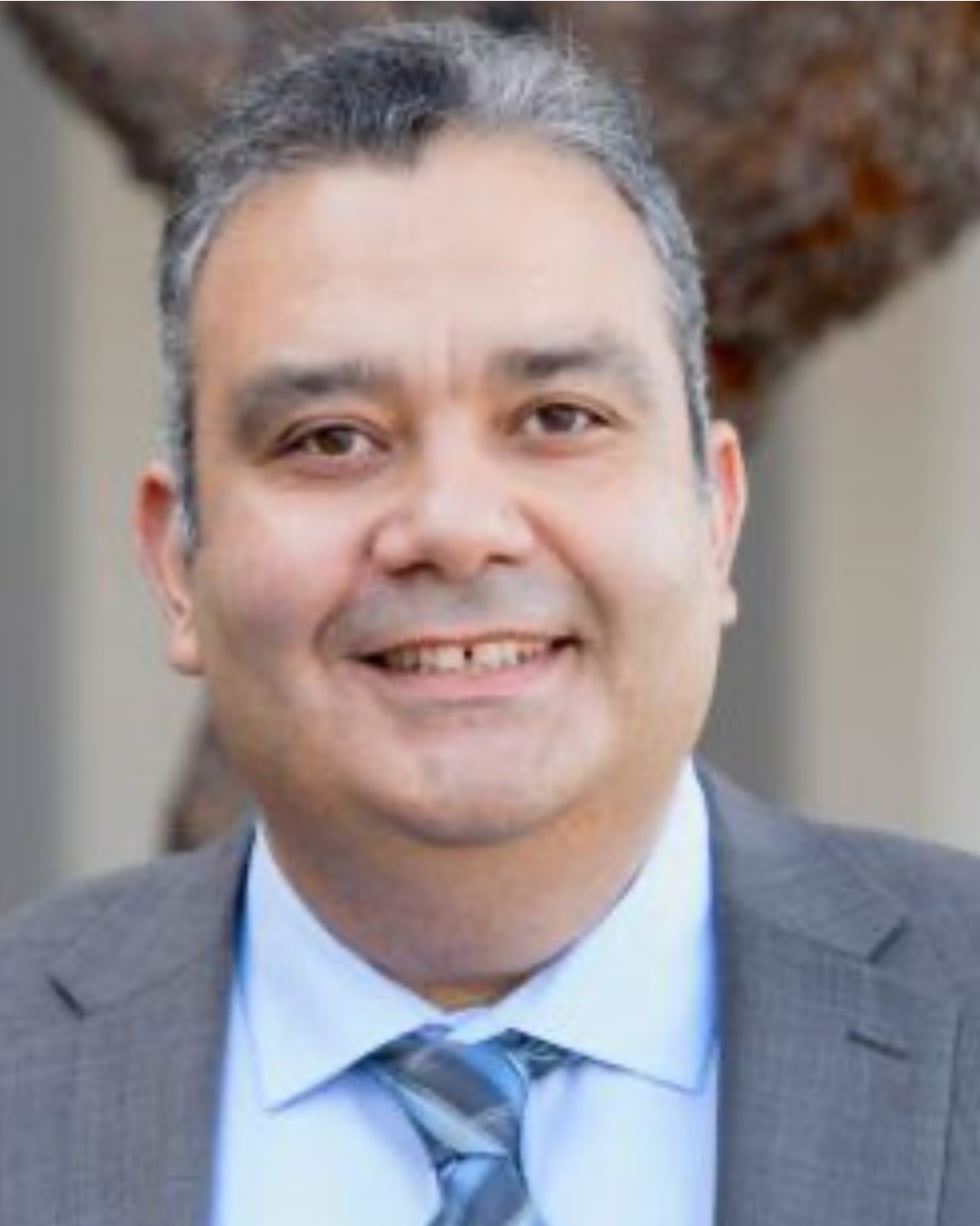}}]{AHMED M. ELTAWIL} (Senior Member, IEEE) received the M.Sc. and B.Sc. degrees (Hons.) from Cairo University, Giza, Egypt, in 1999 and 1997, respectively, and the Ph.D. degree from the University of California, Los Angeles, CA, USA, in 2003. Since 2019, he has been a Professor with the Computer, Electrical and Mathematical Science and Engineering Division (CEMSE), King Abdullah University of Science and Technology (KAUST), Thuwal, Saudi Arabia. Since 2005, he has been with the Department of Electrical Engineering and Computer Science,
University of California at Irvine, where he founded the Wireless Systems and Circuits Laboratory. His research interests are in the general area of low power digital circuit and signal processing architectures with an emphasis on mobile systems. He has been on the technical program committees and steering committees for numerous workshops, symposia, and conferences in the areas of low power computing and wireless communication system design. He received several awards, as well as distinguished grants, including the NSF CAREER Grant supporting his research in low power systems.
\end{IEEEbiography}	

\end{document}